\documentclass[twocolumn]{aastex6}

\usepackage{amssymb}
\usepackage{amsfonts}
\usepackage{amsmath}
\usepackage{natbib}
\usepackage{color}
\usepackage{graphicx}
\usepackage{hyperref}

\newcommand{\defeq}{\equiv}

\newcommand{\Especific}[0]{\mathcal{E}}
\newcommand{\Zspecific}[0]{\mathcal{Z}}
\newcommand{\ud}[0]{\mathrm{d}}
\newcommand{\vect}[1]{\boldsymbol{#1}}
\newcommand{\Lnu}[0]{L_{\nu,52}}

\shorttitle{Resolution Study with Prometheus-Vertex}
\shortauthors{T.~Melson, D.~Kresse, \& H.-T.~Janka}

\begin{document}

\title{Resolution Study for three-dimensional Supernova Simulations with the Prometheus-Vertex Code}

\author{Tobias Melson\altaffilmark{1}, Daniel Kresse\altaffilmark{1,2}, and
        Hans-Thomas Janka\altaffilmark{1}}
\altaffiltext{1}{Max-Planck-Institut f\"ur Astrophysik,
       Karl-Schwarzschild-Str.~1, 85748 Garching, Germany}
\altaffiltext{2}{Physik-Department, Technische Universit\"at M\"unchen, James-Franck-Str.~1, 85748 Garching, Germany}

\begin{abstract}
    We present a carefully designed, systematic study of the angular resolution
    dependence of simulations with the \textsc{Prometheus-Vertex} 
    neutrino-hydrodynamics code. Employing a simplified neutrino heating-cooling
    scheme in the \textsc{Prometheus} hydrodynamics module allows us to sample 
    the angular resolution between
    4$^\circ$ and 0.5$^\circ$. With a newly-implemented static mesh refinement
    (SMR) technique on the Yin-Yang grid, the angular coordinates can be refined
    in concentric shells, compensating for the diverging structure of the
    spherical grid. In contrast to previous studies with \textsc{Prometheus}
    and other codes, we find that higher angular resolution and therefore lower
    numerical viscosity provides more favorable explosion conditions and faster
    shock expansion. We discuss the possible reasons for the discrepant results. The
    overall dynamics seem to converge at a resolution of about $1^\circ$.
    Applying the SMR
    setup to marginally exploding progenitors is disadvantageous for the shock
    expansion, however, because kinetic energy of downflows is dissipated to
    internal energy at resolution interfaces, leading to a loss of turbulent
    pressure support and a steeper temperature gradient. We also present a way to estimate the
    numerical viscosity on grounds of the measured turbulent kinetic-energy spectrum,
    leading to smaller values that are better compatible with the flow behavior
    witnessed in our simulations than results following calculations in previous
    literature. Interestingly, the numerical Reynolds numbers in the turbulent, 
    neutrino-heated postshock layer
    (some 10 to several 100) are in the ballpark of expected
    neutrino-drag effects on the relevant length scales.
    We provide a formal derivation and quantitative assessment of the 
    neutrino drag terms in an appendix.
\end{abstract}


\section{Introduction}

The explosion mechanism of core-collapse supernovae, whether energized by
neutrino heating or mediated by magnetic fields, is a generically
multi-dimensional phenomenon, in which hydrodynamic instabilities play a crucial
role. They do not only foster the explosion but also create initial asymmetries
that determine the emerging geometry of the stellar blast. Three-dimensional
(3D) hydrodynamical simulations are therefore indispensable, and because of the
increasing power of modern massively parallel supercomputers they have recently
become possible in combination with elaborate energy-dependent three-flavor
neutrino transport \citep[for a review of recent developments,
see][]{Janka2016}.  Correspondingly, the pool of 3D stellar core-collapse and
explosion simulations with sub\-stan\-tial\-ly different treatments of neutrino
transport and physics is growing rapidly
\citep[e.g.,][]{Takiwaki2012,Takiwaki2014,Melson2015b,Melson2015a,Lentz2015,Roberts2016,Mueller2017,Mueller2019,Summa2018,Ott2018,OConnor2018,Kuroda2018,Glas2018,Vartanyan2019,Burrows2019}.

These simulations were conducted with different grid geometries (Cartesian grids
with static or adaptive mesh refinement, cubed-sphere multi-block grids, polar
coordinates, Yin-Yang spherical grids, spherical grids with polar mesh
coarsening) and with largely different numerical resolutions. While a number of
resolution studies have already been performed by varying, within rather close
limits, the mesh spacing for fixed grid geometries
\citep[e.g.,][]{Hanke2012,Couch2014,Abdikamalov2015,Roberts2016,Radice2016,Summa2018,OConnor2018},
possible numerical artifacts of the grid geometries themselves are still
completely unexplored in 3D supernova calculations.

The recognition that nonradial mass motions and buoyant bubble rise in the
neutrino-heating layer have an impact on the shock evolution that can be coined
in terms of turbulent (pressure, energy transport, dissipation) effects
\citep[e.g.,][]{Murphy2013,Couch2015,Mueller2015a,Radice2016,Mabanta2018}, has
led to increasing interest in the resolution dependence of the turbulent
kinetic-energy cascade for the postshock flow in supernova models
\citep{Abdikamalov2015,Radice2015,Radice2016,Radice2018}.  Low resolution can be
imagined to cause enhanced numerical viscosity, which might quench the growth of
convective buoyancy, damp nonradial flows, and lead to viscous dissipation of
kinetic energy and associated numerical heating.  In the regime where growth of
the standing accretion-shock instability
\citep[SASI;][]{Blondin2003,Blondin2007} rather than neutrino-driven convection
is favored \citep[see][]{Foglizzo2006,Foglizzo2007}, low radial resolution can
suppress the growth of SASI \citep{Sato2009}, whereas reduced angular resolution
can strengthen SASI activity because of weaker or absent parasitic
Kelvin-Helmholtz and Rayleigh-Taylor instabilities, which would redistribute
kinetic energy from the large SASI scales to vortex flows on smaller scales
\citep{Guilet2010}.  These expectations from theoretical and toy-model
considerations are in line with full-fledged supernova simulations
\citep{Summa2018}.

Moreover, \citet{Radice2015,Radice2016} diagnosed the so-called bottleneck
effect, where the lack of resolution prevents kinetic-energy cascading to small
scales and leads to an accumulation of kinetic energy on scales larger than the
dissipation scale. For this reason the turbulent energy spectra exhibit enhanced
power on these scales and display a more shallow decline than expected in the
inertial range from Kolmogorov's classical theory. The authors argued that this
circumstance might explain why in previous studies by \citet{Hanke2012} with the
\textsc{Prometheus} supernova code---and confirmed by others
\citep[e.g.,][]{Couch2014,Abdikamalov2015,Roberts2016} with different numerical
methods---lower resolution had fostered earlier and stronger explosions in 3D
simulations.

In the present work we report results of a re\-cent, carefully designed resolution
study with the \textsc{Prometheus-Vertex} code, sampling angular cell sizes
between $4^\circ$ and $0.5^\circ$ in full-$4\pi$ simulations of the post-bounce
evolution of collapsing and exploding 9\,$M_\odot$ and 20\,$M_\odot$ stars. We
employ full-fledged ray-by-ray neutrino transport with state-of-the-art neutrino
interactions, or, alternatively for systematic resolution variations, a simple
neutrino-heating and cooling scheme in the form of an improved version of the
treatment by \citet{Hanke2012}.

Our results reveal exactly the opposite resolution dependence compared to
previous investigations, name\-ly that better resolution leads to improved
explosion conditions and faster shock expansion in 3D.  Our results challenge
the interpretation of the previous findings as discussed by \citet{Couch2014},
\citet{Abdikamalov2015}, and \citet{Radice2016}. We understand the behavior
witnessed in our models as a consequence of the fact that lower numerical
viscosity in the case of higher resolution permits an enhanced level of
nonradial (turbulent) kinetic energy in the postshock layer. The conflict with
the simulations by \citet{Hanke2012} can be resolved when one takes into account
the numerical artifacts associated with the polar axis of the spherical
coordinate grid used in those older simulations. In contrast, the supernova
models generated in the present work employed the axis-free Yin-Yang grid
\citep{Kageyama2004,Wongwathanarat2010}. The Yin-Yang grid reduces numerical
artifacts to a much lower level by avoiding axis features as well as effects
caused by the nonuniform sizes of azimuthal grid cells at different latitudes,
which implies finer spatial resolution near the poles. The question arises why
Cartesian 3D simulations reproduced the polar-axis-triggered resolution trend
seen by \citet{Hanke2012}. We speculate, however without having any foundation
by own results, that numerical noise induced on radial flows by Cartesian grids
could play an important role.  A decreased level of such perturbations when the
resolution was improved, might have delayed the onset of explosions in the
better-resolved simulations performed by \citet{Couch2014},
\citet{Abdikamalov2015}, \citet{Roberts2016} and \citet{OConnor2018}.

We also discuss results obtained with a static mesh refinement (SMR) technique
that we implemented on the Yin-Yang grid of the \textsc{Prometheus-Vertex} code.
Despite offering enhanced angular resolution in the turbulent postshock layer,
it turned out to have an unfavorable influence on the onset and development of
explosions in cases that were marginally hitting success. We could trace this
back to a conversion of kinetic energy to thermal energy (with the sum of both
conserved) in mass flows crossing the borders from grid domains with better to
those with coarser resolution. This might suggest additional artifacts that
could be associated with the use of static or adaptive mesh refinement
procedures used in most applications of Cartesian grids.

Our resolution study indicates that convergence of the shock evolution
might be
approached at an angular resolution of about $1^\circ$.  An even more detailed
re\-al\-iza\-tion of the turbulent power spectrum than it is possible with this
resolution seems to have a minor impact on the overall post-bounce dynamics in
the supernova core, as most of the energy is contained on the largest scales.
Increasing the resolution beyond this point to accurately
follow the turbulent energy cascade needs to take into account the effects of
neutrino drag, because
even for our lowest-resolved simulations
neutrino-drag effects begin to compete with the consequences of numerical
viscosity on all scales that yield relevant contributions to the turbulent
kinetic energy.
We provide detailed estimates of the numerical viscosity and
Reynolds number (deduced from the numerically realized turbulent kinetic energy
spectrum) as well as, in an extended appendix, a detailed discussion of the
neutrino-drag effects in the gain layer behind the supernova shock.

Our paper is structured as follows. In Section~\ref{sec:numerics}, we briefly
describe the numerical setup of our models. The results of our 3D simulations
with \textsc{Vertex} neutrino transport are discussed in
Section~\ref{sec:models_transport}. Our systematic resolution study employing
the simple neutrino-heating and cooling scheme is presented in
Section~\ref{sec:models_htcl}. In Sec\-tion~\ref{sec:turbulence}, we focus on a
discussion of turbulence in our models and present our method of deducing the
numerical viscosity. An assessment of our results in a broader context is the
contents of Sect.~\ref{sec:discussion}, followed by our summary and conclusions
in Section~\ref{sec:conclusions}.  Appendix~\ref{sec:smr} provides a concise
description of our SMR technique, and Appendix~\ref{sec:drag} contains a
detailed derivation of the neutrino-drag terms with order-of-magnitude estimates
for supernova conditions
as well as an evaluation based on simulation results with full neutrino transport.

\section{Numerical setup}
\label{sec:numerics}

We performed one-dimensional (1D), two-dimensional (2D), and three-dimensional
(3D) core-collapse super\-nova simulations using two neutrino treatments. The
first set of models was computed with the \textsc{Prometheus-Vertex} code
\citep{Rampp2002,Buras2006a}, whereas the second model set was simulated using
only the hydrodynamics module \textsc{Prometheus} 
\citep{Fryxell1989,Fryxell1991,Mueller1991} with a
simplified heating-cooling (HTCL) scheme, based on an improved version of the
treatment applied by \citet{Hanke2012}. \textsc{Prometheus} is an implementation
of the Piecewise Parabolic Method \citep[PPM;][]{Colella1984}---a Godunov-type
scheme being second-order accurate in space and second-order in time.

In all simulations presented in this work, the collapse phase was computed in
spherical symmetry with the \textsc{Prometheus-Vertex} code using multi-group
neutrino transport and state-of-the-art neutrino interactions. Gravity was
treated in spherical symmetry with general-relativistic corrections \citep[Case
A]{Marek2006}. The high-density equation of state by \citet{Lattimer1991} with a
nuclear incompressibility of $K=220\,\mathrm{MeV}$ was employed.

For the present study, we selected various angular grid resolutions, while the
radial grid was kept unchanged for a given progenitor for comparison.  In
addition to 3D simulations with uniform angular resolution in the whole
computational domain, we also made use of a newly-implemented static mesh
refinement (SMR) procedure. It allows us to change the angular resolution in
radial layers of the spherical grid. A detailed description of this method can
be found in Appendix~\ref{sec:smr}.

\begin{table*}
    \caption{Overview of the simulations discussed in this work.}
\centering
\begin{tabular}{llllll}
\hline
\hline
    Model & Progenitor mass & Neutrino scheme & Dimensionality & Angular
    resolution & 1D core \\
\hline
    s9.0 & $9\,M_\odot$ & Full transport & 3D (Yin-Yang grid) & $3.5^\circ$ & 1.6\,km \\
     &  &  & 3D (Yin-Yang grid) & SMR (up to $0.5^\circ$) & 1.6\,km \\
\hline
    s20 & $20\,M_\odot$ & Full transport & 3D (Spherical polar grid) & $2^\circ$ & 10\,km \\
     &  &  & 3D (Yin-Yang grid)  & SMR (up to $0.5^\circ$) & 1.6\,km \\
\hline
    HTCL\_4.0 & $20\,M_\odot$ & HTCL with $\Lnu = 4$ & 1D & - & -\\
     &  &  & 2D & $0.5^\circ$ to $4^\circ$ & $10^{12}\,\mathrm{g\,cm^{-3}}$ \\
     &  &  & 3D (Yin-Yang grid) & $0.5^\circ$ to $4^\circ$ & $10^{12}\,\mathrm{g\,cm^{-3}}$ \\
     &  &  & 3D (Yin-Yang grid) & SMR (up to $0.5^\circ$) & $10^{12}\,\mathrm{g\,cm^{-3}}$ \\
\hline
    HTCL\_3.96 & $20\,M_\odot$ & HTCL with $\Lnu = 3.96$ & 1D & - &- \\
     &  &  & 2D & $0.5^\circ$ to $4^\circ$ & $10^{12}\,\mathrm{g\,cm^{-3}}$ \\
     &  &  & 3D (Yin-Yang grid) & $0.5^\circ$ to $4^\circ$ & $10^{12}\,\mathrm{g\,cm^{-3}}$ \\
     &  &  & 3D (Yin-Yang grid) & SMR (up to $0.5^\circ$) & $10^{12}\,\mathrm{g\,cm^{-3}}$ \\
\hline
\end{tabular}
\begin{minipage}{\textwidth}
	\tablecomments{
		For all models, we list the model name, the zero-age main sequence mass of the
		progenitor, the treatment of the neutrinos, the dimensionality including the
		grid setup (for 3D cases), the angular resolution, and the criterion (either
		radius or density) for the outer boundary of the spherically symmetric inner
		core employed for relaxing the time step constraints near the grid origin.}
\end{minipage}
\label{tab:models}
\end{table*}

In Table~\ref{tab:models}, we present an overview of all simulations discussed
in this work.

\subsection{Models with neutrino transport}

The first model set was computed using the \textsc{Prometheus-Vertex} code with
three-flavor, energy-dependent, ray-by-ray-plus neutrino transport including
state-of-the art neutrino interactions. We simulated the post-bounce evolution
of a $9\,M_\odot$ star \citep[s9.0;][]{Woosley2015} and a $20\,M_\odot$ progenitor
\citep[s20;][]{Woosley2007}. When mapping from 1D to 3D at 10\,ms after bounce,
random cell-to-cell density perturbations of 0.1\,\% were imposed to break
spherical symmetry.

Besides a setup with uniform angular resolution of $3.5^\circ$ and $2^\circ$,
respectively, we also used an SMR setup with a first refinement step from
$2^\circ$ to $1^\circ$ at the bottom of the gain layer.
To maintain a high resolution in the gain layer over time,
this SMR interface follows the contraction of the gain radius.
A second refinement re\-gion with a resolution of $0.5^\circ$ was
added at 70\,ms after bounce exterior to a fixed radius of 160\,km.
Thus, the resolution was improved twice by a factor of two.

The s20 simulation with a constant $2^\circ$ angular resolution was published
before in \citet{Melson2015b} and computed on a spherical polar grid with a core
of 10\,km treated spherically symmetrically. All other runs in this model set
were evolved on the Yin-Yang grid \citep{Kageyama2004,Wongwathanarat2010} 
with a smaller 1D core of
1.6\,km radius. The radial grid was gradually refined from 400 zones to more
than 600 to account for the steepening of the density gradient at the
proton-neutron star surface.

\subsection{Models with simplified heating-cooling scheme}
\label{sec:numericshtcl}

For the HTCL model set, the $20\,M_\odot$ progenitor from \citet{Woosley2007} was
selected. The multi-di\-men\-sion\-al simulations were started from a 1D collapse run
and mapped to 2D/3D at 15\,ms after bounce. At this stage, random cell-to-cell density
perturbations were imposed with an amplitude of 0.1\,\%.

The radial grid was set to 400 zones in all models and kept constant in time.
Since the proto-neutron star remains larger in the HTCL simulations and no
transport is applied, this radial grid yields sufficient resolution, compatible with what is used in \textsc{Prometheus-Vertex} for neutron stars of similar radius. Note that the simplified HTCL scheme is not subject to the same resolution constraints as detailed neutrino transport. Moreover, the purpose of our simulations with the HTCL treatment is not a most accurate description of the neutron star and its near-surface layers, but instead the goal is a study of turbulence and resolution effects in the postshock domain.

The central
region enclosed by the density contour of $10^{12}\,\mathrm{g\,cm^{-3}}$ was
treated in spherical symmetry, corresponding to a radius of 42--46\,km. This
radius is considerably larger than in the simulations with full-fledged neutrino
transport, because the simplified HTCL scheme weakens the contraction of the
proto-neutron star. We will comment on the consequences of this fact at several
places in the discussion of our results.

All 3D simulations in this model set were computed on the Yin-Yang grid. Besides
the runs with a constant angular resolution in the entire computational domain,
we used the SMR grid with a resolution of $2^\circ$ up to a radius of 123\,km
and $1^\circ$ outside. At 150\,ms after bounce, an additional layer of $0.5^\circ$
angular resolution was added outside a radius of 162\,km, thus improving the resolution a second time by a factor of two. Note that the imposed initial
perturbation patterns were identical in the SMR case and the $2^\circ$ run.

The HTCL scheme was used already in a 3D study by \citet{Hanke2012}. However,
our implementation of this scheme and other aspects of our presented study
differ in some details. First, \citet{Hanke2012} used a spherical polar grid
instead of the Yin-Yang grid. Second, they employed a Newtonian gravitational
potential without general-relativistic corrections.  Third, the high-density
equation of state of \citet{Lattimer1991} with $K=180\,\mathrm{MeV}$ was used in
their work. Also our formulation of the heating and cooling terms was modified
compared to theirs. Our heating and cooling terms read, respectively,
\begin{equation}
\begin{split}
    \dot{q}_\mathrm{heat} = & 1.544\times 10^{20} \left( \frac{L_\nu}{10^{52}
    \,\mathrm{erg\,s^{-1}}} \right) \left(\frac{100\,\mathrm{km}}{r} \right)^2 \\
    &\times \left( \frac{T_\nu}{4\,\mathrm{MeV}} \right)^2 \left( Y_n + Y_p \right) e^{-\tau_\mathrm{eff} / 2}
    \,\mathrm{erg\,g^{-1}\,s^{-1}},
\end{split}
\end{equation}
and
\begin{equation}
\begin{split}
    \dot{q}_\mathrm{cool} = & 1.399\times 10^{20} \left( \frac{T}{2\,\mathrm{MeV}}
    \right)^6 \left( Y_n + Y_p \right) \\
    &\times e^{-\tau_\mathrm{eff} / 2}\,\mathrm{erg\,g^{-1}\,s^{-1}},
\end{split}
\end{equation}
where $r$ is the radius and $T$ is the temperature. $Y_n$ and $Y_p$ are the
neutron and proton number fractions, respectively. $\Lnu \defeq L_\nu /
(10^{52}\,\mathrm{erg\,s^{-1}})$ is a free parameter with two different values
of 3.96 and 4.0 in this work. The neutrino temperature $T_\nu$ is set to 4\,MeV.
The exponential term $\tau_\mathrm{eff}$ is given by
\begin{equation}
    \tau_\mathrm{eff}(r) = \int_r^\infty \ud r'\, \kappa_\mathrm{eff}(r'),
\end{equation}
with
\begin{equation}
    \kappa_\mathrm{eff} = 7.5 \times 10^{-8} \left(
    \frac{\rho}{10^{10}\,\mathrm{g\,cm^{-3}}} \right)\left( \frac{T_\nu}{4\,\mathrm{MeV}} \right)^2\,\mathrm{cm^{-1}},
\end{equation}
where $\rho$ is the mass density.

\begin{figure*}
	\centering
	\includegraphics[width=0.95\linewidth]{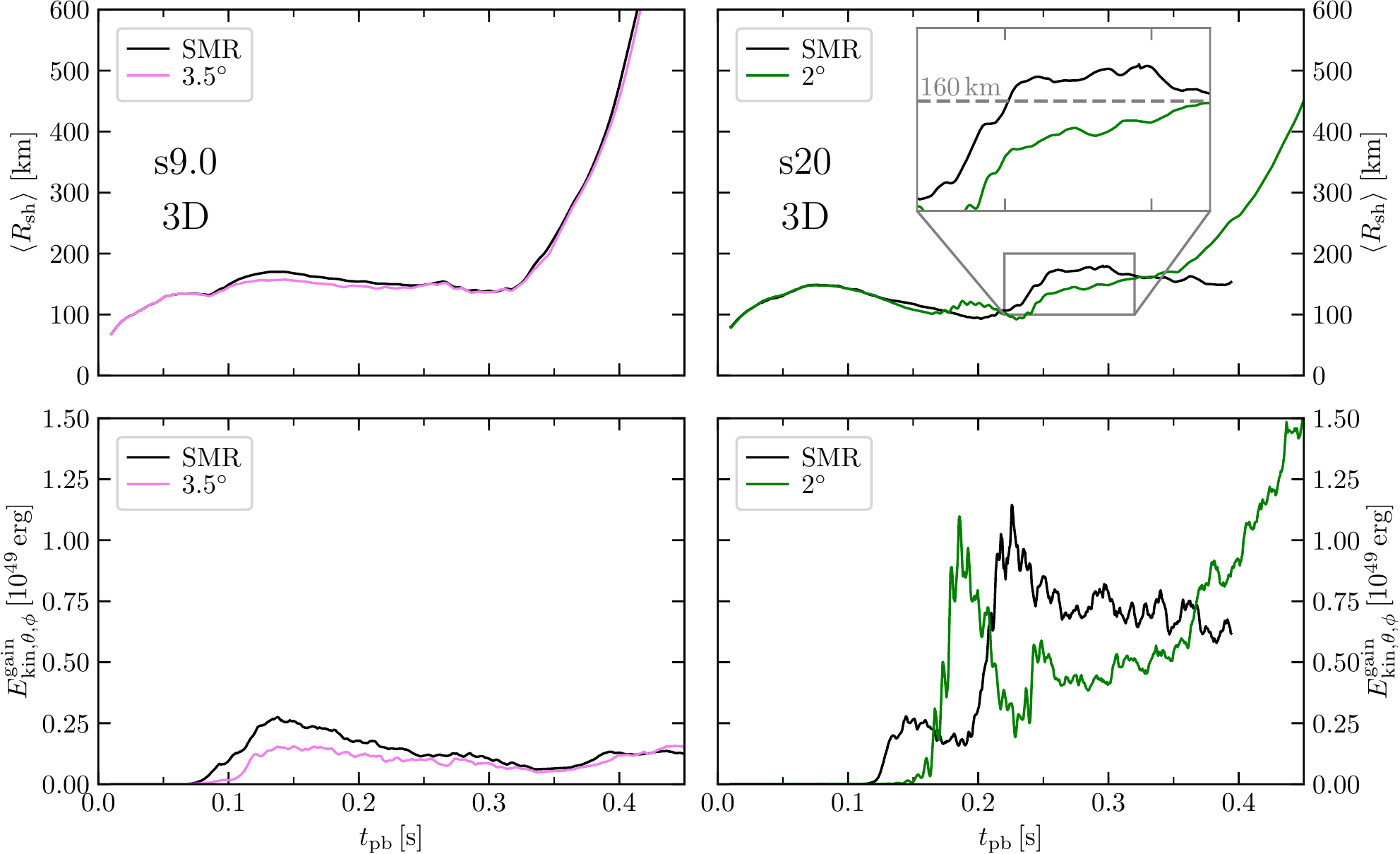}
	\caption{
		Post-bounce evolution of the 3D models computed with full neutrino transport,
		s9.0 (\emph{left}) and s20 (\emph{right}), for different angular grid
		resolutions. \emph{Top:} Angle-averaged shock radius,
		$\left<R_\mathrm{sh}\right>$.
		The gray dashed line in the zoom inset of the right panel indicates the location of the outer SMR refinement interface.
		\emph{Bottom:} Nonradial kinetic energy in
		the gain layer, $E^\mathrm{gain}_{\mathrm{kin},\theta,\phi}$.
	}
	\label{fig:s9s20_shock}
\end{figure*}

Because of the combination of all 
differences, the parameter $\Lnu$ in our scheme is about
a factor of four higher than in the work by \citet{Hanke2012} in order to
trigger shock revival.

\newpage
\section{Models with neutrino transport}
\label{sec:models_transport}

In Fig.~\ref{fig:s9s20_shock}, we show the angle-averaged shock radii as
functions of post-bounce time for the two progenitors evolved with full neutrino
transport. In the s9.0 case, the temporal evolution of the shock remains nearly
unaffected by a change of the angular resolution from a uniform $3.5^\circ$ grid
to the SMR setup. Only between $\sim$100\,ms and $\sim$250\,ms after bounce, the
shock in the SMR case has a slightly larger radius. However, the time of shock
revival and also the shock expansion velocity are nearly identical in both
setups. The reason for this is the robustness of the explosion in the s9.0
model. It is well beyond the critical explosion threshold, because the
mass-accretion rate in this low-mass progenitor decreases rapidly at the
silicon/silicon+oxygen interface leading to a significant drop in the ram
pressure at the shock.

In the s20 model, the simulation with a uniform grid behaves entirely
differently from the SMR case. The former explodes, while the latter does not
experience shock revival until we stopped the simulation at
400\,ms after core bounce. Between 200 and
300\,ms, the SMR model seems to have more favorable explosion conditions because
of a larger shock radius. Also the nonradial kinetic energy in the gain layer,
defined by
\begin{equation}
    E_{\mathrm{kin},\theta,\phi}^\mathrm{gain} \defeq \int_{R_\mathrm{gain} < r
    < R_\mathrm{sh}(\theta,\phi)} \ud V \, \frac{1}{2} \rho \left( v_\theta^2 + v_\phi^2 \right)
\end{equation}
and shown in the lower row of Fig.~\ref{fig:s9s20_shock}, is much higher during
this time interval, because of a strong SASI spiral mode being present in the
SMR model. $R_\mathrm{gain}$ is the angle-averaged gain radius.
At about 350\,ms, however, this picture changes and the simulation
with a fixed resolution of $2^\circ$ explodes, whereas the kinetic energy in the
gain layer decreases continuously in the SMR case.

In contrast, the s9.0 simulations do not differ much in their lateral kinetic
energies in the gain layer at the time when the explosions set it. Although the
SMR model develops higher values transiently, convective overturn becomes
similar in both simulations after 250\,ms.

\begin{figure}
    \centering
    \includegraphics[width=0.95\linewidth]{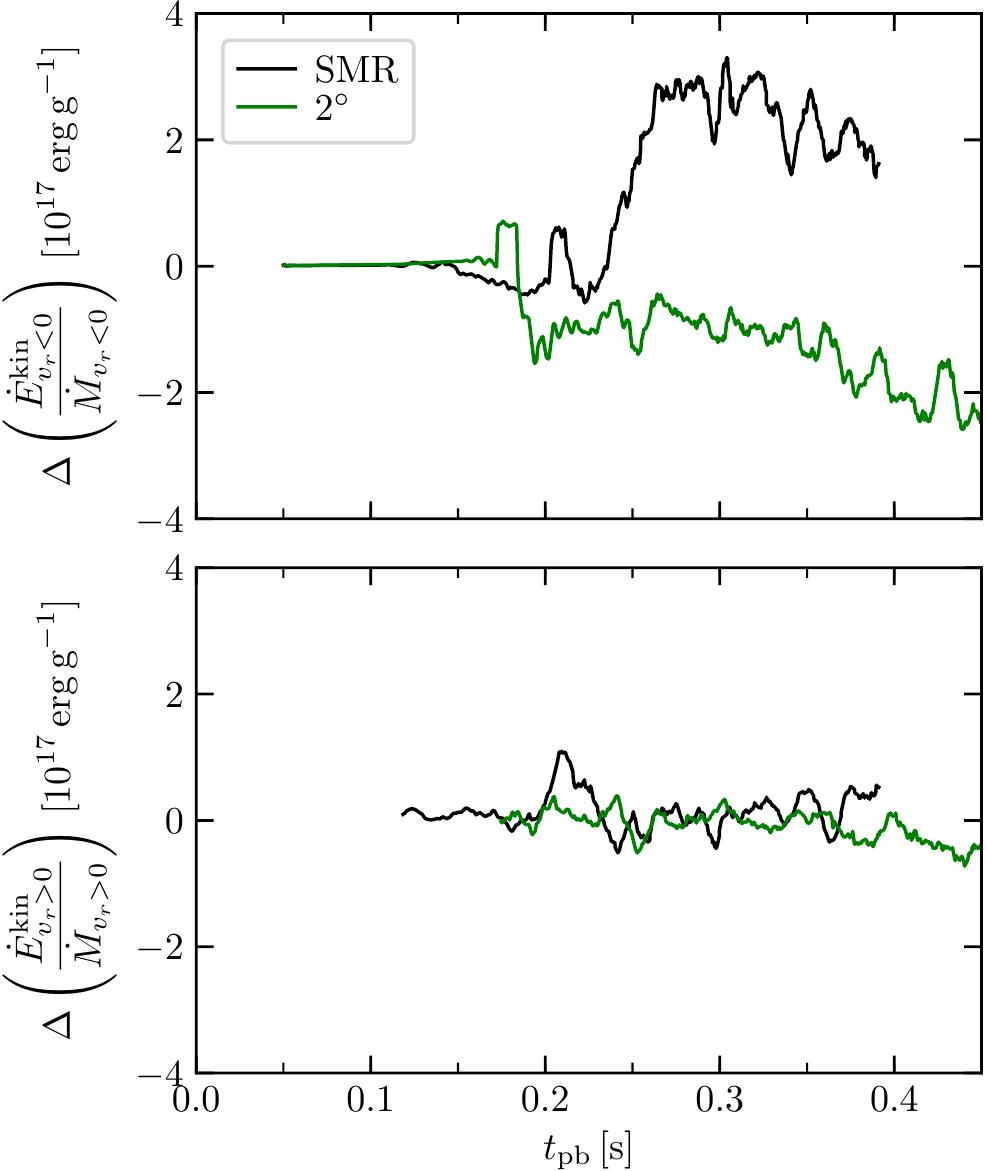}
    \caption{Difference of the radial kinetic energy flux divided by the radial
    mass flux as a function of post-bounce time, $t_\mathrm{pb}$, for the s20
    models computed with full neutrino transport. The values at a radius 1\,km
    below the inner SMR resolution interface are subtracted from the values at
	a radius 1\,km above it, thus probing the flux conservation at this
	interface, which is moved inward from initially 105\,km to 64\,km during
	the simulation. At this radius, the angular resolution in\-creas\-es outward
	from $2^\circ$ to $1^\circ$ in the SMR run. Note
    that both the minuend and the subtrahend are always positive.  The plotted
    values are time-averaged with a window of 10\,ms. We differentiate between
    inflows (\emph{top}) and outflows (\emph{bottom}).
	}
    \label{fig:kineticenergyflux_s20}
\end{figure}

In order to understand why the SMR setup with its angular resolution of
$1^\circ$ in the gain layer and even $0.5^\circ$ above 160\,km prevents shock
revival in the s20 model despite the higher nonradial kinetic energy over a
period of 200\,ms, we investigate the dissipation of kinetic energy at the
interfaces between resolution layers.
We speculate that this effect might be crucial especially in regions close
to the gain radius, where neutrino heating is strongest and its interplay with the
turbulent flow dynamics most pronounced.

In Fig.~\ref{fig:kineticenergyflux_s20}, we thus show differences of the radial
fluxes of kinetic energy
across the inner SMR interface (which is located at the bottom of the gain layer)
for inflowing and outflowing material for the s20
models. The kinetic energy fluxes are computed according to
\begin{align}
    \dot{E}^\mathrm{kin}_{v_r<0} &\defeq r^2 \int \ud \Omega \, \Theta(-v_r) \, \frac{1}{2} \rho \left( v_r^2 + v_\theta^2 + v_\phi^2 \right) v_r, \\
    \dot{E}^\mathrm{kin}_{v_r>0} &\defeq r^2 \int \ud \Omega \, \Theta(v_r) \, \frac{1}{2} \rho \left( v_r^2 + v_\theta^2 + v_\phi^2 \right) v_r,
\end{align}
and the mass fluxes are given by
\begin{align}
	\dot{M}_{v_r<0} &\defeq r^2 \int \ud \Omega \, \Theta(-v_r) \, \rho v_r, \\
    \dot{M}_{v_r>0} &\defeq r^2 \int \ud \Omega \, \Theta(v_r) \, \rho v_r,
\end{align}
where $\Theta(x)$ is the Heaviside step function. Both quan\-ti\-ties are evaluated
as integrals over inflows ($v_r<0$) and outflows ($v_r>0$) separately. The
differences are calculated by subtracting the specific kinetic energy fluxes at a radius 1\,km below the inner SMR resolution interface, $r_1$,
from their values at a radius 1\,km above it, $r_2$. For outflows, for example, we get
\begin{equation}
    \Delta \left(\frac{\dot{E}^\mathrm{kin}_{v_r>0}}{\dot{M}_{v_r>0}}\right) =
    \frac{\dot{E}^\mathrm{kin}_{v_r>0}(r_2)}{\dot{M}_{v_r>0}(r_2)} -
    \frac{\dot{E}^\mathrm{kin}_{v_r>0}(r_1)}{\dot{M}_{v_r>0}(r_1)}\,.
    \label{eq:deltaflux}
\end{equation}
This allows us to
investigate the flux conservation at this resolution interface, which is moved inward during the simulation from initially 105\,km down to 64\,km,
roughly following the contraction of the gain radius. The
individual terms in Eq.~(\ref{eq:deltaflux}) are always positive at both radii so that a positive
difference represents a higher flux at the outer radius for both flow directions.

\begin{figure}
    \centering
    \includegraphics[width=0.95\linewidth]{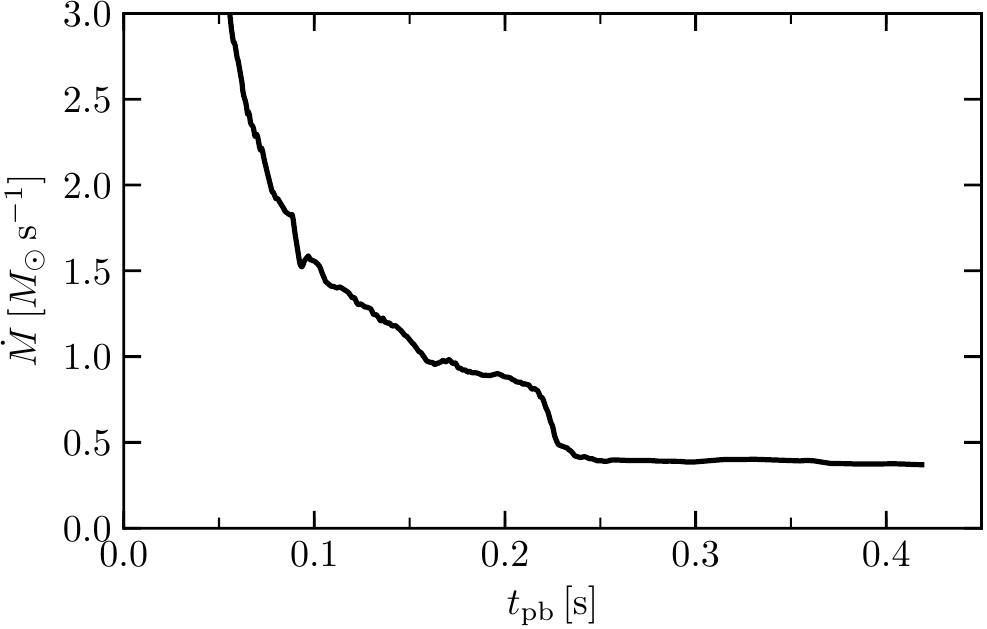}
    \caption{Mass-accretion rate, $\dot{M}$, of the s20 progenitor as a function
    of post-bounce time, $t_\mathrm{pb}$, measured at a radius of
    $400\,\mathrm{km}$.}
    \label{fig:mdot}
\end{figure}

On the SMR grid, matter flowing inwards is passing from the region with an
angular resolution of $1^\circ$ to the layer with a grid spacing of $2^\circ$
at the inner SMR interface. We wonder whether dissipation of kinetic energy occurs as a consequence of the averaging over neighboring grid cells
when the grid resolution decreases (see Appendix~\ref{sec:smr}).
Especially, investigating this effect at the
arrival time of the silicon/silicon+oxygen interface is important, because this
is the crucial phase for shock revival. At that time, the preshock mass-accretion rate
defined (at 400\,km) by
\begin{equation}
    \dot{M} \defeq -r^2 \int \ud \Omega \, \rho v_r
\end{equation}
and thus also the ram pressure at the shock drops significantly, which happens
at about 230\,ms in the s20 progenitor (cf.\ Fig.~\ref{fig:mdot}). Until about
that time,
the energy differences are very similar in both models (Fig.~\ref{fig:kineticenergyflux_s20},
upper panel).
But shortly after that, at about 240\,ms, the difference in the specific kinetic energy fluxes as displayed in the top panel of
Fig.~\ref{fig:kineticenergyflux_s20}
starts to rise steeply in the SMR
simulation.
In contrast, it remains even below zero in
the model with uniform $2^\circ$ resolution, meaning that the absolute
value of the inward flux of kinetic energy per unit of mass at
the inner radius, $r_1$, is higher
than at the outer radius, $r_2$.
The negative value points to a gravitational acceleration of the inward flow.
In the SMR model, the steep rise of the energy difference to positive values implies that
kinetic energy is dissipated in this case at the 
resolution interface, which is crossed by matter flow into the coarser-resolved region.
Such a dissipation of kinetic energy does not happen in the non-SMR case.

In the bottom panel of Fig.~\ref{fig:kineticenergyflux_s20}, we present the same
analysis for outflowing material. Because matter is propagating into the
finer-resolved layer in the SMR model, we do not expect kinetic energy to be
dissipated into thermal energy. Indeed, the difference of the specific kinetic
energy fluxes around the resolution interface
remains close to zero for outflowing material,
both in the SMR and in the $2^\circ$ simulation.

Analogously, dissipation of kinetic energy (of down\-flows) should also occur
at the outer resolution interface of the SMR model, which is located at a fixed radius
of 160\,km. However, performing the same analysis at this position is complicated by
the presence of the deformed shock over an extended period of time.\footnote{In
	the SMR run, the maximum shock radius crosses the outer resolution interface at
	around 230\,ms, while the minimum shock radius never reaches 160\,km. In the
	simulation with uniform $2^\circ$ resolution, the deformed shock is passing the
	radius of 160\,km between 240\,ms and 380\,ms after bounce.}
The shock de\-cel\-er\-ates the radially infalling pre-shock material and thus accounts
for most of the reduction in $\dot{E}^\mathrm{kin}_{v_r<0}/\dot{M}_{v_r<0}$ during this
phase, covering the dissipation effect of the grid geometry.\footnote{On the
	contrary, no dissipation of kinetic energy is expected at the outer resolution
	interface as long as the shock did not yet reach it. In regions where pre-shock
	matter is still radially infalling, the flow geometry is spherically symmetric to
	first order, and therefore a change in the angular resolution should not matter.}
Nevertheless, it is sus\-pi\-cious that the average shock radius in the SMR run
stagnates just after the shock has passed the interface at 160\,km (see zoom inset in
Fig.~\ref{fig:s9s20_shock}). This possibly points towards the dissipation effect
associated with the SMR grid.

The question remains, why the dissipation of kinetic energy in the SMR
simulation of the s20 model hampers shock revival. \citet{Radice2015} discussed
the effects of thermal and turbulent pressure contributions in comparison. With
the energy density $e$ and the adiabatic index $\gamma$, the pressure is given
by $p=(\gamma-1) e$. For the thermal pressure in the radiation ($e^\pm$ pairs
and photons) dominated postshock layer, $\gamma \approx 4/3$, whereas for
anisotropic turbulence as characteristic for the postshock layer, one has
$\gamma = 2$. When turbulent kinetic energy is dissipated into thermal energy,
the pressure contribution therefore decreases per unit of energy density. For this
reason, energy dissipation at SMR resolution interfaces reduces the pressure
support behind the shock and possibly prohibits shock revival.

\section{Models with simplified heating-cooling scheme}
\label{sec:models_htcl}

As described in the previous section, the usage of the SMR grid impeded the
revival of the shock in the s20 model, although the angular grid resolution was
effectively enhanced with this setup. It is therefore crucial to disentangle the
influence of a higher uniform grid resolution from the possible detrimental
effects of the SMR procedure. We thus performed simulations of the s20
progenitor with a simplified heating-cool\-ing (HTCL) scheme replacing
the computing-intense neutrino transport.

\begin{figure*}
    \centering
    \includegraphics[width=0.95\linewidth]{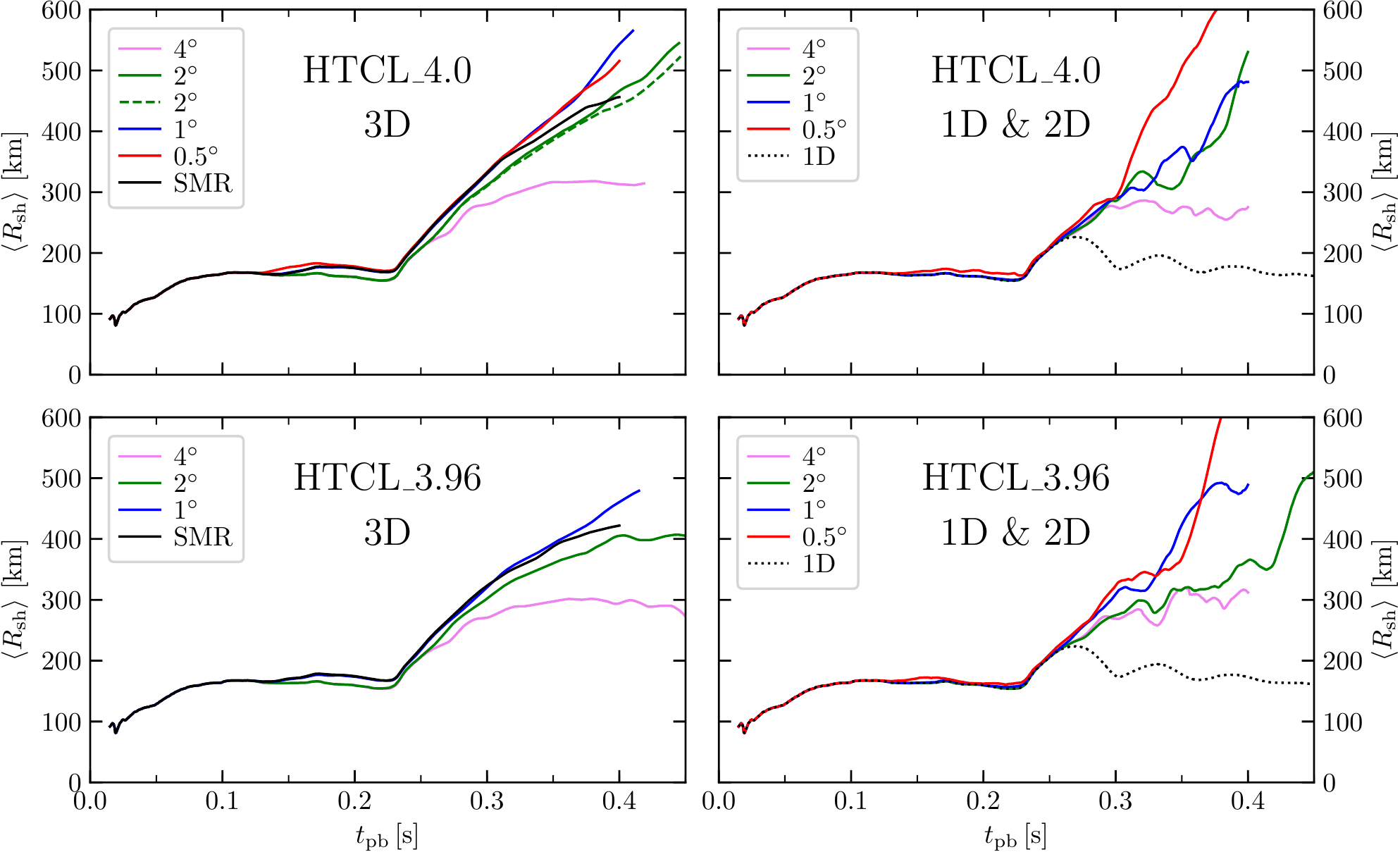}
    \caption{Angle-averaged shock radii of the HTCL models as a function of time
    with parameter choices $\Lnu=4$ (\emph{top}) and $\Lnu=3.96$
    (\emph{bottom}) in 3D (\emph{left}), as well as 1D and 2D (\emph{right}).
    The 3D model with $\Lnu=4$ and an angular resolution of $2^\circ$ was
    repeated twice with different initial random perturbation patterns.
    The model sequences for $\Lnu=3.96$ exhibit a stronger sensitivity and wider
    spread in dependence on the angular resolution (in particular for the 
    2$^\circ$ cases), because the runs are closer to the explosion threshold.}
    \label{fig:shock_HTCL}
\end{figure*}

In Fig.~\ref{fig:shock_HTCL}, we present the angle-averaged shock radii of the
entire model set. We selected two different choices for the HTCL parameter
$\Lnu$, 3.96 and 4.0, to control the tendency to get an explosion, namely the
strength of neutrino heating to overcome the ram pressure of infalling material.

In almost all simulations of this set, shock revival occurs at the arrival time of
the silicon/silicon+oxygen interface (cf.\ Fig.~\ref{fig:mdot}). Only the
simulations with a very coarse angular resolution of $4^\circ$ and the 1D models
do not explode. The latter show the well-known oscillating behavior of the shock
radius during the shock stagnation phase. It needs to be pointed out that the
simple cooling prescription reduces lepton-number and energy losses by the
proto-neutron star and thus weakens its contraction. This, in turn, allows for a
larger shock-stagnation radius than in our full-fledged supernova simulations and
disfavors SASI activity, in particular in 3D.

The shock expansion velocity at $t_\mathrm{pb} \gtrsim 230\,\mathrm{ms}$ is
basically a monotonic function of the angular resolution.  Higher angular
resolution accelerates the propagation of the shock after its revival.  This is
true both in 2D and 3D, although the angle-averaged shock in 2D propagates in a
more oscillatory manner. Note that the presence of the symmetry axis in the 2D
models collimates the flow along this axis and enhances the tendency for
shock-sloshing motions \citep[see, e.g.,][]{Glas2018}. The axial symmetry
fosters shock expansion predominantly along the axis, leading to a prolate shape
of the shock surface, and because of the importance of shock-sloshing 
motions it leads to large statistical variations of the
angle-averaged shock radius in 2D.

Both simulation sets with $\Lnu=4$ and $\Lnu=3.96$ show the same behavior and
resolution dependence. Explosions in the latter runs are weaker with a lower
shock expansion velocity. In the following discussion, we will focus on the
$\Lnu=4$ model set, because we were able to perform a simulation with a uniform
resolution of $0.5^\circ$ for this choice of the heating parameter, which was
not possible for $\Lnu=3.96$ due to computing time limitations.

In the 3D models with $1^\circ$ and $0.5^\circ$ resolution of the $\Lnu=4$ model
set, the shock trajectories behave very similarly until 370\,ms. The difference
between these two cases is much smaller than relative deviations between any other 
simulations.
We recommend not to overinterpret the difference of the shock trajectories in
Fig.~\ref{fig:shock_HTCL} between the $1^\circ$ and $0.5^\circ$ simulations
after 370\,ms. This
difference may be a transient feature connected to the faster rise of a buoyant bubble,
which would be a stochastic phenomenon that can change from model to model. To clarify
this issue, however, the simulations would have to be continued to later times. For
these two cases, we are therefore tempted to conclude that the overall dynamics
in 3D converge at about $1^\circ$ angular resolution, in particular because most
of the kinetic energy is contained in the turbulent flow on the largest scales. However,
a final confirmation of convergence would require simulations with increased radial
resolution and significantly better angular resolution than $0.5^\circ$.

The 3D SMR simulations follow their corresponding highest-resolved uniformly
gridded counterparts for a long time. Only after about 300\,ms for $\Lnu=4$ and
350\,ms for $\Lnu=3.96$, the shock velocity decreases. The reason for this
behavior is the dissipation of kinetic energy at resolution interfaces,
similarly to our findings for the model set with full neutrino transport
discussed above. We will analyze this effect in more detail later.

To prove that our analysis does not suffer from stochastical variations in 3D,
we repeated the $2^\circ$ model in the $\Lnu=4$ set with a different random
cell-to-cell perturbation pattern. Until more than 350\,ms, the shock
trajectories of the two $2^\circ$ simulations remain nearly identical. Only in
the very last phase, they start to differ slightly. The resolution trend we see
in our models is therefore unlikely to be simply a manifestation of stochastical
differences.

\begin{figure*}
    \centering
    \includegraphics[width=0.9\linewidth]{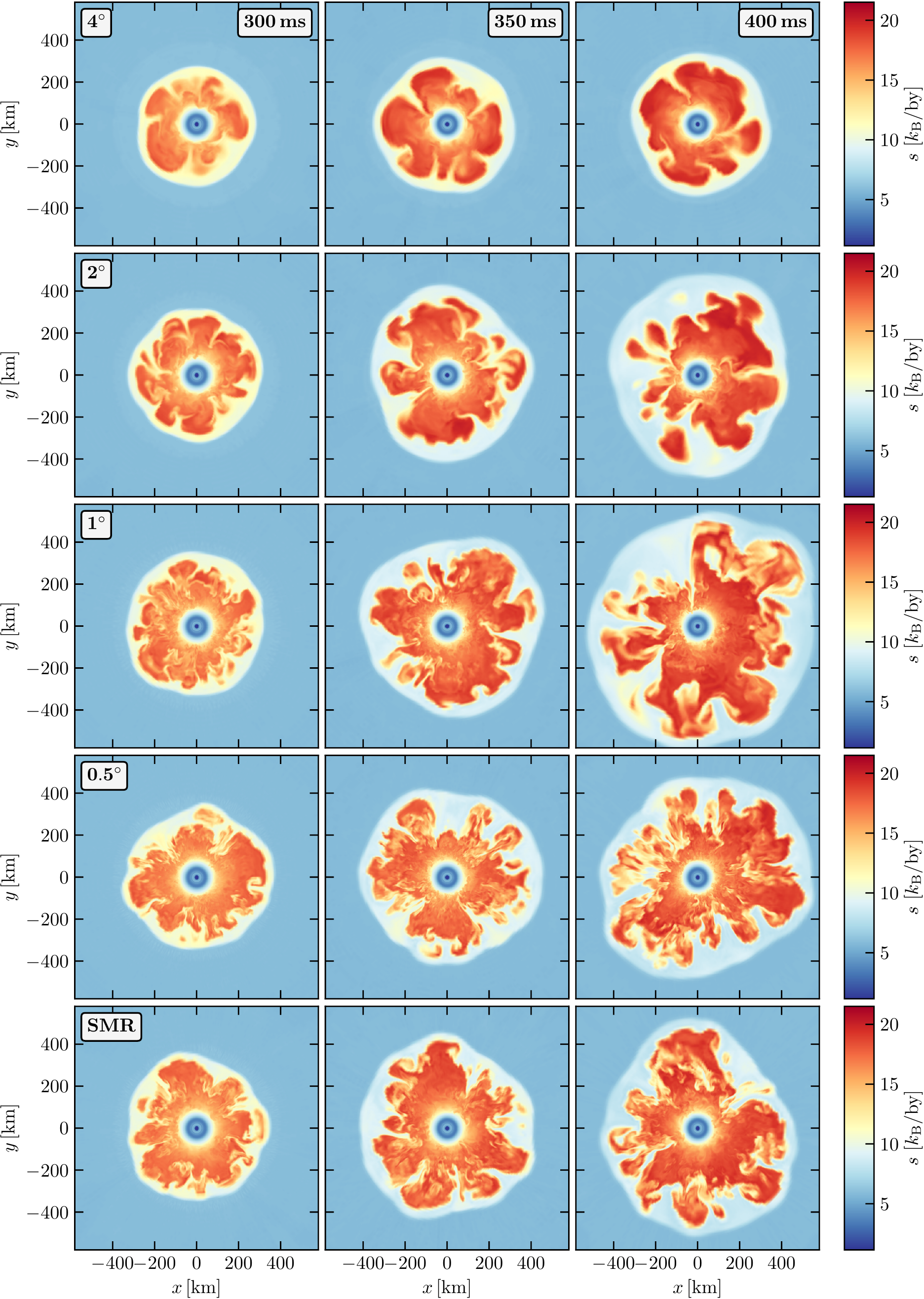}
    \caption{Cross-sectional cuts through the 3D HTCL models of the $\Lnu=4$ model
    set at 300\,ms, 350\,ms, and 400\,ms after bounce with the color-coded entropy per baryon, $s$.}
    \label{fig:cross_sto}
\end{figure*}

\begin{figure*}
	\centering
	\includegraphics[width=0.9\linewidth]{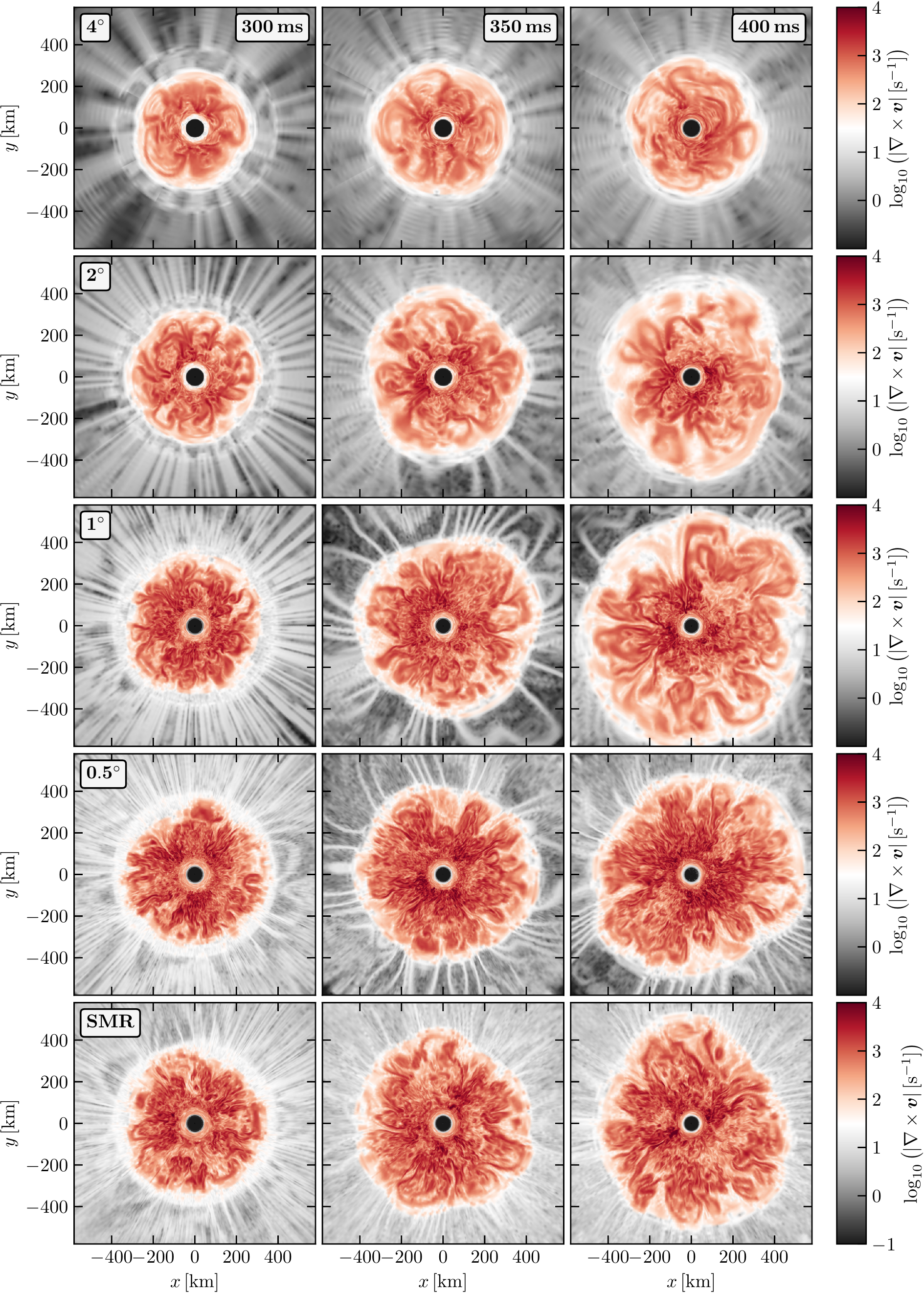}
	\caption{Same as Fig.~\ref{fig:cross_sto}, but with the color-coded
		vorticity, $|\nabla \times \vect{v}|$.}
	\label{fig:vorticity}
\end{figure*}

In Fig.~\ref{fig:cross_sto}, we show cross-sectional slices through the 3D
models with $\Lnu=4$ at 300\,ms, 350\,ms, and 400\,ms after bounce.
All models are dominated by
convection and do not develop visible SASI activity, because the shock does not retreat
far enough during its stagnation phase to provide suitable conditions for SASI
growth.

The color-coded entropy clearly shows that the vortex structures become finer
with increasing angular resolution, and downflows develop smaller-structured
Kelvin-Helmholtz flow patterns. Especially a comparison of the $0.5^\circ$
model with the SMR case does not reveal any noticeable difference. The SMR grid
setup seems to provide enough resolution where necessary to allow for the small
structures to develop.

In all simulations, we observe clearly separated high-entropy plumes. This is in
contrast to the work by \citet{Radice2016}, who argued that high-entropy bubbles
are embedded in a low-entropy surrounding medium only if the resolution is too
low, and that these bubbles should instead form diffuse ``clouds''. In our
models, we also see that the laminar layer behind the shock surface is similarly
structured in all cases and its thickness does not depend on the angular
resolution.

The vorticity, $|\nabla \times \vect{v}|$, shown in Fig.~\ref{fig:vorticity} depicts tur\-bulence in the gain layer, with typical values of $\sim10^2-10^4\,\mathrm{s^{-1}}$ (red colors). With increasing angular resolution, the volume filled with small-scale
turbulent eddies grows.
The magnitude of the vorticity, however, does not depend strongly
on the resolution.
Again, a clear difference between the highest-resolved
uniform grid of $0.5^\circ$ and the SMR setup cannot be spotted.
The radially infalling material ahead of the shock has significantly smaller
values of the vorticity compared to the neutrino-heated postshock matter (grey colors).
The filament-like structures in this region are a consequence of the
random density perturbations of 0.1\,\% amplitude, which are imposed in the whole computational volume at 15\,ms after bounce to break the spherical symmetry of the progenitor model.

\begin{figure}
	\centering
	\includegraphics[width=0.95\linewidth]{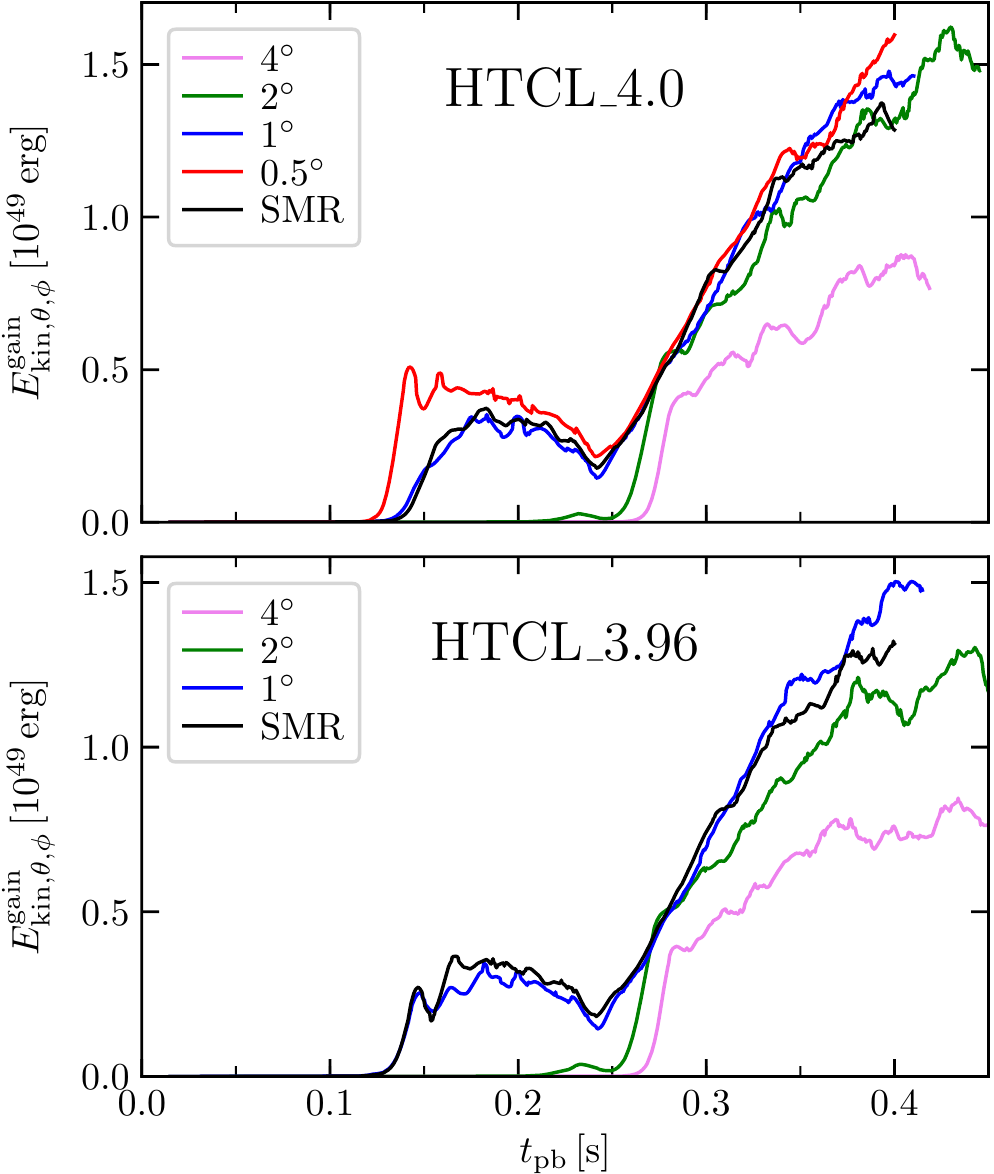}
	\caption{Nonradial kinetic energy in the gain layer as a function of time
		for the two HTCL model sets with $\Lnu=4$ (\emph{top}) and $\Lnu=3.96$
		(\emph{bottom}).}
	\label{fig:ekingain}
\end{figure}

At around 150\,ms, the higher-resolved 3D models experience a phase of slight
shock expansion by about 15\,km on average. This is due to convection in the
neutrino-heating layer, which gains strength at this time. 
The nonradial kinetic energy in
the gain layer plotted in Fig.~\ref{fig:ekingain} shows that postshock
convection starts early at $\sim$120\,ms in the models with an angular
resolution of at least $1^\circ$. In lower-resolved cases, this occurs about
100\,ms later. The onset of turbulent convection depends on the angular
resolution, because low resolution corresponds to a higher numerical viscosity
that dampens the rise of buoyant bubbles. In this context, the SMR models behave
similarly to the cases with $1^\circ$ angular resolution, because this is
precisely the resolution of the gain layer during the shock stagnation phase in
the SMR setup. Note that our simulations do not develop strong SASI activity,
because the shock radius does not retreat, disfavoring SASI growth. This is
another reason why the lower-resolution models do not develop postshock
turbulence before the shock expands after the passage of the
silicon/silicon+oxygen interface.

After shock revival, there remains a less pronounced 
dependence of the lateral kinetic energy
on the angular resolution, except for the lowest-resolved models of $4^\circ$,
which falls clearly behind the others.
This relative insensitivity to the resolution is compatible with the fact that
most of the kinetic energy is contained by vortex flows on the largest scales.

\begin{figure}
    \centering
    \includegraphics[width=0.95\linewidth]{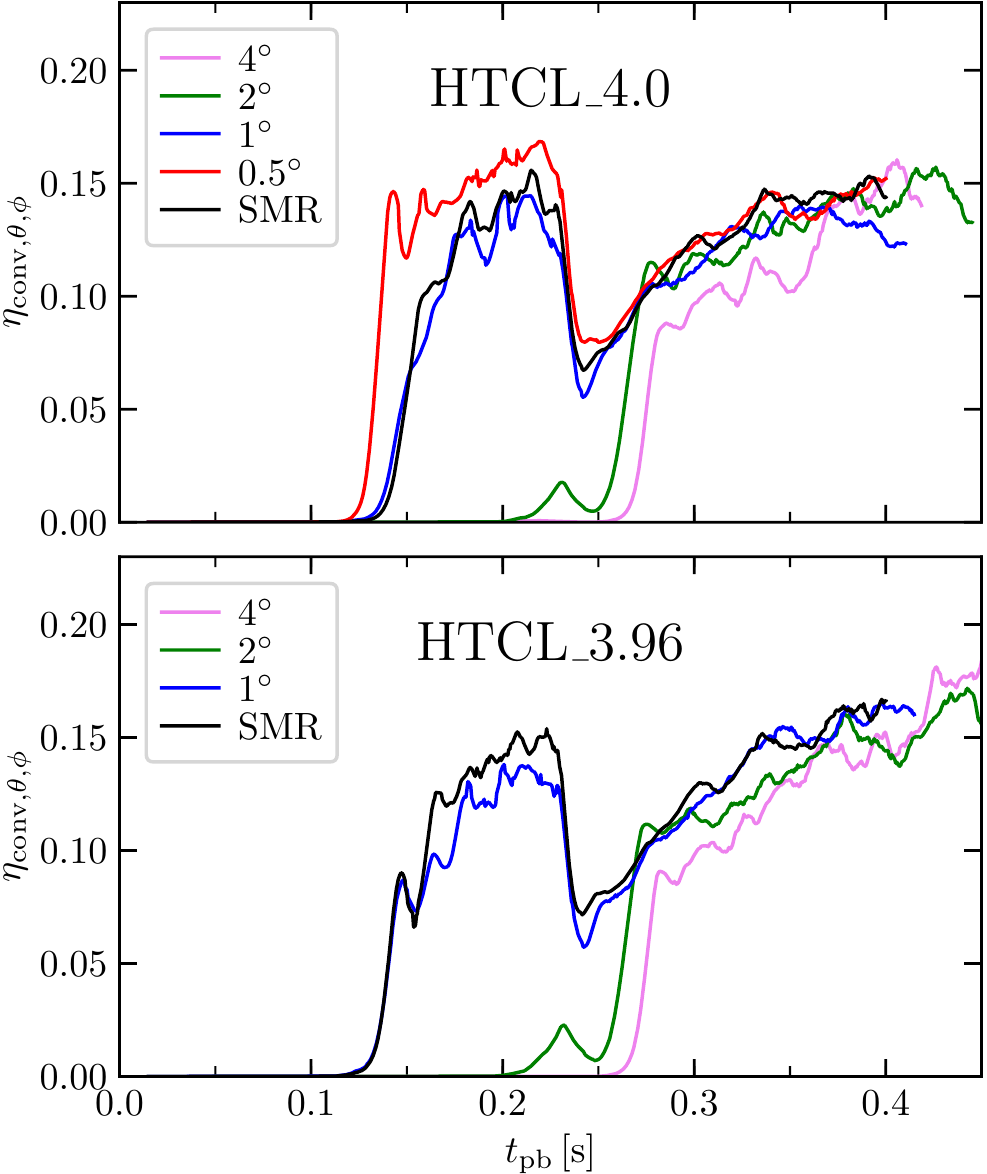}
    \caption{Efficiency factor for the conversion of heating into (nonradial)
    turbulent kinetic energy as a function of time for the two HTCL model sets
    with $\Lnu=4$ (\emph{top}) and $\Lnu=3.96$ (\emph{bottom}).}
    \label{fig:eta}
\end{figure}

\begin{figure*}
	\centering
	\includegraphics[width=0.95\linewidth]{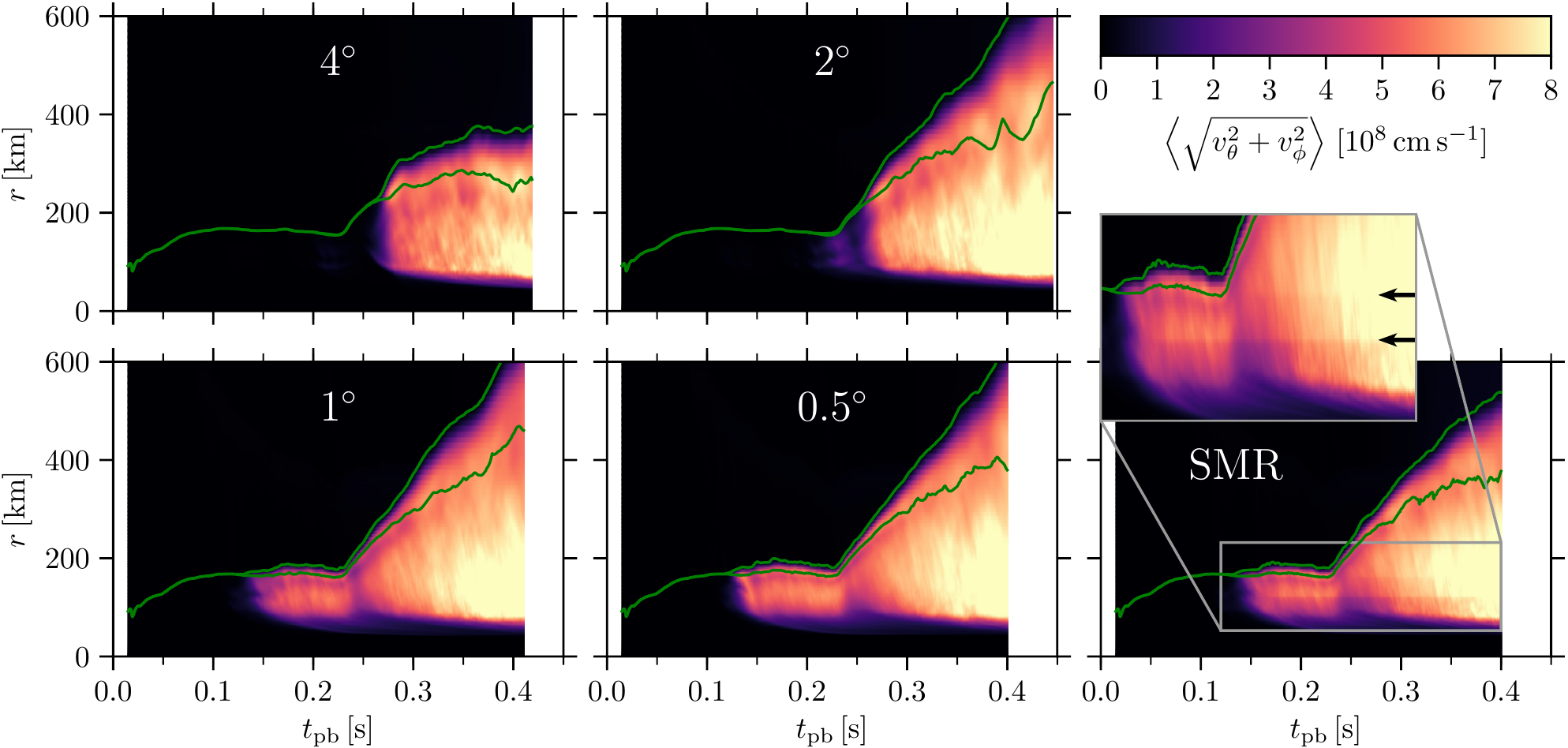}
	\caption{
		Minimum and maximum shock radii as functions of post-bounce time
		(\emph{green lines}) for the HTCL models with $\Lnu=4$. Color-coded is the
		density-weighted angle-average of the lateral velocity,
		$\left< (v_\theta^2 + v_\phi^2)^{0.5} \right>$.
		The zoom inset attached to the lower right panel shows the positions of the two SMR resolution interfaces (123 and 162\,km, \emph{black arrows}), where nonradial kinetic energy of downflows is converted to thermal energy.
		The loss of nonradial kinetic energy is visible as faint discontinuities in the color shading at the radial locations of the two black arrows.}
	\label{fig:vthetaphi}
\end{figure*}

As in \citet{Mueller2017}, we analyze the efficiency for the conversion of
neutrino energy deposited in the gain layer into turbulent kinetic energy,
\begin{equation}
    \eta_{\mathrm{conv},\theta,\phi} = \frac{E_{\mathrm{kin},\theta,\phi}^\mathrm{gain}}{
        \left[ \dot{Q}_\mathrm{gain} \left(\left<R_\mathrm{sh}\right> -
        R_\mathrm{gain}\right)
     \right]^{2/3} M_\mathrm{gain}^{1/3}},
\end{equation}
where $R_\mathrm{gain}$ and $M_\mathrm{gain}$ are the average gain radius and the
gain-layer mass, respectively. The net heating term is given by
\begin{equation}
    \dot{Q}_\mathrm{gain} \defeq \int_{R_\mathrm{gain} < r
    < R_\mathrm{sh}(\theta,\phi)} \ud V \, \rho \left( \dot{q}_\mathrm{heat} -
    \dot{q}_\mathrm{cool} \right). \label{eq:qdotgain}
\end{equation}
The values of $\eta_{\mathrm{conv},\theta,\phi}$ are shown in
Fig.~\ref{fig:eta}. Before the arrival of the silicon/silicon+oxygen interface,
we see the same dependence on the resolution as in Fig.~\ref{fig:ekingain}. The
models with a resolution of at least $1^\circ$ reach an efficiency of about
0.15, while the lower-resolved cases remain convectively less vigorous. After
the onset of shock runaway, the conversion efficiency
$\eta_{\mathrm{conv},\theta,\phi}$ loses its dependency on the angular
resolution if the grid spacing is at least $2^\circ$.

The time evolution of the lateral velocities is presented in
Fig.~\ref{fig:vthetaphi} as a radius-time diagram. It can be clearly seen that
the onset of convection in the gain layer occurs earlier with higher angular
resolution, which we have discussed already before.  The slight shock expansion
at 150\,ms after bounce in the models with at least $1^\circ$ resolution can be
explained by the growing strength of convection at that time.  Models that
remain convectively quiet do not show this effect.  After the revival of the
shock, the convective strength, i.e., the magnitude of the lateral velocity is
roughly equal in all models presented in Fig.~\ref{fig:vthetaphi}. This is in
line with the finding that the nonradial kinetic energy does not depend on the
angular resolution after the arrival of the silicon/silicon+oxygen interface
except for still lower values in the 4$^\circ$ model.

\begin{figure}
    \centering
    \includegraphics[width=0.95\linewidth]{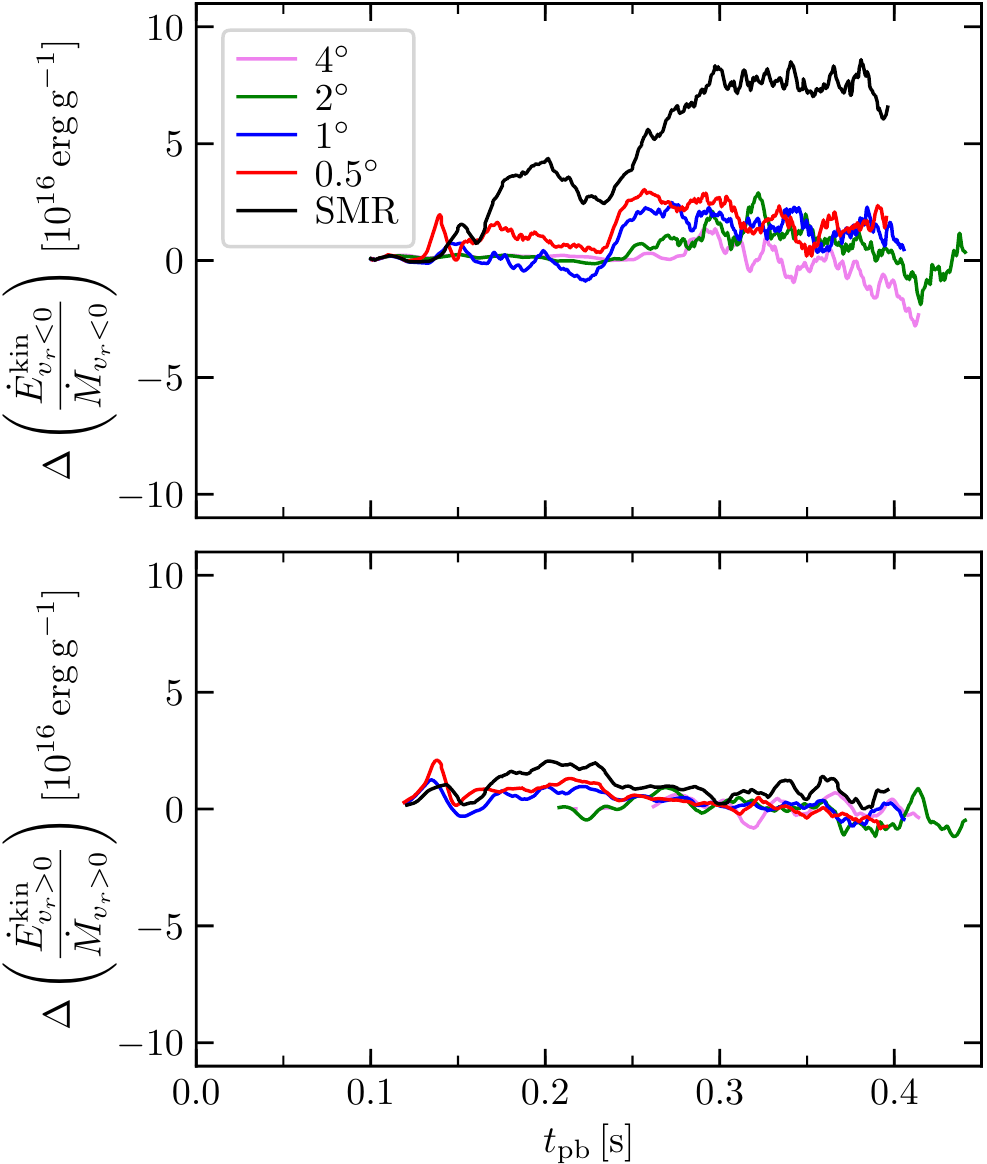}
    \caption{Same as Fig.~\ref{fig:kineticenergyflux_s20}, but for the HTCL
    model set with $\Lnu=4$. The difference of the radial specific kinetic
    energy flux is evaluated around a radius of 123\,km, i.e., the specific
    energy flux at 122\,km is subtracted from its value at 124\,km. 
    We differentiate between
    inflows (\emph{top}) and outflows (\emph{bottom}).}
    \label{fig:kineticenergyflux_HTCL}
\end{figure}

We have shown above that the SMR setup resembles a uniform grid with a
resolution of $0.5^\circ$ in the overall fluid structures. Also the temporal
evolution of the angle-averaged shock is identical until at least 70\,ms after shock
revival. Afterwards, however, the expansion velocity of the shock decreases and falls
below the case of $1^\circ$ resolution. This can again be explained by the
dissipation of kinetic energy at the interfaces between layers of different
angular resolution (see black arrows in the zoom inset of Fig.~\ref{fig:vthetaphi}).

In Fig.~\ref{fig:kineticenergyflux_HTCL}, the radial specific kinetic energy
fluxes are evaluated at two different radii. The value at 122\,km is subtracted
from the value at 124\,km, to investigate the flux conservation at the SMR
interface at 123\,km. This analysis is performed in the same way as for the
model set with full neutrino transport above, but now with $r_1 =
122\,\mathrm{km}$ and $r_2 = 124\,\mathrm{km}$ in Eq.~(\ref{eq:deltaflux}).
Note that the individual values are always positive for infalling and outflowing
fluid elements so that a larger flux at the outer radius results in a positive
flux difference.

Again, we see that the outflowing fluxes do not show any evidence for dissipation of
kinetic energy. This is assuring especially for the SMR model, where matter
flowing outwards is propagating from a coarser grid spacing of $2^\circ$ into a
finer grid of $1^\circ$ resolution.  Infalling material, however, behaves
differently in this comparison. The flux differences in the models with a
uniform grid spacing are rather similar and stay close to zero, whereas in the SMR simulation they are
distinctly higher by 
factors of a few during times of
increased nonradial velocities
in the gain layer. Kinetic energy is therefore dissipated on the SMR grid with its
resolution interface at 123\,km as matter flows from 124\,km to 122\,km.
This effect can also be spotted in the bottom right panel of
Fig.~\ref{fig:vthetaphi}, where the boundaries of the different resolution
layers of the SMR grid display as faint horizontal discontinuities in the color
shading (see associated zoom inset). The magnitude of the lateral velocity decreases visibly at 162 and
123\,km from outside inwards, which would not be the case without kinetic energy
dissipation.

\begin{figure}
    \centering
    \includegraphics[width=0.95\linewidth]{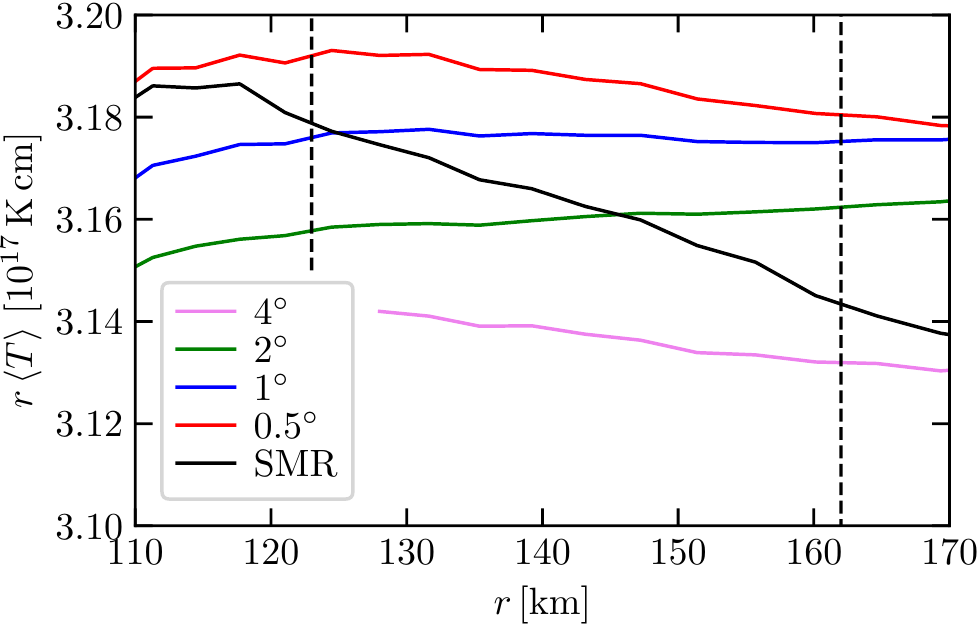}
    \caption{Angle- and time-averaged temperature multiplied with radius for the
    HTCL simulations with $\Lnu=4$. The time average was calculated in the
    interval from 300 to 400\,ms. The angular resolution in the SMR model is
    $2^\circ$, $1^\circ$, and $0.5^\circ$ for $r<123\,\mathrm{km}$,
    $123\,\mathrm{km} < r < 162\,\mathrm{km}$, and $r > 162\,\mathrm{km}$,
    respectively. The boundaries of these regions are indicated by vertical
    dashed lines.}
    \label{fig:temp}
\end{figure}

When kinetic energy is dissipated into thermal energy in the SMR model, the
temperature should increase directly below a resolution interface.\footnote{Our
hydro and SMR scheme are implemented in a conservative form, which means that
fluxes conserve the sum of internal and kinetic energies (see Appendix~\ref{sec:smr}).} To analyze this, we
show radial profiles of the angle- and time-averaged temperature multiplied with
the radius in Fig.~\ref{fig:temp}. This visualization roughly compensates for
the $r^{-1}$ scaling of the temperature and allows us to closely analyze
temperature gradients.

The temperature profile in the SMR case clearly differs from all other models.
It is much steeper and shows an even further increased gradient directly below
the inner resolution interface at a radius of 123\,km. Between the two
resolution layers, i.e.\ between the two dashed lines in Fig.~\ref{fig:temp},
the SMR profile does not flatten, because the dissipation of kinetic energy does
not happen instantaneously as the flow moves inwards. With a radial velocity of
about $-3\times10^8\,\mathrm{cm\,s^{-1}}$, a fluid element needs only about
15\,ms to propagate through the region of $1^\circ$ angular resolution.

We have thus shown that the dissipation of kinetic energy in downflows increases
the thermal energy and changes the average temperature profile at the expense of
turbulent kinetic energy. For the anisotropic postshock turbulence in the
supernova core, kinetic energy and turbulent pressure are coupled by an
effective adiabatic index of $\gamma_\mathrm{turb} = 2$ \citep{Radice2015}. In
contrast, thermal energy of the plasma in the postshock layer, where
relativistic electron-positron pairs and photons dominate the energy density,
provides thermal pressure only with a thermodynamical adiabatic index of
$\gamma_\mathrm{thermal}\approx 4/3$. This suggests that the conversion of
turbulent kinetic energy to thermal energy reduces the ability of the postshock
layer to provide outward push to the supernova shock. This explains why at later
times, $t_\mathrm{pb}\gtrsim 300$\,ms, the expansion of the shock in the SMR
models begins to slightly lag behind the shock of the $1^\circ$ and $0.5^\circ$
simulations (see Fig.~\ref{fig:shock_HTCL}).

\section{Resolution dependence of turbulence}
\label{sec:turbulence}

In this section, we will investigate how the angular grid resolution influences
the turbulent cascade, following the discussion in \citet{Melson2016}.

\subsection{Turbulent kinetic energy spectra}

A fluid transitions from the laminar to the turbulent regime above a certain
critical Reynolds number $\mathrm{Re}$. Turbulence is described phenomenologically and
understood as a superposition of eddies on various scales
\citep{Landau1987,Pope2000}. In the common picture sharpened by
\citet{Kolmogorov1941}, kinetic energy is steadily injected at some large scale
$L$, which is similar to the size of the largest turbulent eddies. These eddies
break up into smaller structures and thus transport energy to successively
smaller scales. Eventually, below some small scale $\mathcal{L}$, kinetic energy
is dissipated into internal energy by viscous effects.

\citet{Kolmogorov1941} assumed that this turbulent energy cascade only depends
on the energy dissipation rate and the viscosity. In the inertial range roughly
between $L$ and $\mathcal{L}$, kinetic energy is carried to smaller scales
without losses. From a self-similarity ansatz, it follows that the kinetic
energy spectrum $E(k)$---with $k$ being the wave number---has a universal
shape of $E(k)\propto k^{-5/3}$ in the inertial range.\footnote{To
    clarify the notation, we write $E_k$ and $E_\ell$ instead of $E(k)$ and
    $E(\ell)$ from now on, because the symbol $E$ is also used for the total
kinetic energy.} These findings only hold if the fluid structures are locally
isotropic, i.e., large-scale anisotropies due to boundary effects can be
neglected for sufficiently small scales below $L$ \citep{Pope2000}.
Moreover, ideal conditions require that the fluid is incompressible and the
flow is stationary. The last two aspects are certainly not fulfilled in the supernova
environment. It is also not clear
whether the first as\-sump\-tion is valid during the shock stagnation phase,
because the accretion flow through the gain layer might impose a preferred
direction not only on the largest turbulent eddies but also on smaller-scale
structures \citep[see, e.g.,][]{Murphy2013,Couch2015}.

Here, we assume that Kolmogorov's theory of turbulence is applicable to the
core-collapse supernova conditions. In order to quantify the turbulent transport
of energy across various scales, the kinetic energy spectra are calculated for
the 3D HTCL models with $\Lnu=4$ at different times after core bounce. Since we
consider stellar cores, which are spherical objects to first order, the
kinetic energy is decomposed into spherical harmonics instead of Cartesian wave
numbers.

Let the complex spherical harmonics be defined as
\begin{equation}
    Y_\ell^m(\theta,\phi) = N_\ell^m \, P_\ell^m(\cos\theta) \, e^{i m \phi}
\end{equation}
with normalization factors
\begin{equation}
    N_\ell^m = \sqrt{\frac{2\ell+1}{4 \pi} \, \frac{(\ell-m)!}{(\ell+m)!}}
\end{equation}
and associated Legendre polynomials $P_\ell^m(\cos\theta)$. The decomposition of
the nonradial kinetic energy density at a given radius is then determined by
\begin{equation}
    E_\ell = \frac{1}{2} \sum_{m=-\ell}^\ell \left| \int \ud \Omega \,
    \sqrt{\rho \left(v_\theta^2 + v_\phi^2 \right)}
    \,Y_\ell^m(\theta,\phi)\right|^2.
\label{eq:energyspectra}
\end{equation}
This spectrum is normalized such that the total nonradial kinetic energy
density on a spherical shell is the sum over all components of the spectrum,
\begin{equation}
    E \defeq \frac{1}{2} \int \ud \Omega \,
    \rho \left( v_\theta^2 + v_\phi^2 \right) = \sum_{\ell=0}^\infty E_\ell.
\label{eq:sumenergyspectra}
\end{equation}
The decomposition applied here was similarly used in other works that analyzed
the turbulent cascade, for example, \citet{Hanke2012}, \citet{Couch2014},
\citet{Hanke2014}, and \citet{Abdikamalov2015}. Note that from a numerical
perspective, accurate results of the integrals in Eq.~(\ref{eq:energyspectra})
can only be achieved by applying Gauss-Legendre quadrature, thus ensuring that
the spherical harmonics are sampled on a finer grid than the computational mesh
of the simulation.

For the discussion later in this chapter, we also define the spectrum of the
specific kinetic energy as
\begin{equation}
    \Especific_\ell = \frac{1}{2} \sum_{m=-\ell}^\ell \left| \int \ud \Omega \,
    \sqrt{v_\theta^2 + v_\phi^2} \,Y_\ell^m(\theta,\phi)\right|^2.
    \label{eq:energyspectraspecific}
\end{equation}
Again, the sum over the coefficients gives the total nonradial specific kinetic
energy on a spherical shell,
\begin{equation}
    \frac{1}{2} \int \ud \Omega \, \left( v_\theta^2 + v_\phi^2 \right) =
    \sum_{\ell=0}^\infty \Especific_\ell.
    \label{eq:nonradspeckinerg}
\end{equation}

\begin{figure*}
    \centering
    \includegraphics[width=0.95\linewidth]{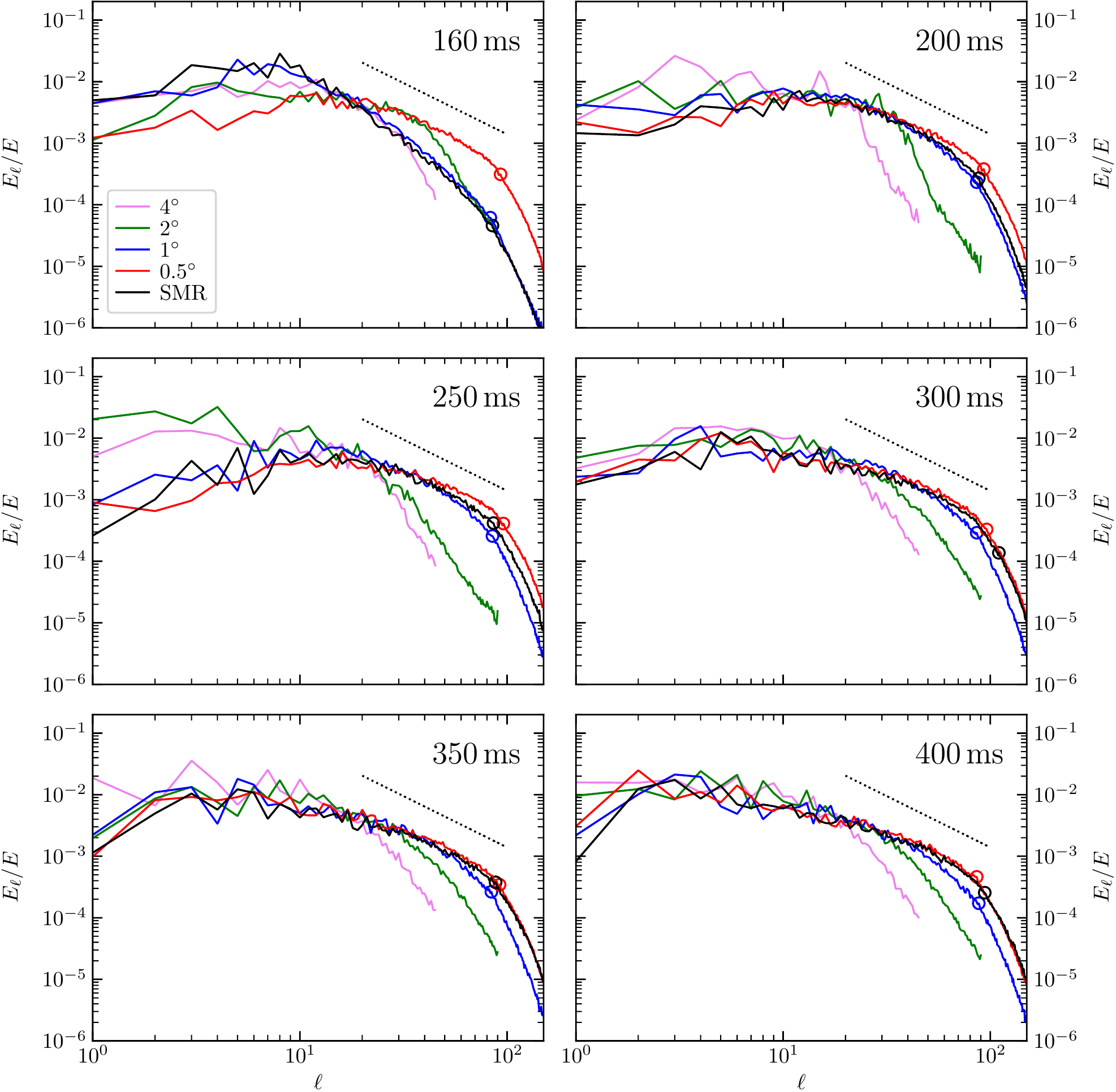}
    \caption{Normalized spectra of the nonradial turbulent kinetic energy for
    the HTCL model set with $\Lnu=4$, measured between the gain radius and the
    minimum shock radius. The spectra are averaged over 10\,km around $R_0$ and
    a time interval of 5\,ms. Open circles mark the beginning of the dissipation range,
    $\ell_\mathrm{diss}$. The dotted lines indicate the $(-5/3)$-power-law slope
    expected for Kolmogorov turbulence in the inertial range.
    At 300\,ms the spectra of the simulations with SMR grid and 0.5$^\circ$
    resolution become very similar, whereas at earlier times the spectra 
    of SMR and 1$^\circ$ simulations are more similar. This is
    fully compatible with the growing resolution of the SMR grid 
    as the shock moves to larger radii.}
    \label{fig:spectra}
\end{figure*}

In Fig.~\ref{fig:spectra}, we present energy spectra for the HTCL simulations
with $\Lnu=4$ at certain times after core bounce. The spectra are measured at a
radius $R_0$ between the angle-averaged gain radius, $R_\mathrm{gain}$, and the
minimum shock radius, i.e.,
\begin{equation}
    R_0 = \frac{1}{2} \left(R_\mathrm{gain} + \min (R_\mathrm{sh}) \right).
    \label{eq:r0}
\end{equation}
This choice assures that we do not include contributions from pre-shock material
in our analysis. In order to smooth the data, we computed volume-weighted
spatial averages of the energy spectra in the range $R_0 \pm 5\,\mathrm{km}$
with an additional time-averaging in the interval $t \pm 2.5\,\mathrm{ms}$. To
guarantee consistency, this procedure is also applied to the total kinetic
energy density $E$ used for the normalization of the spectra.

From the angular grid resolution $\alpha$ of the 3D models, we can calculate the
maximum multipole order $\ell_\mathrm{max}$ roughly according to $\ell_\mathrm{max}
\approx 180^\circ / \alpha$. In the SMR simulation, $\alpha = 1^\circ$
until $\sim$250\,ms and $\alpha = 0.5^\circ$ at later times, depending on
the location of $R_0$ at the time when the spectrum is measured. However, due to
round-off errors caused by limited computational accuracy, we are not able to compute
spherical harmonics for $\ell > 150$.

We can estimate the multipole order of the largest eddies $\ell_L$ from simple
considerations about their size. The largest possible extension of turbulent vortex
structures in the gain layer depends on the thickness of this shell and is
\begin{equation}
    L = \left<R_\mathrm{sh} \right> - R_\mathrm{gain}.
    \label{eq:largesteddysize2}
\end{equation}
As explained, for example, by \citet{Foglizzo2006}, this can be translated into
a multipole order according to
\begin{equation}
    \ell_L = \frac{\pi R_0}{L}.
    \label{eq:largesteddies}
\end{equation}
In the HTCL model set with $\Lnu=4$, we find $\ell_\mathrm{L} \sim 7-10$ during the phase of shock stagnation and $\ell_\mathrm{L} \sim 2-5$
towards the end of the simulations, which roughly coincides with the peak positions, $\ell_\mathrm{peak}$, of the kinetic power spectra.
The multipole order of the largest eddies is nearly the same in all models,
which is expected on grounds of the geometrical
considerations for determining $\ell_L$.

An important aspect of Kolmogorov's theory of turbulence is the presence of a
$k^{-5/3}$ scaling in the inertial range of the energy spectrum. As we will show
below, this translates into an $\ell^{-5/3}$ behavior in our decomposition.
Hence, we added dotted lines of this slope in Fig.~\ref{fig:spectra} to
visualize the inertial range. The power spectra of our highest-resolved
models are indeed close to a $(-5/3)$ power law for $\ell \gtrsim 30$, whereas
for $10 \lesssim \ell \lesssim 30$ the spectra are better described by a
$(-1)$-power-law slope.

The value of $\ell$, above which kinetic energy is dissipated into internal
energy, $\ell_\mathrm{diss}$, depends strongly on the grid resolution.
It is visualized by a circle in Fig.~\ref{fig:spectra} and also given in
Table~\ref{tab:Re}. In models with a resolution of $2^\circ$ or worse, we do not see
any clear Kolmogorov regime. In better resolved models,
the $\ell^{-5/3}$ behavior breaks down near $\ell\sim 100$, i.e.\ on an angular
scale of about $2^\circ$, which means that dissipation sets in at the level of a
few grid cells.  According to \citet{Porter1998} and \citet{Sytine2000}, the
Piecewise Parabolic Method \citep[PPM;][]{Colella1984} that is applied in
\textsc{Vertex-Prometheus} dissipates kinetic energy below 2 to 12 grid cells,
which is roughly consistent with our finding.

To sum up, the similarity of our numerical spectra to Kolmogorov's theory of
steady-state isotropic turbulence can be used as a motivation to (approximately)
apply relations from this theory during phases of shock stagnation in our models.
This seems justified in conditions where SASI does not introduce global
large-amplitude variations by pumping kinetic energy into the lowest modes $\ell
\lesssim 2$. In the high-resolution simulations of at least $1^\circ$, we see a
clear separation between the inertial range with its characteristic slope of
$-5/3$ and the dissipation range with a steeper decay.

\subsection{Numerical viscosity and effective Reynolds number}
\label{sec:viscosity}

The kinematic shear viscosity of the stellar medium in the gain layer is of the
order of $0.1\,\mathrm{cm^2\,s^{-1}}$ and thus extremely low
\citep[e.g.,][]{Abdikamalov2015}. Because the corresponding Reynolds numbers
are extremely high, of order $\sim 10^{17}$, the evolution of the stellar plasma is
described by the Euler equations instead of solving the Navier-Stokes equations,
which include terms that account for viscous effects. However, the viscosity
associated with the numerical scheme is many orders of magnitude larger than the
physical viscosity of the stellar plasma \citep[see., e.g.,][]{Mueller1998}.
Interestingly, the interaction of neutrinos with matter in the gain layer
produces a damping force on fluid motions---a neutrino drag---whose influence on
the plasma flow is in the ballpark of the effects of numerical viscosity on the
relevant scales (see Appendix~\ref{sec:drag} for a derivation and discussion of
the neutrino drag in detail).

As in all other state-of-the-art core-collapse supernova models, we rely on the
implicit large eddy simulation \citep[ILES;][]{Grinstein2007} paradigm. It
assumes that dissipative effects on the smallest scales are implicitly accounted
for by the numerical viscosity.  Instead of solving filtered hydrodynamic
equations and creating a sub-grid model for the dissipation of kinetic energy
\citep{Boris1992}, the ILES approach assumes that such a sub-grid model is
implicitly included at the level of the grid cell size.

Estimating the effective viscosity of a numerical scheme, $\nu_\mathrm{N}$, is
difficult. It depends not only on the algorithm itself, but also on its
implementation.  Even if the numerical viscosity is known for one simulation
code, it is not justified to assume an equal viscosity for all codes that make
use of the same algorithm for treating the hydrodynamics. Consequently, the
numerical viscosity and the effective Reynolds number must be determined for
every code separately in order to estimate the influence of dissipative effects.
This can only be achieved by measuring characteristic quantities from the output
data.

In the following two sections, we will discuss two methods for determining the
numerical viscosity and the effective Reynolds number from properties of the
kinetic energy spectrum. Both approaches will be applied to our 3D simulations.
The first procedure was proposed by \citet{Abdikamalov2015}, while the second
method has been developed by us and was presented in \citet{Melson2016} before.
As in the previous section, the energy spectra are averaged over
$10\,\mathrm{km}$ and $5\,\mathrm{ms}$, and measured at a radius $R_0$ halfway
between the angle-averaged gain radius and the minimum shock radius.

\subsubsection{Based on the Taylor microscale}

The method of \citet{Abdikamalov2015} is based on determining the so-called
Taylor microscale, which is given by \citep{Frisch1995,Pope2000}
\begin{equation}
    \lambda = \sqrt{\frac{5E}{Z}}, \label{eq:lambda1}
\end{equation}
where $E$ is the total kinetic energy density of nonradial fluid motions,
\begin{equation}
    E \approx \sum_{\ell=0}^{\ell_\mathrm{max}} E_\ell,
\end{equation}
and $Z$ is the enstrophy, which can be approximated by
\begin{equation}
    Z \approx \frac{1}{R_0^2} \sum_{\ell=0}^{\ell_\mathrm{max}} \ell (\ell+1) \,
    E_\ell.
    \label{eq:enstrophy1}
\end{equation}
For the upper bound $\ell_\mathrm{max}$, \citet{Abdikamalov2015} picked a value of
$120$, while we calculate it from the angular resolution of the model.

The Taylor microscale $\lambda$ has no direct physical interpretation. It is
situated somewhere between the characteristic scale of the smallest eddies---the
Kolmogorov scale---and the size of the largest structures \citep{Pope2000}.

\citet{Abdikamalov2015} derived a relation for the effective Reynolds number,
\begin{equation}
    \mathrm{Re} = 5 \frac{\tilde{L}^2}{\lambda^2},
    \label{eq:Reynoldsnumber1}
\end{equation}
which is, however, not fully consistent with the literature. Commonly, a factor of
$10$ \citep{Pope2000,Schmidt2014} or even $15$ \citep{Tennekes1972} instead of
$5$ is applied. Nevertheless, in order to compare with the results of
\citet{Abdikamalov2015}, we also use their factor $5$ here. At the end of this
section, we will further discuss this issue.

The size of the energy-containing eddies $\tilde{L}$ is calculated by
\citet{Abdikamalov2015} from the energy spectrum according to (motivated by Eq.~(\ref{eq:largesteddies}))
\begin{equation}
    \tilde{L} = \pi R_0 \left( 1 + \frac{1}{E} \sum_{\ell=0}^{\ell_\mathrm{max}} \ell \,
    E_\ell \right)^{-1}.
    \label{eq:largesteddysize1}
\end{equation}
Finally, the kinematic numerical viscosity can be determined from the fundamental relation
\begin{equation}
    \nu_\mathrm{N} = \frac{v_0 \tilde{L}}{\mathrm{Re}}.
    \label{eq:numericalviscosity1}
\end{equation}
The characteristic velocity $v_0$ of the largest eddies is deduced from the total
kinetic energy density (Eq.~(\ref{eq:sumenergyspectra})) by
\begin{equation}
v_0^2 \defeq \frac{1}{4\pi} \int \ud \Omega \, \left( v_\theta^2 + v_\phi^2 \right) \approx \frac{E}{2\pi\rho_0}.
\label{eq:v01}
\end{equation}
Note that in addition to $E$, also the density $\rho_0$ is averaged over a radial shell
defined by $R_0 \pm 5\,\mathrm{km}$ and a time interval of $t \pm
2.5\,\mathrm{ms}$.

The method of \citet{Abdikamalov2015} suffers from the uncertainty in
Eq.~(\ref{eq:Reynoldsnumber1}) and in the scale defined in
Eq.~(\ref{eq:largesteddysize1}). Their application is debatable, because there
might be factors of 2 or even 3 missing. This issue will be discussed later in
this section. Furthermore, this approach yields Reynolds numbers being
implausibly low and showing only a weak resolution dependence (see
Table~\ref{tab:Re}), which suggests a marginally turbulent flow for all
resolutions tested, in obvious conflict with the situation observed in
Figs.~\ref{fig:cross_sto} and \ref{fig:vorticity}, and the presence of a
Kolmogorov-like power spectrum over roughly one order of magnitude of $\ell$ in
Fig.~\ref{fig:spectra}.

For the reasons mentioned, we have developed a different procedure, which yields
more realistic values of the numerical viscosity and the effective Reynolds
number.

\subsubsection{Based on the energy dissipation rate}
\label{sec:MKJ}

Our method for measuring the numerical viscosity and the Reynolds number is
based on more fundamental properties of the turbulent energy cascade. In the
inertial range, the kinetic energy spectrum only depends on the specific energy
dissipation rate $\varepsilon$ and is given by \citep{Landau1987,Pope2000}
\begin{equation}
    \Especific_k = C \, \varepsilon^{2/3} \, k^{-5/3},
\end{equation}
where $\Especific_k\,\ud k$ is the specific turbulent kinetic energy of the
fluid in the interval $[k, k+\ud k]$.

In order to be consistent with the literature, we employ the spectrum of the
specific kinetic energy as defined in Eq.~(\ref{eq:energyspectraspecific})
rather than that of the kinetic energy density. The factor $C$ is a universal
constant of $C = 1.62$ and independent of the Reynolds number
\citep{Sreenivasan1995,Yeung1997}.

Since we decompose the spectrum by making use of spherical harmonics, it must be
written as a function of the multipole order $\ell$ instead of the wave number
$k$. This transformation reads
\begin{equation}
    k = \frac{\sqrt{\ell(\ell+1)}}{R_0} \approx \frac{\ell}{R_0},
\end{equation}
where the latter approximation is valid for sufficiently high values of $\ell$.
The energy spectrum as a function of $k$,
\begin{equation}
    \Especific_k \, \ud k = C \, \varepsilon^{2/3} \,
      k^{-5/3} \, \ud k,
\end{equation}
can then be written as
\begin{equation}
    \Especific_\ell \, \ud \ell = C \, \varepsilon^{2/3} \,
      R_0^{5/3} \ell^{-5/3} \, \frac{\ud \ell}{R_0}.
\end{equation}
From this relation, we obtain an equation for the specific energy dissipation rate,
\begin{equation}
    \varepsilon(\ell) = \sqrt{\frac{\Especific_\ell^3 \, \ell^5}{R_0^2 \, C^3}},
    \label{eq:energydissipationrate}
\end{equation}
which allows for directly measuring its value from the spectrum, using
Eq.~(\ref{eq:energyspectraspecific}) for $\Especific_\ell$ and Eq.~(\ref{eq:r0})
for $R_0$. Together with the specific enstrophy $\Zspecific$ calculated approximately (becoming exact for $\ell_\mathrm{max} \longrightarrow \infty$)
ac\-cord\-ing to
\begin{equation}
    \Zspecific \approx \frac{1}{R_0^2} \sum_{\ell=0}^{\ell_\mathrm{max}} \ell (\ell+1) \,
    \Especific_\ell,
    \label{eq:enstrophy2}
\end{equation}
we can determine the numerical viscosity from the equation
\citep{Tennekes1972,Pope2000}
\begin{equation}
    \nu_\mathrm{N} = \frac{\varepsilon}{2 \Zspecific}.
    \label{eq:numericalviscosity2}
\end{equation}
Note that the enstrophies defined in Eqs.~(\ref{eq:enstrophy1}) and
(\ref{eq:enstrophy2}) are connected to each other by the relation $Z \approx
\rho_0 \Zspecific$.

The question arises, at which multipole order $\ell$ the energy dissipation rate
should be measured for our purposes. If ideal Kolmogorov turbulence would apply,
$\varepsilon$ would actually by constant in the inertial range. Because of
deviations from this perfect Kolmogorov case, the most conservative approach is
taking the peak value of $\varepsilon$ to evaluate Eq.~(\ref{eq:numericalviscosity2}),
since we want to maximize our estimate of the numerical
viscosity, which is known to be large. In practice, however, the spectra
$\varepsilon(\ell)$ turn out to possess a very broad maximum, compatible with
the Kolmogorov-like behavior.

Ultimately, we can calculate the effective Reynolds number as
\begin{equation}
    \mathrm{Re} = \frac{v_0 L}{\nu_\mathrm{N}},
    \label{eq:Reynoldsnumber2}
\end{equation}
where $L$ is taken from Eq.~(\ref{eq:largesteddysize2}) and $v_0$ is the characteristic
velocity of the largest eddies given by (see Eq.~(\ref{eq:nonradspeckinerg}))
\begin{equation}
v_0^2 = \frac{1}{2\pi} \sum_{\ell=0}^{\infty} \Especific_\ell \approx \frac{1}{2\pi} \sum_{\ell=0}^{\ell_\mathrm{max}} \Especific_\ell.
\label{eq:charcteristicvelocity}
\end{equation}
In contrast to the previous method, $L$ is assumed to be equal to the radial thickness
of the gain layer (Eq.~(\ref{eq:largesteddysize2})). The values of $v_0$ obtained with
Eqs.~(\ref{eq:charcteristicvelocity}) and (\ref{eq:v01}) are extremely similar 
(see Table~\ref{tab:Re}).

\subsubsection{Comparison of the methods}
\label{sec:comparison}

\begin{table*}
    \caption{Numerical Reynolds numbers and corresponding numerical
    	viscosities evaluated using the AOR+ and MKJ methods
    	for the 3D simulations of the HTCL model set with
    	$\Lnu=4$.}
\centering
\newcommand{\ccc}[1]{\multicolumn{1}{c}{#1}}
\begin{tabular}{ll>{\bfseries}r>{\bfseries}rrr@{\extracolsep{15pt}}>{\bfseries}r@{\extracolsep{0pt}}>{\bfseries}rrr@{\extracolsep{5pt}}r@{\extracolsep{0pt}}r}
\hline
\hline
    & & \multicolumn{4}{c}{\citet[AOR+]{Abdikamalov2015}} & \multicolumn{4}{c}{This work (MKJ)} & & \\
    \cline{3-6} \cline{7-10}
    \ccc{$t_\mathrm{pb}$} & \ccc{3D model} & \ccc{Re} & \ccc{$\nu_\mathrm{N}$} &
    \ccc{$\tilde{L}$} & \ccc{$v_0$} &
    \ccc{Re} & \ccc{$\nu_\mathrm{N}$} & \ccc{$L$} & \ccc{$v_0$} &
    \ccc{$\ell_\mathrm{diss}$} & \ccc{$R_0$}\\
    \ccc{$\mathrm{[ms]}$} & & & \ccc{$\mathrm{[10^{13}\,cm^2\,s^{-1}]}$} &
    \ccc{$\mathrm{[km]}$} & \ccc{$\mathrm{[10^8\,cm\,s^{-1}]}$} & &
    \ccc{$\mathrm{[10^{13}\,cm^2\,s^{-1}]}$} & \ccc{$\mathrm{[km]}$} &
    \ccc{$\mathrm{[10^8\,cm\,s^{-1}]}$} & & \ccc{$\mathrm{[km]}$} \\
	\hline
	160 & $4^\circ$\dotfill    & \normalfont(41) & \normalfont(0.05) & 134 & 0.02 & \normalfont(40) & \normalfont(0.02) & 46 & 0.02 & - & 141 \\
	& $2^\circ$\dotfill 	    & \normalfont(50) & \normalfont(0.03) &  87 & 0.02 & \normalfont(60) & \normalfont(0.01) & 46 & 0.02 & - & 141 \\
	& $1^\circ$\dotfill         & 47 &  8.4 &  99 & 3.9 &  87 &  2.6 &  56 & 4.0 & 83 & 140 \\
	& $0.5^\circ$\dotfill       & 58 &  5.7 &  61 & 5.4 & 195 &  1.8 &  65 & 5.5 & 93 & 144 \\
	& SMR ($1^\circ$)\dotfill   & 43 & 11.0 & 103 & 4.6 &  55 &  4.8 &  56 & 4.7 & 85 & 141\\
	\hline
	200 & $4^\circ$\dotfill     & \normalfont(35) &  \normalfont(0.6) & 137 & 0.2 &  \normalfont(12) &  \normalfont(0.6) &  45 & 0.2 & - & 139 \\
	& $2^\circ$\dotfill         & (\normalfont46) &  \normalfont(0.6) &  96 & 0.3 &  \normalfont(21) &  \normalfont(0.6) &  44 & 0.3 & - & 139 \\
	& $1^\circ$\dotfill         & 54 &  6.8 &  68 & 5.3 & 153 &  2.2 &  61 & 5.4 & 86 & 141 \\
	& $0.5^\circ$\dotfill       & 58 &  5.9 &  57 & 5.9 & 199 &  1.9 &  64 & 6.0 & 93 & 142 \\
	& SMR ($1^\circ$)\dotfill   & 59 &  6.4 &  68 & 5.6 & 163 &  2.1 &  61 & 5.7 & 88 & 141 \\
	\hline
	225 & $4^\circ$\dotfill     & \normalfont(18) &  \normalfont(1.2) &  97 & 0.2 &  \normalfont(10) &  \normalfont(0.9) &  41 & 0.2 & - & 134 \\
	& $2^\circ$\dotfill         & 22 &  6.0 &  83 & 1.6 &  13 &  5.6 &  43 & 1.6 & - & 134 \\
	& $1^\circ$\dotfill         & 54 &  6.1 &  67 & 4.9 & 132 &  2.1 &  57 & 5.0 & 81 & 137 \\
	& $0.5^\circ$\dotfill       & 59 &  5.5 &  54 & 6.0 & 197 &  1.8 &  59 & 6.1 & 97 & 138 \\
	& SMR ($1^\circ$)\dotfill   & 57 &  6.4 &  66 & 5.5 & 129 &  2.5 &  57 & 5.6 & 83 & 137 \\
	\hline
	250 & $4^\circ$\dotfill     & \normalfont(39) &  \normalfont(0.8) & 159 & 0.2 &  \normalfont(48) &  \normalfont(0.4) &  92 & 0.2 & - & 161 \\
	& $2^\circ$\dotfill         & 39 &  4.1 & 124 & 1.3 &  46 &  2.6 &  92 & 1.3 & - & 160 \\
	& $1^\circ$\dotfill         & 54 &  6.8 &  77 & 4.7 & 230 &  2.2 & 106 & 4.8 & 85 & 163 \\
	& $0.5^\circ$\dotfill       & 60 &  5.3 &  62 & 5.1 & 306 &  1.8 & 109 & 5.1 & 96 & 165 \\
	& SMR ($0.5^\circ$)\dotfill & 61 &  6.0 &  73 & 5.0 & 279 &  2.0 & 108 & 5.0 & 86 & 163 \\
	\hline
	300 & $4^\circ$\dotfill     & 36 & 25.5 & 173 & 5.3 & 101 &  8.8 & 168 & 5.3 & - & 182 \\
	& $2^\circ$\dotfill         & 47 & 19.6 & 137 & 6.6 & 183 &  7.3 & 198 & 6.7 & - & 197 \\
	& $1^\circ$\dotfill         & 55 & 11.4 &  97 & 6.4 & 368 &  3.8 & 216 & 6.5 & 86 & 206 \\
	& $0.5^\circ$\dotfill       & 59 &  9.3 &  83 & 6.7 & 443 &  3.3 & 219 & 6.7 & 96 & 204 \\
	& SMR ($0.5^\circ$)\dotfill & 62 & 10.2 &  92 & 6.9 & 515 &  3.0 & 217 & 7.0 & 110 & 207 \\
	\hline
	350 & $4^\circ$\dotfill     & 34 & 33.6 & 184 & 6.2 &  97 & 13.2 & 206 & 6.2 & - & 198 \\
	& $2^\circ$\dotfill         & 47 & 23.3 & 150 & 7.3 & 322 &  6.4 & 281 & 7.3 & - & 220 \\
	& $1^\circ$\dotfill         & 56 & 15.9 & 121 & 7.3 & 413 &  5.6 & 312 & 7.4 & 84 & 238 \\
	& $0.5^\circ$\dotfill       & 57 & 12.8 &  96 & 7.6 & 631 &  3.8 & 313 & 7.7 & 92 & 229 \\
	& SMR ($0.5^\circ$)\dotfill & 63 & 13.3 & 109 & 7.7 & 535 &  4.3 & 293 & 7.8 & 88 & 231 \\
	\hline
	400 & $4^\circ$\dotfill     & 35 & 35.8 & 172 & 7.3 & 132 & 11.5 & 208 & 7.3 & -  & 175 \\
	& $2^\circ$\dotfill         & 45 & 27.6 & 168 & 7.3 & 286 &  9.5 & 362 & 7.5 & -  & 240 \\
	& $1^\circ$\dotfill         & 58 & 20.8 & 156 & 7.6 & 549 &  6.2 & 437 & 7.8 & 88 & 279 \\
	& $0.5^\circ$\dotfill       & 59 & 14.5 & 111 & 7.7 & 682 &  4.6 & 400 & 7.9 & 86 & 250 \\
	& SMR ($0.5^\circ$)\dotfill & 63 & 14.6 & 111 & 8.3 & 715 &  4.1 & 348 & 8.4 & 94 & 233 \\
	\hline
\end{tabular}
\begin{minipage}{\textwidth}
\tablecomments{
Numerical Reynolds numbers, Re, corresponding numerical
viscosities, $\nu_\mathrm{N}$, estimates of the energy-containing eddy scales,
$\tilde{L}$ and $L$, respectively, and characteristic velocities of the
largest eddies, $v_0$. 
We show the results of the two evaluation methods under consideration, namely once
based on the Taylor microscale (AOR+) and a second case based on the energy
dissipation rate (MKJ).
Note that the two methods are meaningful only at times when postshock 
turbulence is fully developed. When this is not the case, the values of 
Re and $\nu_\mathrm{N}$ are set in parentheses.
$\ell_\mathrm{diss}$ represents the spherical harmonics mode at which clear
deviations from a Kolmogorov-like energy spectrum become evident and the
dissipation range of kinetic energy sets in (see open circles in
Fig.~\ref{fig:spectra}).
$R_0$ is the radius of evaluation as defined in Eq.~\eqref{eq:r0}. In the 
SMR model, $R_0$ lies in the layer of $1^\circ$ angular resolution until $\sim$250\,ms, 
and in the region with $0.5^\circ$ resolution at later times (see values in parentheses 
in the second column).
}
\end{minipage}
\label{tab:Re}
\end{table*}

We present the Reynolds numbers and numerical viscosities calculated with the
two methods described above in Table~\ref{tab:Re}. The procedure of
\citet{Abdikamalov2015} based on the Taylor microscale is denoted by AOR+ and
compared to our method (denoted by MKJ) based on the energy dissipation rate.

The Reynolds numbers computed with the AOR+ method have only a very weak
resolution dependence and exhibit only marginal changes after 200\,ms post bounce in models with developed postshock turbulence. Note that in Table~\ref{tab:Re}, numbers for Re and $\nu_\mathrm{N}$ are set in parentheses when the corresponding models had not yet developed such turbulent conditions in the postshock layer.

While the grid spacing in our model with $0.5^\circ$
angular resolution is a factor of eight finer
in each angular direction
than in the $4^\circ$ case, the
Reynolds number for the AOR+ estimate
increases only by a factor of $\sim$2
to values around 60. 
Such values of the
Reynolds numbers seem to be too low in view of the well developed
Kolmogorov-like turbulent cascade witnessed over roughly one order of magnitude of
$\ell$ in Fig.~\ref{fig:spectra}, and they appear to be underestimated also in
comparison to other results discussed in the literature. Based on a systematic
study of \citet{Porter1994}, \citet{Keil1996}, for
example, estimated values of $\mathrm{Re} \sim 1000$ for 2D simulations with
$1.5^\circ$ resolution performed with \textsc{Prometheus}.

With our method (MKJ), we obtain values for $\mathrm{Re}$ that are more
consistent with the observed flow behavior in the neutrino-heated postshock
layer. During the shock-stagnation phase (which lasts until $\sim$230\,ms after bounce), Reynolds numbers of up to $\sim$200
are reached in the models with at least $1^\circ$ angular resolution.

From 200\,ms to 225\,ms, the Reynolds numbers remain relatively constant for
fixed resolution in our analysis. This indicates that the turbulent cascade is
in steady-state conditions during this period. We do not witness any evidence that 
the scaling relations of classical turbulence theory might not be applicable here.
After 250\,ms, the shock expansion leads to an increasing value of $L$ and a corresponding increase of the Reynolds number, in contrast to values obtained with the AOR+ approach.

With increasing angular resolution of the simulations, we see also increasing Reynolds numbers because of decreasing values of the numerical viscosity, as expected from the point of view that better resolution should reduce viscous damping of the turbulent flow by dissipative effects associated with the grid discretization.
Models that do not follow this trend have not yet fully developed turbulence in the postshock layer, for which reason the MKJ estimates become unreliable.
The cor\-re\-spond\-ing numerical results are set in parentheses in Table~\ref{tab:Re}.
The numbers of Re and $\nu_\mathrm{N}$ for the SMR models are close to those models with fixed resolution that match the resolution of the SMR models at the radius of evaluation, $R_0$.

The two approaches, AOR+ and MKJ, described above yield Reynolds numbers and
numerical viscosities that differ significantly. In order to analyze why this
is the case, we divide Eqs.~(\ref{eq:Reynoldsnumber1}) and
(\ref{eq:numericalviscosity1}) by Eqs.~(\ref{eq:Reynoldsnumber2}) and
(\ref{eq:numericalviscosity2}), respectively, i.e., we divide the values calculated with the
method of \citet[AOR+]{Abdikamalov2015} by the values obtained with our
procedure (MKJ). The ratio of the Reynolds numbers reads
\begin{equation}
    R_\mathrm{Re} \defeq \frac{\mathrm{Re}^\mathrm{AOR+}}{\mathrm{Re}^\mathrm{MKJ}}
    = \frac{\tilde{L}^2Z/E}{2\Zspecific v_0L/\varepsilon}
    = \frac{\varepsilon \tilde{L}^2}{4\pi v_0^3 L},
\end{equation}
and the ratio of the numerical viscosities is given by
\begin{equation}
    R_{\nu_\mathrm{N}} \defeq
    \frac{\nu_\mathrm{N}^\mathrm{AOR+}}{\nu_\mathrm{N}^\mathrm{MKJ}}
    = \frac{v_0E/\tilde{L}Z}{\varepsilon/2\Zspecific}
    = \frac{4\pi v_0^3}{\varepsilon \tilde{L}},
\end{equation}
where in addition to the mentioned equations we also made use of
Eqs.~(\ref{eq:lambda1}) and (\ref{eq:v01}), and of $Z \approx \rho_0 \Zspecific$.
Ideally, both quantities ---$R_\mathrm{Re}$ and
$R_{\nu_\mathrm{N}}$--- should be unity. However, the ratio $R_\mathrm{Re}$
can be as small as 0.1, while $R_{\nu_\mathrm{N}}$
can reach values of 3--4 (see Table~\ref{tab:Re}). Since the estimates
of $v_0$ from both methods are basically identical and $\tilde{L}$
and $L$
do not differ by more than 2--3 in most cases (see Table~\ref{tab:Re}), we conclude that
our MKJ appoach yields lower estimates for the numerical energy-dissipation rate, $\varepsilon$. In other words,
the method of AOR+ intrinsically overestimates the numerical
energy-dissipation rate.

In our approach, the energy-dissipation rate $\varepsilon$
is directly measured from the energy
spectrum according to Eq.~(\ref{eq:energydissipationrate}). The only term in
Eq.~(\ref{eq:energydissipationrate}) being not precisely known is the constant $C$.
However, it was determined to satisfactory accuracy and even the largest possible
reduction of $C$ within the error bars mentioned by \citet{Sreenivasan1995}
would enhance $\varepsilon$ only by $18\%$. For the proposed best value of $C$,
we have measured the peak amplitude of $\varepsilon(\ell)$ to maximize the
numerical viscosity. Hence, we conclude that a significant underestimation of
the energy dissipation rate is unlikely.

Another ambiguity concerns the length scales of the relevant turbulent eddies.
The values in Table~\ref{tab:Re} sug\-gest that
$\tilde{L}$ according to \citet{Abdikamalov2015} is sometimes too large,
especially in cases where $\tilde{L}$ is larger than the gain-layer width.
Our calculated values of $L$ are often more than a factor of two different from
$\tilde{L}$ (smaller at early times and larger towards the end of our simulations).
These values of $L$ represent the true size of the largest eddies, which marks an upper limit for the energy-containing eddy scale, i.e.\ for $\tilde{L}$.
Determining the relevant length scale from the spectral shape
as done by \citet{Abdikamalov2015} seems problematic, because
$\tilde{L}$ varies significantly with grid resolution, whereas on grounds of physics one would expect that
the size of
the energy-containing eddies, corresponding to the
energy-weighted mean of the multipole order, should hardly depend on the grid resolution.
Since the energy-injection scale is determined by the thickness of the layer of main neutrino-energy deposition, this scale should be very similar between all simulations within the HTCL model set.
Also the location of the peak of the power spectrum is fairly similar in all models with fully developed turbulence, as can be seen in Fig.~\ref{fig:spectra}.
Moreover, solely the radial thickness of the gain layer constrains the diameter of the largest turbulent structures, which is exactly the motivation for our choice of $L$ in Eq.~(\ref{eq:largesteddysize2}).
For all these reasons, it does not seem plausible that the relevant scale for estimating the Reynolds number, as in the approach of \citet{Abdikamalov2015}, exhibits a strong variation with the angular resolution of the model.

We therefore suspect that the misjudgment of the relevant turbulent eddy scale in the AOR+ method might be the underlying cause for the counterintuitively weak variation of the Reynolds number with angular resolution and for the strange fact that Re is basically independent of the growing radial diameter of the postshock layer at later times (Table~\ref{tab:Re}).
Furthermore, note that due to the quadratic dependence in Eq.~(\ref{eq:Reynoldsnumber1}), the Reynolds number in the AOR+ approach is more sensitive to the length scale than in our method.

Besides these considerations of over- and un\-der\-es\-ti\-mat\-ed eddy scales, the numerical
constant used in the calculations of the Reynolds
numbers in the approach of \citet{Abdikamalov2015} appears to be too
low in general. The factor $5$ in Eq.~(\ref{eq:Reynoldsnumber1}) is disputable,
because it is smaller than what is reported in the literature.
\citet{Tennekes1972} formulated Eq.~(\ref{eq:Reynoldsnumber1}) in a different
way, namely as
\begin{equation}
    \mathrm{Re} = \frac{15}{A} \frac{\tilde{L}^2}{\lambda^2},
\end{equation}
where $A$ is an ``undetermined constant'' of order unity. \citet{Pope2000} and
\citet{Schmidt2014} used $A=3/2$, where\-as \citet{Abdikamalov2015} and also
\citet{Couch2015} applied $A=3$. Obviously, there is some am\-bi\-gu\-i\-ty with respect
to the value of this constant. The Reynolds numbers of \citet{Abdikamalov2015}
are therefore likely to be more than a factor of two too small. This highlights
an important aspect of tur\-bu\-lence theory. Many equations are obtained from
self-similarity considerations and therefore based only on proportionalities.
Scaling factors are then derived from further assumptions or they remain
undetermined.

Our approach to calculate the numerical viscosity relies on the
fundamental relations of Kolmogorov's theory and avoids the usage of other
equations. Although our values computed for the Reynolds numbers might be
slightly overestimated due to their sensitive dependence on $v_0$ and $L$, the
numerical viscosities deduced from the measured turbulent power spectra with
our formalism are not subject to
corresponding uncertainties and can therefore be considered as solid 
measures, provided that turbulence is fully developed and that Kolmogorov's 
theory is applicable.

\section{Discussion}
\label{sec:discussion}

The main result of our resolution study, namely that higher angular resolution
is beneficial for stronger shock expansion and shock revival in 3D simulations,
is in contradiction with results of a previous investigation by
\citet{Hanke2012}, whose 3D models with higher angular resolution showed the
tendency to explode later or not at all, in spite of the success of
lower-resolution cases.

Both generations of simulations differ in several aspects (see
Section~\ref{sec:numericshtcl}), namely in the use of slightly different
versions of the high-density equation of state of \citet{Lattimer1991}, minor
modifications in the neutrino-cooling description, general relativistic
corrections in the gravitational potential instead of the Newtonian gravity used
by \citet{Hanke2012}, and the replacement of the previously employed spherical
polar grid by the axis-free Yin-Yang grid. While the first three aspects have
the effect of changing the value of the neutrino luminosity needed to trigger
shock revival, they cannot explain the opposite dependence of the explosion
behavior on resolution.

The crucial change was the introduction of the Yin-Yang grid, which allows for
numerically cleaner resolution tests due to less grid-associated effects.  The
polar axis of regular spherical coordinate grids does not only possess a
coordinate singularity that can induce artifacts, but also the nonuniform cell
sizes of the angular grid with smaller azimuthal cells near the polar axis
(and thus lower numerical viscosity) may have
perturbative effects.  In all 3D simulations with a standard polar grid, we can
observe postshock convection appearing earlier near the polar axis and becoming
first visible by a buoyant plume that expands along the axis and deforms the
shock. This shock deformation creates vorticity and entropy perturbations in the
postshock flow and thus triggers the development of neutrino-driven convection or 
SASI mass motions, depending on which of these instabilities is favored to grow
faster by the physical conditions.  Therefore, in all of the 3D simulations
performed by \citet{Hanke2012}, even in the runs with the lowest angular
resolution of $3^\circ$, shock asphericity and nonradial kinetic energy were
found to rise already at 80--130\,ms after bounce. This is in sharp contrast to
our current set of models, where due to numerical viscosity in the
low-resolution ($2^\circ$ and $4^\circ$) cases, nonradial kinetic energy in the
gain layer does not appear on a visible level before 200\,ms after bounce (see
Figs.~\ref{fig:ekingain}, \ref{fig:eta}, and \ref{fig:vthetaphi}).

For this reason, the shock expansion and revival in the simulations by
\citet{Hanke2012} were strongly influenced by the presence of the polar grid
axis and the variable cell sizes of the angular grid, which had the consequence
of enhancing nonradial mass motions in the postshock region. With higher
resolution (in 3D it could be improved only moderately to 2$^\circ$ instead of
3$^\circ$) this influence seems to have lost strength, which is why the better
resolved models showed a reduced tendency to produce explosions. In the models
of the current study the angular cells of the Yin-Yang grid are basically
uniform and a polar axis is absent.  Therefore, grid-induced irregularities
occur on a much lower level and do not determine the development of nonradial
flows in the postshock layer.  Consequently, our models with higher angular
resolution and correspondingly lower numerical viscosity exhibit stronger
turbulence, which supports shock expansion and fosters explosions.

From this discussion another consequence arises: Comparisons of our results
based on the \textsc{Prometheus} code to resolution studies discussed in the
literature require great caution and are by no means straightforward.
Code-specific aspects such as the order of the employed hydro solver and
different grid setups used by different groups could play a role. It is, for
example, conspicuous that the result of \citet{Hanke2012} of low resolution
yielding more favorable conditions for explosions in 3D, was reproduced by
studies that employed Cartesian grids with static or adaptive mesh refinement
(AMR) \citep{Couch2014,Couch2015,Roberts2016} or with a combination of
overlapping grid blocks in a cubed-sphere multi-block AMR system
\citep{Abdikamalov2015}, or, as in the study by \citet{Radice2016}, with a
spherical mesh but a computational domain that was constrained to an octant with
inner and outer radial boundaries, using periodicity in the angular directions
and a reflecting boundary condition at the inner boundary.  While possible
artificial effects of such a constrained simulation volume with polar
coordinates have not been explored yet (\citealt{Roberts2016} and
\citealt{OConnor2018} also investigated cases with octant symmetry but employed
Cartesian AMR), it is known that Cartesian grids impose perturbations on radial
flows. Even the boundaries between AMR or grid domains with different
resolutions or geometry could have problematic numerical consequences, similar
to what we observed at the resolution boundaries of our SMR grid. One might
speculate that Cartesian grids with higher resolution create a lower level of
numerical noise, thus leading to weaker driving of 
postshock turbulence and therefore less
beneficial conditions for explosions. This might be the reason why Cartesian
setups with lower resolution produced faster explosions, while the authors of
the corresponding papers attributed this result to greater nonradial kinetic
energy on the lowest-order multipolar scales because of suppressed cascading of
turbulent energy to high-$\ell$ scales.

While more extended speculations about the possible impact of grid effects do
not appear very productive in default of investigations of how different codes
with different grid setups perform on the same well-controlled test problems, it
is clear from all of what was said above that Cartesian results cannot be
contrasted with results from polar grids by simply identifying the size of the
Cartesian cells with an effective angular resolution of a spherical grid
\citep[see, e.g.,][]{Couch2013,Ott2013,Abdikamalov2015,Couch2015,OConnor2018}.
Numerical artifacts associated with Cartesian grids and polar grids are too
different and may even govern the solutions. Moreover, changes of the resolution
in radial and angular directions can have different consequences
\citep[see][]{Hanke2012}, but in Cartesian simulations they cannot be varied
independently. Correspondingly, in 3D supernova simulations with Cartesian grids
the minimum cell size in the vicinity of the steep density decline near the
surface of the proto-neutron star is typically around 500\,m or even more
\citep[e.g.,][]{Couch2013,Couch2014,Couch2015,Dolence2013,Ott2013,Kuroda2016,Roberts2016,OConnor2018},
similar to what was employed in recent 3D calculations with the \textsc{Fornax}
code using spherical (dendritic) coordinates \citep{Vartanyan2019,Burrows2019}.
In contrast, in applications of the \textsc{Prometheus} code with simplified
neutrino treatment as well as the \textsc{Prometheus-Vertex} code with elaborate
and computationally expensive neutrino transport, the radial resolution in the
same region is chosen to be much finer, and it is improved with time as the density
gradient gradually steepens, to become as good as 50--100\,m after 500\,ms post
bounce.  It is evident that more studies, including direct comparisons of
different codes with different grids, applied on the same test problems, are
needed to disentangle numerical and physical effects in the growing suite of 3D
supernova models.

\section{Summary and Conclusions}
\label{sec:conclusions}

\subsection{Summary}

In this paper, we investigated the resolution dependence and convergence
properties of 3D simulations with the \textsc{Prometheus-Vertex} supernova code.
Because of limited computational resources, previous neutrino-hydrodynamics
calculations with this code, in particular also the successful 3D explosion
models reported by \citet{Melson2015b,Melson2015a} and \citet{Summa2018}, were
conducted with an angular cell size of $2^\circ$ for the employed polar and
Yin-Yang grids.  However, in regions where hydrodynamic instabilities and
turbulent effects play a role, in particular in the convectively unstable
neutrino-heating layer behind the stalled supernova shock, more angular
resolution is desirable. Therefore we introduced a new static mesh refinement
(SMR) procedure in our code, which can compensate the decreasing resolution (in
terms of absolute scales) associated with the geometrical widening of the
lateral and azimuthal grid zones with growing distance from the coordinate
center. This SMR grid allows us to increase the number of angular grid cells in
defined radial regions without equally increasing the number of angular zones
(also termed radial ``rays'') for the ray-by-ray-plus neutrino transport. In the
neutrino-heating layer and farther outside, where neutrinos are nearly decoupled
from the stellar background (the optical depth of these layers is typically
below 0.2), the use of less transport rays than angular zones in the
hydrodynamics solver is a viable approximation.  Such an approach saves
considerable amounts of computer time because the transport module accounts for
the dominant part of the required computational resources.

The results presented here show, however, that the SMR technique comes with some
downsides. While in the case of a robustly exploding 9\,$M_\odot$ progenitor we
did not observe any significant differences between simulations with uniformly
spaced low-resolution ($3.5^\circ$) grid and a high-resolution SMR setup, a
$20\,M_\odot$ model that evolved along the borderline between explosion and
failure showed undesirable sensitivity to the chosen grid setup. It developed an
explosion with uniform $2^\circ$ angular resolution, whereas it did not succeed
to blow up when the SMR grid was used. The SMR model failed despite the fact
that its average shock radius was transiently larger (reaching up to
170--180\,km) than in the case with uniform angular grid, where it was only
$\sim$150\,km.

We took this finding as a motivation for a systematic study that was intended to
clarify the underlying numerical reasons and to disentangle the consequences of
higher angular grid resolution from effects associated with the SMR method.  To
achieve this goal with acceptable investment of computing time, we replaced the
\textsc{Vertex} neutrino-transport treatment by a simplified heating and cooling
(HTCL) scheme and set up 20\,$M_\odot$ simulations such that the supernova shock
reached a stagnation radius of about 170--180\,km as it did temporarily in the
20\,$M_\odot$ SMR model with full-fledged neutrino transport. All exploding
models in this carefully controlled study with SMR grid and uniform resolutions
of $4^\circ$, $2^\circ$, $1^\circ$, and $0.5^\circ$ experienced shock revival
(or temporary shock expansion) at the same time. This fact enabled a
particularly clean and conclusive investigation of the influence of varied
angular resolution on the shock evolution.

As in the simulations with full neutrino transport, we observed that higher
resolution leads to a slightly larger shock stagnation radius. Moreover, better
resolved simulations do not only display a considerably earlier onset of
postshock convection in the neutrino-heating layer but also show more fine
structure in the postshock flow once turbulent convection has developed. This
corresponds to differences in the normalized turbulent power spectra with
relatively less kinetic energy on large multipolar scales ($\ell \lesssim 10$)
and relatively more power on small scales, roughly following a Kolmogorov-like
$\ell^{-5/3}$ power law from $\ell \sim 30$ up to the dissipation scale around
$\ell \sim 100$ in all cases where the resolution is better than
$\sim$1$^\circ$. Low resolution obviously delays the growth of nonradial
postshock instabilities, and the associated higher numerical viscosity prevents
cascading of kinetic energy from the largest scales to turbulent vortex flows on
smaller scales.  This goes hand in hand with lower nonradial kinetic energy and
a reduced efficiency for conversion of neutrino heating to turbulent kinetic
energy. During this phase the highest-resolved $0.5^\circ$ model still exhibits
noticeable differences compared to the SMR and $1^\circ$ simulations.

Shock expansion in reaction to the arrival of the infalling
silicon/silicon+oxygen interface at the stagnant shock finally enables the onset
of postshock convection in all of our HTCL models, also in the coarse-resolved
ones, because the decreased accretion velocity in the postshock flow allows
buoyant plumes to rise outward.  In this phase, the shock develops runaway
expansion in all cases except the $4^\circ$-degree runs. The expansion velocity
of the shock clearly shows a monotonic dependence on the resolution with possible
convergence at about $1^\circ$.  Simulations that are closer to the explosion
threshold are more sensitive to resolution changes. Models with a  resolution of
at least $1^\circ$ develop similar fluid structures. The SMR run resembles
simulations with a uniform resolution of $0.5^\circ$ in this respect. Only
towards the end of the SMR simulation, the expansion velocity of the shock drops
slightly below the value of the $1^\circ$ and $0.5^\circ$ models. Our analysis
revealed that this effect is caused by the dissipation of kinetic energy at the
interfaces of layers with different angular resolutions in the SMR setup.
Downflows propagating from the finer grid to the layer with coarser resolution
under the constraint of total energy conservation lose kinetic energy that is
transformed into internal energy. This leads to reduced pressure support of the
expanding shock because thermal energy provides pressure with an adiabatic index
of $4/3$, whereas turbulent pressure connects to turbulent kinetic energy with
an equivalent adiabatic index of 2 \citep[see][]{Radice2015}.

The bottom line is that the dynamical evolution of our 20\,$M_\odot$ HTCL models
during the shock stagnation and shock revival phases is basically identical in
3D simulations with $1^\circ$, $0.5^\circ$, and SMR grid, despite remaining
resolution-dependent differences in various measures of turbulence (e.g., the
initial growth of convective activity, the saturation level of the nonradial
kinetic energy during shock stagnation, and the detailed shape of the normalized
mode spectrum of the kinetic energy).  Even the $2^\circ$ run follows closely
when the shock expansion sets in, while the $4^\circ$ model exhibits a visibly
weaker shock expansion because higher numerical viscosity attenuates turbulent
mass motions and reduces the turbulent kinetic energy. A similar effect, though
less extreme, could also be witnessed at the interfaces of domains of different
angular resolution in our SMR setup, where kinetic energy of flows crossing
boundaries from higher to lower resolution is converted to thermal energy.  The
grid resolution has a more sensitive influence in cases that marginally overcome
the explosion threshold.

In order to quantify the influence of numerical viscosity, we introduced a
method of calculation that is based on the energy dissipation rate measured
directly from the turbulent energy spectrum. Our method differs from the
approach used by \citet{Abdikamalov2015}, who derived results based on the
Taylor scale. We determined effective numerical Reynolds numbers for the
postshock flow in our high-resolution ($0.5^\circ$, $1^\circ$, and SMR) runs of
up to several hundred, and in our $2^\circ$ and $4^\circ$ models of a few dozens
to a few hundred. Such values agree with previous rough estimates for
simulations of proto-neutron star convection with the \textsc{Prometheus} code
\citep{Keil1996}, and they are consistent with those reported by
\citet{Handy2014}. Our numerical viscosities, however, are lower by a factor of
$\sim$2--4 compared to the results we obtained with the formalism of
\citet{Abdikamalov2015}, and our numerical Reynolds numbers are considerably
higher (by up to a factor of $\sim$10)
than those computed ac\-cord\-ing to these authors.  We consider our
estimates as better compatible with the fine-structured vortex pattern witnessed
in the convective postshock flow of our high-resolution models, which is
mirrored by a near-Kolmogorov spectrum of the turbulent kinetic energy over an
order of magnitude in the spherical harmonics modes $\ell$.

We also presented a detailed evaluation of neutrino-drag terms in the
hydrodynamics equations for the conditions between neutrinosphere and supernova
shock in Appendix~\ref{sec:drag}. Interestingly, the numerical Reynolds numbers
of postshock turbulence
even in our lowest-resolved simulations, which range between
several 10 and some 100 on the relevant scales, are in the ballpark of the
damping effects associated with neutrino drag acting on the flow in the gain
layer. Concerns were expressed that current full-scale supernova models are
severely underresolved and that much higher grid resolution is needed to
describe the turbulent energy cascading in the convective postshock layer in
order to reproduce the self-similar power-law spectrum of Kolmogorov's classical
theory in the inertial range despite numerical viscosity
\citep{Couch2015,Radice2016,Radice2018}. Such reservations, however, must be
confronted with the presence of neutrino drag in this region.  Neutrino drag
has a non-negligible influence on all structures in the postshock flow that are
responsible for significant contributions to the turbulent kinetic energy.
It is neither accounted for by the leakage schemes applied for
neutrino energy and lepton sources in all previous resolution studies
\citep[e.g.,][]{Couch2014,Couch2015,Abdikamalov2015,Radice2016} nor by the
simple heating and cooling treatment used in our work, but it requires
the inclusion of neutrino-momentum transfer terms in the equation of motion
(see Appendix~\ref{sec:drag}).

\subsection{Conclusions}

In our systematic resolution study, designed in a very careful way to avoid
numerical perturbations (mainly grid-induced ``noise'' and fluctuations
associated with neutrino effects) as much as possible, we witnessed a beneficial
effect of higher angular resolution on shock revival. This can be understood by
a lower level of numerical viscosity, allowing for higher turbulent kinetic
energy because of reduced viscous damping and dissipation of nonspherical flows.
Convergence of the overall shock dynamics in 3D seems to be approached at an
angular resolution of about $1^\circ$, but simulations with a resolution of
$2^\circ$ are not far off, despite the fact that properties characterizing
turbulence, for example the exact shape of the turbulent energy spectrum and the
scale when dissipation sets in, still exhibit differences when improving the
resolution to $0.5^\circ$. Also in this context, our simulations with $1^\circ$
angular resolution match the requirements of convergence to a Kolmogorov-like
behavior (at least for $\ell \gtrsim 30$) by displaying a clear separation of inertial range and dissipation
range in the turbulent kinetic-energy spectra, contradicting a proposition by
\citet{Radice2018} that these two length scales are usually merged in
core-collapse supernova simulations and therefore misidentified.

The bottleneck effect pointed out by \citet{Radice2015,Radice2016,Radice2018},
which is present in the turbulent energy spectra even for the highest feasible
resolutions, seems to have a relatively minor influence on the overall shock
evolution. Such a possibility was also admitted by \citet{Radice2016}.  This
may be explained by the fact that most of the turbulent kinetic energy, which
produces turbulent pressure to support shock expansion, is carried by vortex
motions on large scales but not on the smallest scales near the dissipation regime.

The convergence of the supernova dynamics seen around $1^\circ$ angular
resolution provides some back-up to the successful 3D explosion models computed
with the \textsc{Prometheus-Vertex} and \textsc{Alcar} codes by
\citet{Melson2015b,Melson2015a}, \citet{Summa2018}, and \citet{Glas2018} with a
resolution of $2^\circ$.  Higher angular resolution has been found to be
supportive of an explosion, and resolution-dependent differences are mostly
relevant for cases whose postbounce evolution proceeds very close to the
threshold between successful explosion and failure. Based on our estimates of
the magnitudes of numerical viscosity and of neutrino drag acting on the flow in
the gain layer, we infer that both effects are in the same ballpark
for the numerical resolutions applied in the models of this work.
Increasing the resolution significantly beyond an angular
resolution of 1$^\circ$--2$^\circ$ is an exercise of direct relevance for the case of
core-collapse supernovae only when neutrino viscosity (at high densities where
neutrinos diffuse) and neutrino drag (outside of the diffusion regime) are taken
into account.

Using our new SMR grid setup for improving the angular resolution in defined
computational domains had the drawback that kinetic energy was converted to
internal energy in flows crossing the borders from finer to coarser grid under
the constraint of total energy conservation. In models evolving very close to the
explosion threshold, this undesirable effect was found to weaken the explosion,
whereas robustly exploding models remained unaffected. Finally, our result of
higher angular resolution being favorable for explosion in 3D contradicts the
trend seen by \citet{Hanke2012}.  The reason for this difference is the use of a
polar grid in the previous simulations, whereas an axis-free Yin-Yang grid was
applied in the current work.  The presence of the polar-axis singularity and
smaller azimuthal grid cells in the vicinity of the polar axis caused numerical
artifacts that were stronger for lower-resolution runs. Correspondingly enhanced
turbulent activity in the postshock region produced more favorable conditions
for explosion in the lower-resolved 3D models of \citet{Hanke2012}. In contrast,
the cleaner setup of the resolution study presented here revealed the opposite
behavior.  Our findings that grid effects can have a severe influence on the
solutions should also be taken as a clear warning that great caution is demanded
when resolution studies based on Cartesian grids with AMR
\citep[e.g.,][]{Couch2014,Couch2015,Abdikamalov2015} or based on constrained
volumes with radial and angular boundaries \citep[e.g., the semiglobal setup
considered by][]{Radice2016} are interpreted.  Boundary artifacts and
unavoidable numerical noise imposed on radial flows on Cartesian grids, whose
scale and amplitude differ when the resolution is varied, could affect the
results and might provoke misleading trends.

Our work constitutes a very careful investigation of
resolution effects in 3D supernova simulations with the
\textsc{Prometheus} code for comparison with results and
their interpretation in the literature. It suggests that
previous studies revealed incorrect trends, namely opposite
to our resolution-dependent results, because of numerical
artifacts associated with the existence of a polar
grid axis or numerical perturbations induced by Cartesian
grids. ``Realistic'' simulations of stellar core collapse
and explosions, however, are probably not seriously jeopardized
by any of these numerical shortcomings, maybe
not even by a moderate overestimation of numerical viscosity
in the neutrino-heating layer due to the choice of
modest spatial resolution, enforced by limited computational 
resources. The pre-collapse perturbations in the
infalling stellar matter that are caused by fluctuations
associated with convective shell burning in the progenitor
stars \citep[e.g.,][]{Couch2013a,Mueller2015a,Couch2015a,Mueller2016,
Yoshida2019,Yadav2019},
provide a strong
driving force of postshock turbulence, which is likely to
easily dominate the damping effects of numerical viscosity
in the currently best-resolved 3D supernova models
\citep[e.g.,][]{Mueller2017,Mueller2019}.

\acknowledgments
We thank Alexander Summa for stimulating discussions,
Robert Glas for improving the numerical accuracy
of our analysis of the turbulent energy spectra, and
Bernhard M\"uller and Ewald M\"uller for their valuable comments on the manuscript.
The 3D simulations were performed on \emph{SuperMUC} at the Leibniz
Supercomputing Centre with resources granted by the Gauss Centre for
Supercomputing (LRZ project IDs: pr48ra, pr53yi). For the computation of the 1D and 2D
simulations and for the postprocessing of the data, we employed the \emph{Hydra}
system of the Max Planck Computing and Data Facility (MPCDF). The project was
supported by the European Research Council through grant ERC-AdG No.\
341157-COCO2CASA and by the Deutsche For\-schungs\-ge\-mein\-schaft through
Son\-der\-for\-schungs\-be\-reich SFB 1258 ``Neutrinos and Dark Matter in Astro- and
Particle Physics'' (NDM).

\software{\textsc{Prometheus-Vertex} \citep{Fryxell1989,Rampp2002,Buras2006a},
NumPy and Scipy \citep{Oliphant2007}, IPython \citep{Perez2007}, Matplotlib
\citep{Hunter2007}.}


\section*{Appendix}
\renewcommand\thesection{\Alph{section}}
\renewcommand\thefigure{\thesection\arabic{figure}}
\renewcommand\theequation{\thesection\arabic{equation}}
\setcounter{section}{0}
\setcounter{figure}{0}
\setcounter{equation}{0}


\section{Static mesh refinement}
\label{sec:smr}

Both the spherical polar grid and the Yin-Yang grid have a common disadvantage.
The surface element $\mathrm{d} A = r^2 \, \sin \theta \, \mathrm{d} \theta \,
\mathrm{d} \phi$ is proportional to the radius squared.  For a given constant
angular resolution, the effective size of the grid cells grows with increasing
radius. Fine angular resolution near the grid center can impose severe
constraints on the Courant-Friedrichs-Lewy (CFL) time step, while coarse
resolution at large radii (behind the supernova shock) limits the possibility to
resolve turbulent flows. Here, we present a static mesh refinement (SMR) technique, which
compensates for the diverging structure of these spherical grids.

\begin{figure}
    \centering
    \includegraphics[height=14\baselineskip]{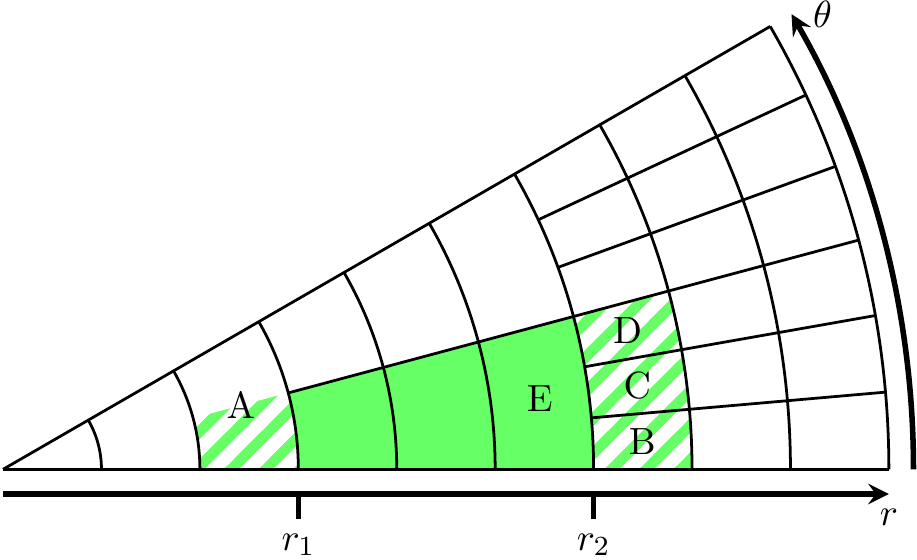}
    \caption{Example of a setup with static mesh refinement (SMR) in two
dimensions. The angular resolution increases by a factor of $6$ from the center
to the outer edge of the grid. The ghost cells for the green-shaded radial sweep
between $r_1$ and $r_2$ are illustrated as green hatched areas. Note that unlike the exemplary setup shown in this figure, all supernova applications with SMR grid discussed in the main text were performed with doubling of the resolution in each angular direction at both SMR interfaces.}
    \label{fig:smr}
\end{figure}

The SMR setup allows us to define certain radial intervals with different
angular resolutions. An example for such a setup is given in Fig.~\ref{fig:smr}
in two dimensions for coordinates $(r,\theta)$. Although this was chosen for the
sake of simplicity here, the discussion is completely analogue to the 3D case,
where it applies to both angular directions, $\theta$ and $\phi$.  Let us define
the angular resolution in Fig.~\ref{fig:smr} for $r<r_1$ as $\psi$. Then, the
resolution between $r_1$ and $r_2$ is $2\psi$, and further refined to $6\psi$
for $r>r_2$.

Generally, arbitrary integer refinement steps and an arbitrary number of
concentric layers can be chosen in our implementation. Note that the inner
spherically symmetric volume, which is used in our simulations to allow for
larger hydrodynamical time steps, can be understood as a special case with the
lowest possible angular resolution corresponding to the whole sphere.

The radial locations of the refinement boundaries can be adjusted manually in our
implementation at each restart of the code. Such a shift is motivated, for
example, by a contraction of the gain radius that should roughly be followed by
the grid setup.

\citet{Mueller2015} and \citet{Skinner2019} also implemented a similar setup,
however, with the motivation of avoiding time-step constraints as the grid cells
become smaller towards the grid singularities at the axis and at the origin. In
their ``dendritic'' grid, they coarsen only the polar grid ($\theta$) to
compensate for the diverging cell size with increasing radius. The azimuthal
mesh ($\phi$) is coarsened towards the axis, which is not required in our setup
when we use the axis-free Yin-Yang grid.

\subsection{Treatment of the hydrodynamics}

Generally, the SMR procedure only affects the radial direction. The computations
of $\theta$ and $\phi$ sweeps remain unchanged. Besides the spherical polar
grid, using the Yin-Yang grid is thus also possible and only requires a slight
modification. As the angular resolution changes in the computational domain, the
Yin-Yang ghost cell positions providing the boundary conditions for the angular
sweeps are different for each refinement layer, which has to be considered when
data is exchanged between both grid patches.

Radial sweeps are computed separately for every resolution layer. For an
example, consider the green-filled grid cells in Fig.~\ref{fig:smr}, for which
the radial sweep should be calculated. Let us assume that one ghost cell is
needed below and one above this sweep, depicted as green-hatched areas. On the
lower side, the ghost cell data can directly be taken from cell A. This is
because finite-volume methods generally assume that values in a cell always
represent averages. At the upper end of the sweep, we need to average over
cells B, C, and D to get the required data.  Generally, averaging is required
for the ghost cells located in the finer neighboring layer. This is the reason
why only integer refinement steps are allowed in the SMR construction.
Otherwise, computing the averages would be difficult and introduce additional
interpolation errors.

\subsection{Flux correction}

If we naively used the averaged ghost cell data from the finer layer at the outer
end of a radial sweep, numerical inaccuracies would occur resulting in the
violation of conservation laws.  In order to ensure exact conservation of the
conserved quantities in our hydrodynamical scheme, a ``flux correction''
algorithm is applied acting on the outermost cell of each refinement region. We
will explain this procedure with the aid of Fig.~\ref{fig:smr}.  Note that the
flux correction is also done for the spherically symmetric innermost volume,
which is usually employed in 3D to mitigate the time step constraints at the grid
origin.

In our example illustrated in Fig.~\ref{fig:smr}, the flux correction considers the
interface between the cell E and the cells B, C, and D at $r_2$. After all
radial sweeps have been calculated, the quantity $\mathcal{X}$---a placeholder for a
conserved quantity in our hydrodynamical scheme---should be exactly conserved. The
considered interface at $r_2$ is part of four different radial sweeps: the sweep
between $r_1$ and $r_2$ containing the cell E, and the three sweeps beyond $r_2$
containing the cells B, C, and D. In the following, we discuss the different
Riemann fluxes of $\mathcal{X}$ at the interfaces between these four cells and
their corresponding ghost zones. Let $\mathcal{F}_\mathrm{E}$ be the flux density
of $\mathcal{X}$ at the \textit{upper} edge of the cell E. Likewise,
$\mathcal{F}_\mathrm{B}$, $\mathcal{F}_\mathrm{C}$, and $\mathcal{F}_\mathrm{D}$, denote
the flux densities at the \textit{lower} boundaries of the cells B, C, and D,
respectively. All these fluxes are evaluated at the interface $r_2$ with
positive values corresponding to the radially outward direction. The areas of
the cell interfaces are labeled $A_\mathrm{E}$, $A_\mathrm{B}$, $A_\mathrm{C}$, and
$A_\mathrm{D}$. Theoretically, it should hold
\begin{equation}
\mathcal{F}_\mathrm{E} A_\mathrm{E} \stackrel{\text{!}}{=} \mathcal{F}_\mathrm{B}
A_\mathrm{B} + \mathcal{F}_\mathrm{C} A_\mathrm{C} + \mathcal{F}_\mathrm{D} A_\mathrm{D}.
\end{equation}
This is, however, not guaranteed numerically. In order to cure this problem, the
value of $\mathcal{X}$ in cell E is updated to
\begin{equation}
\mathcal{X}_\mathrm{new} = \mathcal{X} + \Delta t \, \frac{\mathcal{F}_\mathrm{E}
A_\mathrm{E} - \mathcal{F}_\mathrm{B} A_\mathrm{B} - \mathcal{F}_\mathrm{C} A_\mathrm{C} -
\mathcal{F}_\mathrm{D} A_\mathrm{D}}{V_\mathrm{E}},
\end{equation}
where $\Delta t$ is the time step length and $V_\mathrm{E}$ the volume of cell E.
In this way, differences of Riemann fluxes are converted into updates of
conserved quantities. This is done for all variables at all interfaces between
layers of different angular resolution to ensure that global conservation laws
are fulfilled numerically.

\subsection{Treatment of the neutrino transport}

The implementation of the neutrino transport requires some modifications when the
SMR procedure is used. Generally, the computational grids, on which the
hydrodynamics and the neutrino transport are computed, should be identical in
the regime where neutrinos are tightly coupled to the stellar matter. Otherwise,
the source terms given by the transport solution are not well-balanced with the
internal energy density of the fluid, for which reason deviations from thermodynamic
equilibrium can arise \citep{Rampp2002}. These, in turn, can cause numerical
fluctuations perturbing the simulation.

In the context of the SMR grid, the coarsest angular grid of the innermost
multi-dimensional refinement layer is chosen to coincide with the transport angular
grid so that the number of transport rays can remain unchanged throughout the
radial grid. For example, the
complete wedge shown in Fig.~\ref{fig:smr} would be one transport ray with the
first refinement step at $r_1$ being placed in a region where the neutrino opacity is
already sufficiently low and neutrinos are not expected to reach equilibrium
with the fluid. For $r>r_1$, the input for the neutrino transport solver is
obtained by angular averages of the hydrodynamical quantities, similar to the
filling procedure of the ghost cells of different angular resolution patches.
The computed source terms for the hydrodynamical equations are finally applied to
all hydrodynamic cells crossed by the particular transport ray.

\section{Neutrino drag in the gain layer}
\label{sec:drag}

In the high-density regime of a hot neutron star, 
neutrino diffusion applies because the neutrino
mean free paths are much shorter than the length scales of
structures and gradients of the stellar medium. In this
limit neutrino momentum transfer by scattering and absorption
creates viscosity 
\citep[e.g.,][and references therein]{Keil1996,Guilet2015}.
Thus it affects shear flows, has a
damping influence on velocity fluctuations, and creates viscous
dissipation of kinetic energy. In the turbulent neutrino-heating
layer, however, the optical depth for neutrinos is typically less 
than $\sim$0.2. The neutrino mean free paths are much longer than
the scales of velocity perturbations. In this case neutrino 
momentum transfer cannot be treated as a viscous process but 
neutrinos exert a drag force on the stellar plasma. The drag
is connected to Doppler effects seen from the comoving frame
of the fluid and acts opposite to the direction
of the plasma motion. Since the neutrino-radiation field is 
highly anisotropic in the gain region, drag calculations
available in the literature \citep[e.g.,][]{Subramanian1998,Nio1998,Agol1998,Guilet2015},
which assume isotropic radiation in the laboratory, 
are not directly applicable.
We therefore provide a more detailed assessment for the 
relevant SN conditions in the following.

\subsection{Fundamental equations}
\label{sec:equations}

In the fluid momentum equation the radiation force appears as a
source term \citep{Mihalas1984,Hubeny2014}:
\begin{equation}
\frac{\partial(\rho\vect{v})}{\partial t} + \nabla\cdot\left[\,
\rho(\vect{v}\otimes\vect{v}) + p\,\mathbf{I} - \mathbf{S}\,\right]
= \rho\vect{g} + \vect{G}\,,
\label{eq:momeq}
\end{equation}
where $\rho$, $\vect{v}$, $p$, $\mathbf{S}$, and $\vect{g}$ are
the stellar-fluid density, velocity, pressure, viscous stress
tensor, and gravitational acceleration, $\mathbf{I}$ is the unit
matrix, $(\vect{a}\otimes\vect{b})_{ij} = a_ib_j$ the outer
(dyadic) product of two vectors, and $\vect{G}$ the net radiative 
force density on the material, i.e., the rate with which radiation
and matter exchange momentum (in both directions) per unit volume:
\begin{equation} 
\vect{G} = \frac{1}{c}\int_0^\infty\mathrm{d}\epsilon
\oint\mathrm{d}\omega\,\, \vect{n}\, 
(\chi_{\vect{n},\epsilon}I_{\vect{n},\epsilon}-\eta_{\vect{n},\epsilon})\,.
\label{eq:g}
\end{equation}
Here $I_{\vect{n},\epsilon} = I(\vect{r},t;\vect{n},\epsilon)$
denotes the radiation intensity, which depends on time $t$, spatial
position $\vect{r}$, neutrino energy $\epsilon$, and neutrino momentum
direction $\vect{n}$, $\chi_{\vect{n},\epsilon} = 
\chi(\vect{r},t;\vect{n},\epsilon)$ is the extinction coefficient
(opacity) containing true absorption and scattering contributions,
and $\eta_{\vect{n},\epsilon} = \eta(\vect{r},t;\vect{n},\epsilon)$
is the total emissivity including a thermal-emission part and
a scattering part. $\vect{G} = \vect{G}(\vect{r},t)$ also appears
with the opposite sign on the right-hand side (rhs) of the total radiation momentum
equation:
\begin{equation}
\frac{1}{c^2}\,\frac{\partial\vect{F}}{\partial t} + 
\nabla\cdot \mathbf{P} = -\,\vect{G}
\label{eq:radmom}
\end{equation}
with $\vect{F} = \vect{F}(\vect{r},t)$ being the radiation flux
density and $\mathbf{P} = \mathbf{P}(\vect{r},t)$ the radiation 
pressure tensor. 

All radiation quantities in these equations (radiation intensity,
radiation moments, energy, direction of motion) are measured in
the laboratory frame. For evaluating the radiation force density,
it is most convenient to consider $\vect{G}$ in its mixed-frame
form where the material coefficients are computed in the comoving
(rest) frame of the stellar fluid while radiation quantities
and energies are in the inertial (laboratory) frame. With 
$\chi_0$ and $\eta_0$ denoting the fluid-frame quantities,
one gets to first order in $(v/c)$ \citep{Mihalas1984}:
\begin{eqnarray}
\vect{G} = &\phantom{-}& \frac{1}{c}\int_0^\infty\mathrm{d}\epsilon\,
\chi_0^\mathrm{t}(\epsilon)\vect{F}(\epsilon) \nonumber \\
&-& \frac{\vect{v}}{c}\int_0^\infty\mathrm{d}\epsilon\,
\frac{4\pi}{c}\,\eta_0(\epsilon)   \nonumber \\
&-& \frac{\vect{v}}{c}\cdot \int_0^\infty\mathrm{d}\epsilon\,
\left[\chi_0(\epsilon) + 
\epsilon\,\frac{\partial\chi_0}{\partial\epsilon}\right]
\,\mathbf{P}(\epsilon) = \nonumber \\
=&& \vect{G}^{(1)} + \vect{G}^{(2)} + \vect{G}^{(3)},
\label{eq:radforce}
\end{eqnarray}
where in contrast to \citet{Mihalas1984} we have 
accounted for the anisotropy of neutrino-nucleon and neutrino-nuclei
scattering, which leads to the transport opacity 
(indicated by the superscript ``t'' and defined in
Section~\ref{sec:evaluation})
appearing in the first term on the rhs instead of the
angle-integrated opacity. The first term on the rhs
of Eq.~(\ref{eq:radforce}) corresponds to the force density
of the radiation acceleration, while the fluid-velocity dependent
terms denote the force density for the radiation 
drag.\footnote{In Newtonian hydrodynamics the term $\vect{G}^{(2)}$ 
should be omitted, because it would be compensated in the special 
relativistic modeling by a reduction of the relativistic mass (i.e., no 
change of the velocity would take place).}
Equation~(\ref{eq:radforce}) is compatible with expressions
for the radiation drag applied, e.g., by 
\citet{Agol1998,Nio1998}.
While these works focused on isotropic
radiation fields and (isotropic) Thomson scattering of photons off
electrons, we will evaluate Eq.~(\ref{eq:radforce}) in 
Section~\ref{sec:evaluation} for neutrino absorption, emission and
scattering in reactions with nucleons and nuclei in the gain layer,
taking into account the anisotropy of the neutrino momentum 
distribution in this region.
In Section~\ref{sec:evaluation_s20}, we will quantify the 
neutrino-drag effects in more detail by evaluating results from a 
20\,$M_\odot$ core-collapse simulation with \textsc{Prometheus-Vertex}.

\subsection{Order of magnitude estimates}
\label{sec:estimates}

The force density for the neutrino radiation acceleration
(first term) on the rhs of Eq.~(\ref{eq:radforce}) scales with
the product of the total neutrino opacity, $\chi_0^\mathrm{t} = 
\kappa^\mathrm{t} = \kappa_\mathrm{a} + \kappa_\mathrm{s}^\mathrm{t}$
(absorption opacity plus transport opacity for scattering;
see Section~\ref{sec:evaluation}), 
and the total (energy integrated)
neutrino flux, $F(r)$, which can be written also in terms of
the luminosity, $F = L/(4\pi\,r^2)$:
\begin{equation}
G_\mathrm{acc} \propto \frac{1}{c}\,\langle\kappa^t\rangle\,F
= \frac{1}{c}\,\frac{\langle\kappa^t\rangle\,L}{4\pi\,r^2}\,,
\label{eq:radacc1}
\end{equation}
where the angle brackets indicate a suitable average over the
radiated neutrino spectrum. Summing up the contributions of all
neutrino species $\nu_i$ ($i$ running from 1 to 6 for 
$\nu_e$, $\bar\nu_e$ and the four types of heavy-lepton
neutrinos), and setting the neutrino acceleration force in 
relation to the gravitational force by the neutron star of 
mass $M$, ${\cal G}M\rho/r^2$ (${\cal G}$ being the 
gravitational constant), one gets
\begin{equation}
\frac{\sum_iG_{\mathrm{acc},\nu_i}}{{\cal G}M\rho/r^2} = 
\sum_i \frac{L_{\nu_i}\langle\kappa^t\rangle_{\nu_i}}{4\pi c\,{\cal G}M\rho}
\sim\sum_iL_{\nu_i}\,\frac{\langle\kappa^t\rangle}{4\pi c\,{\cal G}M\rho}\,.
\label{eq:radacc2}
\end{equation}
In the transformation to the final form
we have assumed that the opacities
off all kinds of neutrinos are (very roughly) similar. This is a valid
assumption because the dominant interactions of $\nu_e$ and 
$\bar\nu_e$ are charged-current absorption as well as neutral-current
scattering with free nucleons (and, if present, with nuclei),
whereas heavy-lepton neutrinos only undergo scatterings but
have significantly higher root mean square energies (the cross sections of all
mentioned processes scale with the square of the neutrino energy). 
Now considering that the optical depth of the gain layer is
typically $\tau_\mathrm{gain} \approx \langle\kappa^t\rangle(R_\mathrm{sh}-R_\mathrm{gain})$
and the width of the gain layer
(difference between average shock radius, $R_\mathrm{sh}$, and 
average gain radius, $R_\mathrm{gain}$) is typically around $10^7$\,cm,
we estimate that $\langle\kappa^t\rangle\sim 10^{-8}$\,cm$^{-1}$,
i.e., the average neutrino mean free path is roughly ten times
larger than the diameter of the gain layer. With that we obtain
\begin{equation}
\frac{G_\mathrm{acc}}{{\cal G}M\rho/r^2} \lesssim
\frac{1}{40}\,\frac{L_{\nu,53}\langle\kappa^t\rangle_{-8}}{M_{1.5}
\,\rho_9}\,,
\label{eq:radacc3}
\end{equation}
where $G_\mathrm{acc} = \sum_iG_{\mathrm{acc},\nu_i}$,
$L_\nu = \sum_iL_{\nu_i}$, and $L_\nu$, $\langle\kappa^t\rangle$,
$M$, and the density in the gain region, $\rho$, have been 
normalized by $10^{53}$\,erg\,s$^{-1}$, $10^{-8}$\,cm$^{-1}$,
1.5\,$M_\odot$, and $10^9$\,g\,cm$^{-3}$, respectively.
We thus confirm the general understanding that neutrino
momentum transfer is a small effect in the gain layer because
the neutrino luminosity is far below the Eddington limit.
This is just a manifestation of the fact that neutrino-driven
supernova explosions are powered by neutrino heating rather
than being caused by neutrino-momentum transfer.

In the second and third terms 
on the rhs of Eq.~(\ref{eq:radforce}) the emissivity
$\frac{4\pi}{c}\eta_0$ as well as the integrand depending on
$\mathbf{P}$ scale similarly with the product of neutrino
opacity and the specific neutrino-energy density
$E(\epsilon) = \partial E/\partial\epsilon$ 
(see Section~\ref{sec:evaluation}). Thus we can recover the
scaling relation used by \citet{Guilet2015} 
for the damping rate $\Gamma$ associated with the neutrino drag:
\begin{equation}
G_\mathrm{drag} \propto -\,\rho\,\Gamma\,v
\label{eq:dragacc}
\end{equation}
with\footnote{Note that $\kappa$ in Eqs.~(\ref{eq:gamma1}), (\ref{eq:gamma2}), and (\ref{eq:gamma3}) stands for $\chi_0(\epsilon) + \partial\chi_0/\partial\epsilon$ (see third term in Eq.~(\ref{eq:radforce})), with $\chi_0 = \kappa = \kappa_\mathrm{a} + \kappa_\mathrm{s}$.}
\begin{equation}
\Gamma \sim \frac{E\langle\kappa\rangle}{\rho\,c}
= \frac{F\langle\kappa\rangle}{\langle\xi\rangle\rho\,c^2}
= \frac{L\langle\kappa\rangle}{4\pi r^2\langle\xi\rangle\rho\,c^2}
\,,
\label{eq:gamma1}
\end{equation}
where
$\langle\xi\rangle=\langle\xi(r)\rangle=F/(Ec)$ is the spectral
average of the flux factor $\xi$. Using 
$\langle\xi\rangle=\langle\xi(r) \rangle\sim 1$ for the gain
layer and applying the same assumptions as employed in the 
case of the neutrino acceleration term, we derive
\begin{equation}
\Gamma \sim \frac{\sum_iL_{\nu_i}\langle\kappa\rangle_{\nu_i}}{4\pi
\,r^2\langle\xi\rangle\rho\,c^2} \sim \left(1\,\mathrm{s}^{-1}\right) 
\frac{L_{\nu,53}\langle\kappa\rangle_{-8}}{r_7^2\,\rho_9}\,,
\label{eq:gamma2}
\end{equation}
where $r_7 = r/(10^7\,\mathrm{cm})$ is a radial location between
$R_\mathrm{gain}$ and $R_\mathrm{sh}$. This result can be easily 
understood when Eq.~(\ref{eq:gamma1}) is slightly rewritten by
introducing $\rho = n_\mathrm{B}m_\mathrm{B}$ ($n_\mathrm{B}$ is
the baryon number density and $m_\mathrm{B}$ the average baryon
mass):
\begin{equation}
\Gamma \sim \frac{\sum_i F_{\nu_i}\langle\kappa\rangle_{\nu_i}}
{\langle\xi\rangle n_\mathrm{B}}
\cdot \frac{1}{m_\mathrm{B}\,c^2} \,.
\label{eq:gamma3}
\end{equation}
Here, the first factor is a multiple of the neutrino-heating
rate per baryon in the gain layer, because it includes not only 
the heating reactions of $\nu_e$ and $\bar\nu_e$ absorptions
by nucleons but also the scattering processes of all kinds of 
neutrinos with nucleons, which transfer a similar amount of momentum
per interaction. The total effect is therefore typically on the 
order of 1000\,MeV\,s$^{-1}$ per nucleon, to be compared with a 
nucleon rest mass-energy of roughly 
$m_\mathrm{B}c^2\approx 940$\,MeV. 

The Reynolds number for a viscous medium is defined as
\begin{equation}
\mathrm{Re} \defeq \frac{v\,l}{\mu_\mathrm{vis}} \,,
\label{eq:re1}
\end{equation}
which weighs the importance of the inertial term 
$\nabla\cdot(\rho\vect{v}\otimes\vect{v})$ relative to the
viscous term $\nabla\cdot\mathbf{S}$ with components
$S_{ij}\propto\rho\mu_\mathrm{vis}(\partial v_i/\partial x^j)$.
In Eq.~(\ref{eq:re1}) $v$ and $l$ are typical amplitudes and 
length scales of velocity 
perturbations\footnote{\label{fn:naming_convention}In this appendix, 
in contrast to the naming convention employed in Section~\ref{sec:turbulence} 
(see, e.g., Eq.~(\ref{eq:Reynoldsnumber2})), we use $l$ for the 
typical length scales in order to avoid confusion with the luminosity, 
which is labeled with $L$ here.} and $\mu_\mathrm{vis}$
is the kinematic shear viscosity (with units [cm$^2$\,s$^{-1}$]).
In analogy to Re one can also define a ``drag number'', Dr, to
quantify the ratio of inertial term and drag term. With
Eq.~(\ref{eq:dragacc}) this yields
\begin{equation}
\mathrm{Dr} \defeq \frac{\rho\,v\,v}{l\,|G_\mathrm{drag}|} = 
\frac{v}{l\,\Gamma}\,.
\label{eq:dr1}
\end{equation}
With the value of $\Gamma$ estimated for the gain layer
(Eq.~(\ref{eq:gamma2})) and typical flow velocities of 
$v \sim 10^9$\,cm\,s$^{-1}$ on the largest spatial scales,
$l \sim R_\mathrm{sh}-R_\mathrm{gain} \sim 10^7$\,cm, we 
obtain a value for Dr of the order of
\begin{equation}
\mathrm{Dr} \sim 100\,\frac{v_9}{l_7}
\left(\frac{\Gamma}{1\,\mathrm{s}^{-1}}\right)^{-1}\,.
\label{eq:dr2}
\end{equation}
This means that the neutrino drag can be expected to
cause a damping influence on the fluid motions on relevant
scales roughly in the ballpark of the numerical viscosity
effects discussed in Section~\ref{sec:viscosity}. However, in contrast
to the Reynolds number, which decreases with smaller spatial
scales, the drag number behaves in the opposite way. Assuming
a turbulent flow with a Kolmogorov spectrum, i.e.\
with a velocity $v_\lambda = v(\lambda/l)^{1/3}$ on spatial
scales $\lambda$, one can write
\begin{equation}
\mathrm{Re}(\lambda) = \frac{v_\lambda\,\lambda}{\mu_\mathrm{vis}}
= \frac{v\,l}{\mu_\mathrm{vis}}\,
\left(\frac{\lambda}{l}\right)^{4/3}\,,
\label{eq:re2}
\end{equation}
whereas we get
\begin{equation}
\mathrm{Dr}(\lambda) = \frac{v_\lambda}{\lambda\,\Gamma}
= \frac{v}{l\,\Gamma}\,
\left(\frac{l}{\lambda}\right)^{2/3}\,.
\label{eq:dr3}
\end{equation}
On small scales, i.e.\ for $\lambda$ decreasing, the neutrino drag loses its
damping influence because the long neutrino mean free paths prevent frequent
neutrino interactions in small volumes, different from the action of viscous
shear in the fluid. While on the largest scales in the gain layer Dr $\sim$ 100
(Eq.~(\ref{eq:dr2})), one expects for $l/\lambda \sim 100$, i.e.\ on the smallest
well-resolved scales in the current simulations (represented by a few grid
cells), values of $\mathrm{Dr}(\lambda) \sim 2000$. Overall, these numbers are
compatible with numerical Reynolds numbers in our best-resolved 3D simulations
(see Section~\ref{sec:viscosity} and Table~\ref{tab:Re}).
We will evaluate the neutrino-drag more quantitatively on grounds of results from full-fledged neutrino transport calculations in Section~\ref{sec:evaluation_s20} and compare it with the magnitude of numerical-viscosity effects.

\subsection{Detailed evaluation}
\label{sec:evaluation}

For a more accurate quantitative analysis we will now
evaluate the terms on the rhs of Eq.~(\ref{eq:radforce}) 
with the conditions in the gain layer in greater detail.  

\subsubsection{Interaction coefficients}
\label{sec:iacoeffs}

The most relevant neutrino interaction processes for 
momentum transfer around the gain layer (dominant 
on a level of $>$95\%) are $\nu_e$ and $\bar\nu_e$ 
absorption on free neutrons and protons, respectively,
and, involving neutrinos of all species, neutrino-nucleon 
scattering, as well as coherent neutrino scattering off
nuclei. The last process is relevant only at conditions
where nuclei are present, i.e., in the undissociated 
material in the infall region ahead of the supernova shock
and at temperatures $T\lesssim 1$\,MeV behind the shock,
where nucleons begin to recombine to $\alpha$-particles
and later to heavier nuclei. 
 
Accordingly, $\chi_0$ and $\chi_0^\mathrm{t}$ include
additive contributions from all of the mentioned reactions,
evaluated in the rest frame of the stellar fluid.
The corresponding opacities, to lowest order in terms of 
the ratios of neutrino energy $\epsilon$ to charged-fermion
rest-mass energies, are 
\citep[e.g.,][]{Tubbs1975,Freedman1977,Bruenn1985,Janka1991}: 
\begin{eqnarray}
\kappa_\mathrm{a}^\ast(\epsilon) &\approx& 
\kappa_\mathrm{a}(\epsilon) = \frac{1+3g_\mathrm{A}^2}{4}\,
\sigma_0 \left(\frac{\epsilon}{m_ec^2}\right)^{\!2} n_{\{n,p\}}\,, 
\label{eq:kapa}\\
\kappa_{\mathrm{s},n}(\epsilon) &=& \frac{1+3g_\mathrm{A}^2}{16}\,
\sigma_0 \left(\frac{\epsilon}{m_ec^2}\right)^{\!2} n_n \,,
\label{eq:kapsn}\\
\kappa_{\mathrm{s},p}(\epsilon) &=& 
\frac{4(2\sin^2\!\theta_\mathrm{W}\!-\!
\frac{1}{2})^2+3g_\mathrm{A}^2}{16}\,
\sigma_0\!\left(\!\frac{\epsilon}{m_ec^2}\!\right)^{\!2}\! n_p \,,
\label{eq:kapsp}\\
\kappa_{\mathrm{s},A}(\epsilon) &=& 
\frac{A^2}{16}\!\left[2\sin^2\!\theta_\mathrm{W}
\!-\!(1\!-\!2\sin^2\!\theta_\mathrm{W})\!
\left(\!2\frac{Z}{A}\!-\!1\!\right)\right]^{\!2}\!\times\nonumber\\
&\phantom{=}&\phantom{\frac{A^2}{16}}\!\times
\sigma_0 \left(\frac{\epsilon}{m_ec^2}\right)^{\!2} n_A \,,
\label{eq:kapsA}
\end{eqnarray}
where $m_ec^2 \approx 0.511$\,MeV is the rest-mass energy
of electrons, $\sigma_0 \approx 1.76\times 10^{-44}$\,cm$^2$,
$g_\mathrm{A}\approx 1.26$, $\theta_\mathrm{W}$ the Weinberg
angle with $\sin^2\theta_\mathrm{W}\approx 0.2315$, 
and $n_n$, $n_p$, and $n_A$ are
the number densities of free neutrons, free protons, and 
nuclei with mass number $A$ and charge number $Z$, respectively.
In Eq.~(\ref{eq:kapa}) the target particles are neutrons, $n$,
for $\nu_e$ and protons, $p$, for $\bar\nu_e$. In the same 
equation, $\kappa_\mathrm{a}$ and $\kappa_\mathrm{a}^\ast$
are distinguished by a factor $\Psi(\epsilon)$ appearing in 
the latter quantity. It is defined by 
\citep[e.g.,][]{Cernohorsky1989}
\begin{equation}
\Psi(\epsilon) = \frac{1-f_{\{e^-,e^+\}}(\epsilon)}
{1-f_{\{\nu_e,\bar\nu_e\}}(\epsilon)} \,,
\label{eq:fermiblock}
\end{equation}
where $f_i(\epsilon) = [1+\exp(\epsilon/T-\psi_i)]^{-1}$ 
are the equilibrium phase-space distributions of the
fermions $i$ with degeneracy parameters $\psi_i$ and
gas temperature $T$ (measured in energy units). This factor
accounts for fermion-phase space blocking (and ensures
detailed balance). Because of the high temperatures and
relatively low densities, fermions in the gain layer are
nondegenerate and therefore $\Psi \approx 1$. For this 
reason we can use $\kappa_\mathrm{a}$ instead of
$\kappa_\mathrm{a}^\ast$.

For the scattering processes, the opacities for momentum
transfer (``transport opacities''), which appear in the 
first term on the rhs of Eq.~(\ref{eq:radforce}), are
defined by the angle integral of the differential opacity
including the additional factor $(1-\mu)$ in the 
integrand ($\mu = \cos\vartheta_\mathrm{s}$ with 
$\vartheta_\mathrm{s}$ being the scattering angle between
ingoing and outgoing neutrino):
\begin{equation}
\kappa_\mathrm{s}^\mathrm{t} \defeq \oint\mathrm{d}\omega
\,\frac{\mathrm{d}\kappa_\mathrm{s}}{\mathrm{d}\omega}\,
(1-\mu) 
\label{eq:traop}
\end{equation} 
\citep{Cernohorsky1989,Janka1991}.
Performing this integration one gets:
\begin{eqnarray}
\kappa_{\mathrm{s},n}^\mathrm{t}(\epsilon) &=& 
\frac{1+5g_\mathrm{A}^2}{24}\,
\sigma_0 \left(\frac{\epsilon}{m_ec^2}\right)^{\!2} n_n \,,
\label{eq:kapstn} \\
\kappa_{\mathrm{s},p}^\mathrm{t}(\epsilon) &=& 
\frac{4(2\sin^2\!\theta_\mathrm{W}\!-\!
\frac{1}{2})^2+5g_\mathrm{A}^2}{24}\,
\sigma_0\!\left(\!\frac{\epsilon}{m_ec^2}\!\right)^{\!2}\! n_p \,,
\label{eq:kapstp}\\
\kappa_{\mathrm{s},A}^\mathrm{t}(\epsilon) &=& 
\frac{A^2}{24}\!\left[2\sin^2\!\theta_\mathrm{W}
\!-\!(1\!-\!2\sin^2\!\theta_\mathrm{W})\!
\left(\!2\frac{Z}{A}\!-\!1\!\right)\right]^{\!2}\!\times\nonumber\\
&\phantom{=}&\phantom{\frac{A^2}{24}}\!\times
\sigma_0 \left(\frac{\epsilon}{m_ec^2}\right)^{\!2} n_A \,,
\label{eq:kapstA}
\end{eqnarray}
Expressing the number densities $n_i$ in terms of the number
fractions $Y_i$ or the mass fractions $X_i$, i.e.,
\begin{equation}
n_i = \frac{\rho}{m_\mathrm{B}}\,Y_i = 
\frac{\rho}{m_\mathrm{B}}\,\frac{X_i}{A_i}\,,
\label{eq:massfrac}
\end{equation}
and using $\sin^2\theta_\mathrm{W}\approx 0.25$, we can 
write the total scattering opacity $\kappa_\mathrm{s}$ as
\begin{equation}
\kappa_\mathrm{s}(\epsilon) = \kappa_{\mathrm{s},n} 
+ \kappa_{\mathrm{s},p} + \kappa_{\mathrm{s},A} 
= K_\mathrm{s}\,\sigma_0\,
\left(\frac{\epsilon}{m_ec^2}\right)^{\!2}\frac{\rho}{m_\mathrm{B}}
\label{eq:kapstot}
\end{equation}
with 
\begin{equation}
K_\mathrm{s} = \frac{1}{16}\left[(1+3g_\mathrm{A}^2)X_n+
3g_\mathrm{A}^2X_p + \sum_iN_i^2\frac{X_i}{A_i}\right]\,,
\label{eq:kstot}
\end{equation}
where the sum runs over all nuclear species including
and heavier than helium with neutron numbers $N_i=A_i-Z_i$.
Analogously, the total transport opacity for neutrino
scattering is
\begin{equation}
\kappa_\mathrm{s}^\mathrm{t}(\epsilon) = 
\kappa_{\mathrm{s},n}^\mathrm{t} 
+ \kappa_{\mathrm{s},p}^\mathrm{t} + 
\kappa_{\mathrm{s},A}^\mathrm{t} 
= K_\mathrm{s}^\mathrm{t}\,\sigma_0
\left(\frac{\epsilon}{m_ec^2}\right)^{\!2}\frac{\rho}{m_\mathrm{B}}
\label{eq:kapsttot}
\end{equation}
with 
\begin{equation}
K_\mathrm{s}^\mathrm{t} = \frac{1}{24}\left[(1+5g_\mathrm{A}^2)X_n+
5g_\mathrm{A}^2X_p + \sum_iN_i^2\frac{X_i}{A_i}\right]\,.
\label{eq:ksttot}
\end{equation}
Introducing the definition
\begin{equation}
K_{\mathrm{a},\{n,p\}} = \frac{1+3g_\mathrm{A}^2}{4}\,X_{\{n,p\}}\,,
\label{eq:ka}
\end{equation}
the absorption opacity is
\begin{equation}
\kappa_\mathrm{a}(\epsilon) = K_{\mathrm{a},\{n,p\}}\,\sigma_0
\left(\frac{\epsilon}{m_ec^2}\right)^{\!2}\frac{\rho}{m_\mathrm{B}}
\,,
\label{eq:kapak}
\end{equation}
where $K_{\mathrm{a},n}$ applies for $\nu_e$ and
$K_{\mathrm{a},p}$ for $\bar\nu_e$. Finally, the total opacity
$\chi_0$ and transport opacity $\chi_0^\mathrm{t}$, respectively, 
in the comoving frame of the fluid are
\begin{equation}
\chi_0^{(\mathrm{t})}(\epsilon) = K_\mathrm{tot}^{(\mathrm{t})}\,
\sigma_0
\left(\frac{\epsilon}{m_ec^2}\right)^{\!2}\frac{\rho}{m_\mathrm{B}}
\,,
\label{eq:kaptot}
\end{equation}
where
\begin{equation}
K_\mathrm{tot}^{(\mathrm{t})} = K_{\mathrm{a},\{n,p\}} +
K_\mathrm{s}^{(\mathrm{t})}\,.
\label{eq:ktot}
\end{equation}
With this we can straightforwardly obtain
\begin{equation}
\chi_0(\epsilon) + \epsilon\,
\frac{\partial\chi_0(\epsilon)}{\partial\epsilon}
= 3\,K_\mathrm{tot}\,\sigma_0
\left(\frac{\epsilon}{m_ec^2}\right)^{\!2}\frac{\rho}{m_\mathrm{B}}
\,,
\label{eq:chider}
\end{equation}
which is needed to evaluate the third integrand on the rhs of
Eq.~(\ref{eq:radforce}). 

In terms of the absorption and scattering opacities,
the true emission and scattering contributions to the 
emissivity coefficient in the second integral of 
Eq.~(\ref{eq:radforce}) read as follows 
\citep{Mihalas1984}:
\begin{eqnarray}
\frac{4\pi}{c}\,\eta_0(\epsilon) &=& 
\kappa_\mathrm{a}^\ast(\epsilon)\,E_0^\mathrm{eq}(\epsilon) + 
\kappa_\mathrm{s}(\epsilon)\,E_0(\epsilon) \nonumber \\
&\approx&
\kappa_\mathrm{a}(\epsilon)\,E_0^\mathrm{eq}(\epsilon) + 
\kappa_\mathrm{s}(\epsilon)\,E(\epsilon) \,,
\label{eq:emissivity}
\end{eqnarray}
where in the final expression we have again assumed 
that $\kappa_\mathrm{a}^\ast\approx \kappa_\mathrm{a}$
for the absorption opacity of $\nu_e$ and $\bar\nu_e$. 
$E(\epsilon) = \mathrm{d}E/\mathrm{d}\epsilon$ is the
differential energy density in the lab frame and 
connected to the energy density in the comoving frame
of the fluid by $E_0(\epsilon)=E(\epsilon)+{\cal O}(v/c)$.
The additional term of order $(v/c)$ is omitted because
it leads to a second-order correction in $(v/c)$. The
equilibrium energy density of $\nu_e$ or $\bar\nu_e$
is computed as
\begin{equation}
E_0^\mathrm{eq}(\epsilon) = 
\frac{\mathrm{d}E_0^\mathrm{eq}(\epsilon)}
{\mathrm{d}\epsilon} = \frac{4\pi}{(hc)^3}\,
\frac{\epsilon^3}{1 + \exp(\epsilon/T - \psi_\nu)}
\label{eq:energyeq}
\end{equation}
with the degeneracy parameter $\psi_\nu$ of the 
Fermi-Dirac spectrum ($h$ is Planck's constant).

\subsubsection{Emission model}
\label{sec:emmmodel}

In the following we assume that the neutron star 
radiates neutrinos as a spherical source. In this
case the radiation field is azimuthally invariant
around the radial direction and depends only on the 
radius as spatial coordinate and on the cosine of 
the angle $\theta$ relative to the radial direction,
$\mu = \cos(\theta)$. Moreover, the pressure
tensor $\mathbf{P}$ of the radiation field is 
diagonal with components $P_{rr}$, $P_{\theta\theta}$
and $P_{\phi\phi}$. This is a good approximation for
the situation in the gain layer even in the general
3D case, because the diagonal elements of $\mathbf{P}$
dominate the off-diagonal ones by at least an order
of magnitude \citep[see][]{Richers2017}.

With the radiation intensity $I(r;\epsilon,\mu)$ the
radiation moments (energy density, flux, and pressure)
can be computed as
\begin{eqnarray}
E(r;\epsilon) &=& 
\frac{\mathrm{d}E(\epsilon)}{\mathrm{d}\epsilon}
= \frac{1}{c}\oint\mathrm{d}\omega\,I = 
\frac{2\pi}{c}\int_{-1}^{+1}\!\!\mathrm{d}\mu\,I ,
\label{eq:erad1}\\
F(r;\epsilon) &=&
\frac{\mathrm{d}F(\epsilon)}{\mathrm{d}\epsilon}
= \oint\mathrm{d}\omega\,\mu\,I = 
2\pi\int_{-1}^{+1}\!\!\mathrm{d}\mu\,\mu\,I ,
\label{eq:frad1}\\
P(r;\epsilon) &=& 
\frac{\mathrm{d}P(\epsilon)}{\mathrm{d}\epsilon}
= \frac{1}{c}\oint\mathrm{d}\omega\,\mu^2 I = 
\frac{2\pi}{c}\!\!\int_{-1}^{+1}\!\!\!\!\mathrm{d}\mu\,\mu^2 I,~~~
\label{eq:prad1}
\end{eqnarray}
and the nonvanishing diagonal elements of the pressure
tensor are
\begin{eqnarray}
P_{rr}(\epsilon) &=& P(\epsilon)\,,\label{eq:prr} \\
P_{\theta\theta}(\epsilon) &=& P_{\phi\phi}(\epsilon) 
= \frac{1}{2}\,[E(\epsilon)-P(\epsilon)] \,.
\label{eq:pnornor}
\end{eqnarray}

For simplicity, let us also assume that the angular and 
energy distributions of the radiated neutrinos can be
separated, i.e., we make the following ansatz for the 
neutrino intensity:
\begin{equation}
I(r;\epsilon,\mu) \defeq \frac{c}{(hc)^3}\,{\cal N}(r)\,
\epsilon\,f_\alpha(\epsilon)\,g(\mu)\,.
\label{eq:nuintensity1}
\end{equation}
Here ${\cal N}(r)$ is a normalization factor and
$f_\alpha(\epsilon)$ is the well established normalized 
alpha-fit for the radiated neutrino-number spectrum
\citep[][and references therein]{Tamborra2014a}:
\begin{equation}
f_\alpha(\epsilon) = \frac{\epsilon^\alpha}{\Gamma_{\alpha+1}}
\left(\frac{\alpha + 1}{\langle\epsilon\rangle}\right)^{\!\alpha+1}
\exp\left[-(\alpha+1)\,\frac{\epsilon}{\langle\epsilon\rangle}\right]
\label{eq:alphafit}
\end{equation}
with the mean energy $\langle\epsilon\rangle$ and the gamma
function
\begin{equation}
\Gamma_{\alpha+1} \defeq \int_0^\infty\mathrm{d}x\,x^\alpha\,e^{-x}
= \alpha\,\Gamma_\alpha \,.
\label{eq:gamma}
\end{equation}
The first and second energy moments of $f_\alpha$ yield the
normalization relation and the mean energy:
\begin{equation}
\int_0^\infty \mathrm{d}\epsilon\,f_\alpha(\epsilon) = 1
\quad\mathrm{and}\quad
\int_0^\infty \mathrm{d}\epsilon\,\epsilon\,f_\alpha(\epsilon) = 
\langle\epsilon\rangle \,.
\label{eq:alphamoms}
\end{equation}

To approximate the angular distribution function $g(\mu)$
we assume that the neutrinos stream off isotropically
and freely from a sharp
neutrinosphere of radius $R_\nu$. Therefore at radius $r$ 
the emitted neutrinos fill a cone with half-opening angle
$\theta_\mathrm{max}$ isotropically. This angle is 
subtended by the radiating sphere when observed from 
distance $r$ and obeys the relation
\begin{equation}
\mu_\mathrm{min}(r) = \cos\theta_\mathrm{max}(r) =
\sqrt{1-\left(\frac{R_\nu}{r}\right)^{\! 2}} 
\label{eq:mumin}
\end{equation}
for $r\ge R_\nu$. We therefore construct $g(\mu)$ by
using the Heaviside function $\Theta(\mu-\mu_\mathrm{min})$
as
\begin{equation}
g(\mu) = \frac{\Theta(\mu-\mu_\mathrm{min})}
{2\pi\,(1-\mu_\mathrm{min})}\,,
\label{eq:gfunc}
\end{equation}
which satisfies the normalization condition
\begin{equation}
\oint\mathrm{d}\omega\,g(\mu) = \int_0^{2\pi}\mathrm{d}\phi
\int_{-1}^{+1}\mathrm{d}\mu\,g(\mu) = 1\,.
\label{eq:gnorm}
\end{equation}

The normalization factor ${\cal N}(r)$ in 
Eq.~(\ref{eq:nuintensity1}) can be obtained by making
use of the constraint that the angle-energy integral of
the intensity must yield the total luminosity $L$. In
our approximative emission model we assume that
the lab-frame luminosity $L$ as well as the 
corresponding neutrino energy spectrum and thus 
$\langle\epsilon\rangle$ are conserved quantities
exterior to the neutrinosphere:
\begin{equation}
L(r) = 4\pi\,r^2 \int_0^\infty \mathrm{d}\epsilon\,
F(r;\epsilon) = L(R_\nu) = L = \mathrm{const}\,.
\label{eq:lcons}
\end{equation}
Using now Eq.~(\ref{eq:frad1}) with Eqs.~(\ref{eq:nuintensity1}), 
(\ref{eq:alphafit}), and (\ref{eq:gfunc}), this yields:
\begin{equation}
{\cal N}(r) = \frac{L}{2\pi\,r^2\,c\,(hc)^{-3}\,
(1+\mu_\mathrm{min})\,\langle\epsilon\rangle}\,.
\label{eq:nnorm}
\end{equation}
Introducing this and Eq.~(\ref{eq:gfunc}) into 
Eq.~(\ref{eq:nuintensity1}), we finally obtain
\begin{equation}
I(r;\epsilon,\mu) = \frac{L}{4\pi^2r^2}\,\,
\frac{\epsilon}{\langle\epsilon\rangle}\,\,f_\alpha(\epsilon)\,\,
\frac{\Theta(\mu-\mu_\mathrm{min})}{1-\mu_\mathrm{min}^2(r)}\,.
\label{eq:nuintensity2}
\end{equation}

With the result of Eq.~(\ref{eq:nuintensity2}) we can now
compute the angular integrals of the radiation 
intensity in Eqs.~(\ref{eq:erad1})--(\ref{eq:prad1}),
\begin{eqnarray}
E(r;\epsilon) &=& \frac{L}{4\pi\,r^2c}\,
\frac{\epsilon}{\langle\epsilon\rangle}\,f_\alpha(\epsilon)\,
\frac{2}{1+\mu_\mathrm{min}} \,,
\label{eq:erad2}\\
F(r;\epsilon) &=& \frac{L}{4\pi\,r^2}\,
\frac{\epsilon}{\langle\epsilon\rangle}\,f_\alpha(\epsilon)\,,
\label{eq:frad2}\\
P(r;\epsilon) &=& \frac{L}{4\pi\,r^2c}\,
\frac{\epsilon}{\langle\epsilon\rangle}\,f_\alpha(\epsilon)\,\,
\frac{2}{3}\,\frac{1+\mu_\mathrm{min}+\mu_\mathrm{min}^2}
{1+\mu_\mathrm{min}} \,,
\label{eq:prad2}
\end{eqnarray}
as well as the corresponding energy-integrated radiation 
moments:
\begin{eqnarray}
E(r) &=& \frac{L}{4\pi\,r^2c}\,\frac{2}{1+\mu_\mathrm{min}}\,,
\label{eq:erad3}\\
F(r) &=& \frac{L}{4\pi\,r^2}\,,
\label{eq:frad3}\\
P(r) &=& \frac{L}{4\pi\,r^2c}\,\,
\frac{2}{3}\,\frac{1+\mu_\mathrm{min}+\mu_\mathrm{min}^2}
{1+\mu_\mathrm{min}} \,.
\label{eq:prad3}
\end{eqnarray}
Because of the separation of energy and angle dependence in
$I(r;\epsilon,\mu)$ the flux factor is energy independent,
\begin{eqnarray}
\xi(r) &=& \frac{F(r;\epsilon)}{E(r;\epsilon)\,c} =
\frac{F(r)}{E(r)\,c}=\frac{1}{2}\,\left[1+\mu_\mathrm{min}(r)\right]
\nonumber \\
&=& \frac{1}{2}
\left[1+\sqrt{1-\left(\frac{R_\nu}{r}\right)^{\!2}}\right] \,,
\label{eq:fluxfac}
\end{eqnarray}
with the limits $\xi(R_\nu) = \frac{1}{2}$ and $\xi(\infty) = 1$.

In order to evaluate the different terms of the radiation
force density, Eq.~(\ref{eq:radforce}), we also need
to compute the third energy moment of $f_\alpha(\epsilon)$:
\begin{equation}
\langle\epsilon^3\rangle =
\int_0^\infty\mathrm{d}\epsilon\,\epsilon^3\,f_\alpha(\epsilon)
= \frac{(\alpha+3)(\alpha+2)}{(\alpha+1)^2}\,
\langle\epsilon\rangle^3 \,.
\label{eq:eps3}
\end{equation}

\begin{table*}
        \caption{Numerical Reynolds numbers and estimates of gain-layer averaged 
neutrino-drag numbers for model s20.}
        \centering
        \newcommand{\ccc}[1]{\multicolumn{1}{c}{#1}}
        \begin{tabular}{lcccccc>{\bfseries}cr>{\bfseries}rrrrrrr}
                \hline
                \hline
                \ccc{$t_\mathrm{pb}$} & \ccc{$R_\mathrm{gain}$} & \ccc{$R_\mathrm{sh,min}$} & \ccc{$R_0$} & \ccc{$l$} &
\ccc{$v$} & \ccc{$\nu_\mathrm{N}$} & \ccc{Re} & \ccc{$\Gamma$} & \ccc{Dr} & \ccc{$L_{\nu_e}$} & \ccc{$L_{\bar{\nu}_e}$} & 
\ccc{$L_{\nu_x}$} & \ccc{$\kappa_{\nu_e}$} & \ccc{$\kappa_{\bar{\nu}_e}$} & \ccc{$\kappa_{\nu_x}$} \\
                \ccc{$\mathrm{[ms]}$} & \ccc{$\mathrm{[km]}$} & \ccc{$\mathrm{[km]}$} & \ccc{$\mathrm{[km]}$} & 
\ccc{$\mathrm{[km]}$} & \ccc{$\mathrm{[10^9\,cm/s]}$} & \ccc{$\mathrm{[10^{13}\,cm^2\,s^{-1}]}$} & & \ccc{$\mathrm{[s^{-1}]}$} &
& \multicolumn{3}{c}{$\mathrm{[10^{52}\,erg/s]}$} & \multicolumn{3}{c}{$\mathrm{[10^{-8}\,cm^{-1}]}$} \\
                \hline
                200 & 66 & 88 & 77 & 53 & 1.19 & 4.0 & 158 & 16.2 & 14 & 6.58 & 6.23 & 4.01 & 5.99 & 4.98 & 2.36 \\
                300 & 54 & 131 & 93 & 94 & 1.11 & 4.4 & 238 & 9.1 & 13 & 4.42 & 4.56 & 3.43 & 1.51 & 1.37 & 0.58 \\
                400 & 46 & 198 & 122 & 216 & 1.28 & 7.1 & 389 & 4.8 & 12 & 4.02 & 4.32 & 3.20 & 0.62 & 0.56 & 0.24 \\
                500 & 40 & 513 & 277 & 722 & 1.09 & 15.4 & 511 & 0.5 & 31 & 3.29 & 3.58 & 2.86 & 0.05 & 0.05 & 0.02 \\
                \hline
        \end{tabular}
        \begin{minipage}{\textwidth}
                \tablecomments{
$t_\mathrm{pb}$ is the post-bounce time, $R_\mathrm{gain}$ the average gain radius, $R_\mathrm{sh,min}$
the minimum shock radius, $R_0$ the arithmetic mean of $R_\mathrm{gain}$ and $R_\mathrm{sh,min}$,
$l$ the length scale of the largest turbulent vortices (Eq.~(\ref{eq:largesteddysize3})),
$v$ the characteristic nonradial velocity measured at $R_0$, $\nu_\mathrm{N}$ the kinematic 
numerical viscosity, Re the corresponding Reynolds number, $\Gamma$ the neutrino-damping rate
(Eq.~(\ref{eq:gamma4})), Dr the neutrino-drag number (Eq.~(\ref{eq:dr1}), using the listed
values of $\Gamma$, $v$ and $l$), $L_{\nu_i}$ are the luminosities
of neutrino species $\nu_i = \nu_e, \bar{\nu}_e, \nu_x$ in the lab frame, and 
$\kappa_{\nu_i} \equiv \langle\chi_0\rangle_{\nu_i}$ the 
spectrally-averaged neutrino opacities in the comoving frame of the fluid.
The numerical viscosity, $\nu_\mathrm{N}$, and corresponding Reynolds number,
Re, are evaluated at radius $R_0$ (as in Section~\ref{sec:turbulence}), whereas the 
neutrino-related quantities, $\Gamma$, $L_{\nu_i}$, and $\kappa_{\nu_i}$, are time-averaged
over $t_\mathrm{pb}\pm2.5\,\mathrm{ms}$ and spatially averaged over a radial shell
extending from $R_\mathrm{gain}+10\,\mathrm{km}$ to $R_\mathrm{sh,min}-10\,\mathrm{km}$.
}
        \end{minipage}
        \label{tab:drag1}
\end{table*}

\subsubsection{Radiation drag terms}
\label{sec:raddragtrms}

Using the opacities discussed in Section~\ref{sec:iacoeffs} 
and the neutrino-emission model introduced in 
Section~\ref{sec:emmmodel}, we now provide the final forms
for the three terms of the radiation force density on the 
rhs of Eq.~(\ref{eq:radforce}):
\begin{equation} 
\vect{G} = \vect{G}^{(1)} + \vect{G}^{(2)} + \vect{G}^{(3)}\,.
\label{eq:gsum}
\end{equation}
Each of the three summands contains components from all 
neutrino species:
\begin{equation}
\vect{G}^{(j)} = \vect{G}_{\nu_e}^{(j)} + \vect{G}_{\bar\nu_e}^{(j)}
+ 4\,\vect{G}_{\nu_x}^{(j)} \,,
\label{eq:gnusum}
\end{equation}
where $j = 1,\,2,\,3$ and we multiply the contribution of an 
individual type of heavy-lepton neutrino ($\nu_x$) by a factor
of 4 because of the near-equality of the emission
and interaction properties of the four species.

Because of the symmetry assumptions employed by us, the only
nonvanishing vector component of the radiation acceleration
is the radial one, $G_r^{(1)}$. It is given by
\begin{eqnarray}
G_{r,\nu_i}^{(1)} &=& K_\mathrm{tot,\nu_i}^\mathrm{t}\,
\frac{\rho}{m_\mathrm{B}}\,\sigma_0\,\frac{L_{\nu_i}}{4\pi\,r^2c}\times
\nonumber \\
&\phantom{=}&
\times
\left(\frac{(\alpha+3)(\alpha+2)}{(\alpha+1)^2}\right)_{\!\nu_i}\!
\left(\frac{\langle\epsilon\rangle_{\nu_i}}{m_ec^2}\right)^{\!2}\,
\label{eq:g1},
\end{eqnarray}
where $K_\mathrm{tot,\nu_e}^\mathrm{t} = K_{\mathrm{a},n}
+K_\mathrm{s}^\mathrm{t}$, $K_\mathrm{tot,\bar\nu_e}^\mathrm{t} = 
K_{\mathrm{a},p} +K_\mathrm{s}^\mathrm{t}$, and 
$K_\mathrm{tot,\nu_x}^\mathrm{t} = K_\mathrm{s}^\mathrm{t}$
with $K_{\mathrm{a},\{n,p\}}$ from Eq.~(\ref{eq:ka})
and $K_\mathrm{s}^\mathrm{t}$ from Eq.~(\ref{eq:ksttot}).

For the radiation drag associated with the emission term 
(true emission and outgoing scattering) we obtain 
\begin{eqnarray}
\vect{G}_{\nu_i}^{(2)} = &-&\frac{\vect{v}}{c}\,K_{\mathrm{a},\nu_i}\,
\frac{\rho}{m_\mathrm{B}}\,\frac{\sigma_0}{(m_ec^2)^2}\,\frac{4\pi}{(hc)^3}\,
T^6\,{\cal F}_5(\psi_{\nu_i}) \nonumber\\
&-&\frac{\vect{v}}{c}\,K_\mathrm{s}\,
\frac{\rho}{m_\mathrm{B}}\,\sigma_0\,\frac{L_{\nu_i}}{4\pi\,r^2c}\,
\frac{2}{1+\mu_\mathrm{min}(r)}\times \nonumber\\
&\phantom{-}&
\phantom{\frac{\vect{v}}{c}\,}
\times
\left(\frac{(\alpha+3)(\alpha+2)}{(\alpha+1)^2}\right)_{\!\nu_i}\!
\left(\frac{\langle\epsilon\rangle_{\nu_i}}{m_ec^2}\right)^{\!2}
\label{eq:g2}
\end{eqnarray}
with the Fermi integral 
${\cal F}_5(\psi_{\nu_i}) = \int_0^\infty\mathrm{d}x\,x^5
[1+\exp(x-\psi_{\nu_i})]^{-1}$ and the constants 
$K_{\mathrm{a},\nu_e} = K_{\mathrm{a},n}$, 
$K_{\mathrm{a},\bar\nu_e} = K_{\mathrm{a},p}$,
$K_{\mathrm{a},\nu_x} = 0$, and 
$K_\mathrm{s}$ from Eq.~(\ref{eq:kstot}).

Finally, making use of the diagonal nature of the radiation
pressure tensor in our emission model (Eqs.~(\ref{eq:prr}) and 
(\ref{eq:pnornor}) with Eq.~(\ref{eq:prad2})), we find for the force 
density of the radiation drag caused by momentum transfer through
neutrino absorption and ingoing scattering:
\begin{eqnarray}
\vect{G}_{\nu_i}^{(3)} = &-&\frac{\vect{v}}{c}\cdot \mathbf{M}\,\,
K_{\mathrm{tot},\nu_i}\,
\frac{\rho}{m_\mathrm{B}}\,\sigma_0\,\frac{L_{\nu_i}}{4\pi\,r^2c}\,
\frac{1}{1+\mu_\mathrm{min}(r)}\times \nonumber\\
&\phantom{-}&
\phantom{\frac{\vect{v}}{c}\,\frac{\rho}{m_\mathrm{B}}\,}
\times
\left(\frac{(\alpha+3)(\alpha+2)}{(\alpha+1)^2}\right)_{\!\nu_i}\!\!
\left(\frac{\langle\epsilon\rangle_{\nu_i}}{m_ec^2}\right)^{\!2}\!\!,
\label{eq:g3}
\end{eqnarray}
where $K_\mathrm{tot,\nu_e} = K_{\mathrm{a},n}
+K_\mathrm{s}$, $K_\mathrm{tot,\bar\nu_e} =
K_{\mathrm{a},p} +K_\mathrm{s}$, and
$K_\mathrm{tot,\nu_x} = K_\mathrm{s}$
with $K_{\mathrm{a},\{n,p\}}$ from Eq.~(\ref{eq:ka})
and $K_\mathrm{s}$ from Eq.~(\ref{eq:kstot}).
Only the diagonal elements of $\mathbf{M}$ do not vanish; they are
%
\begin{eqnarray}
M_{rr} &=& 2(1+\mu_\mathrm{min}+\mu_\mathrm{min}^2)\,,
\label{eq:mrr} \\
M_{\theta\theta} &=& M_{\phi\phi} = 
2-\mu_\mathrm{min}-\mu_\mathrm{min}^2 \,.
\label{eq:mnornor}
\end{eqnarray}

\begin{figure*}
        \centering
        \includegraphics[width=0.85\linewidth]{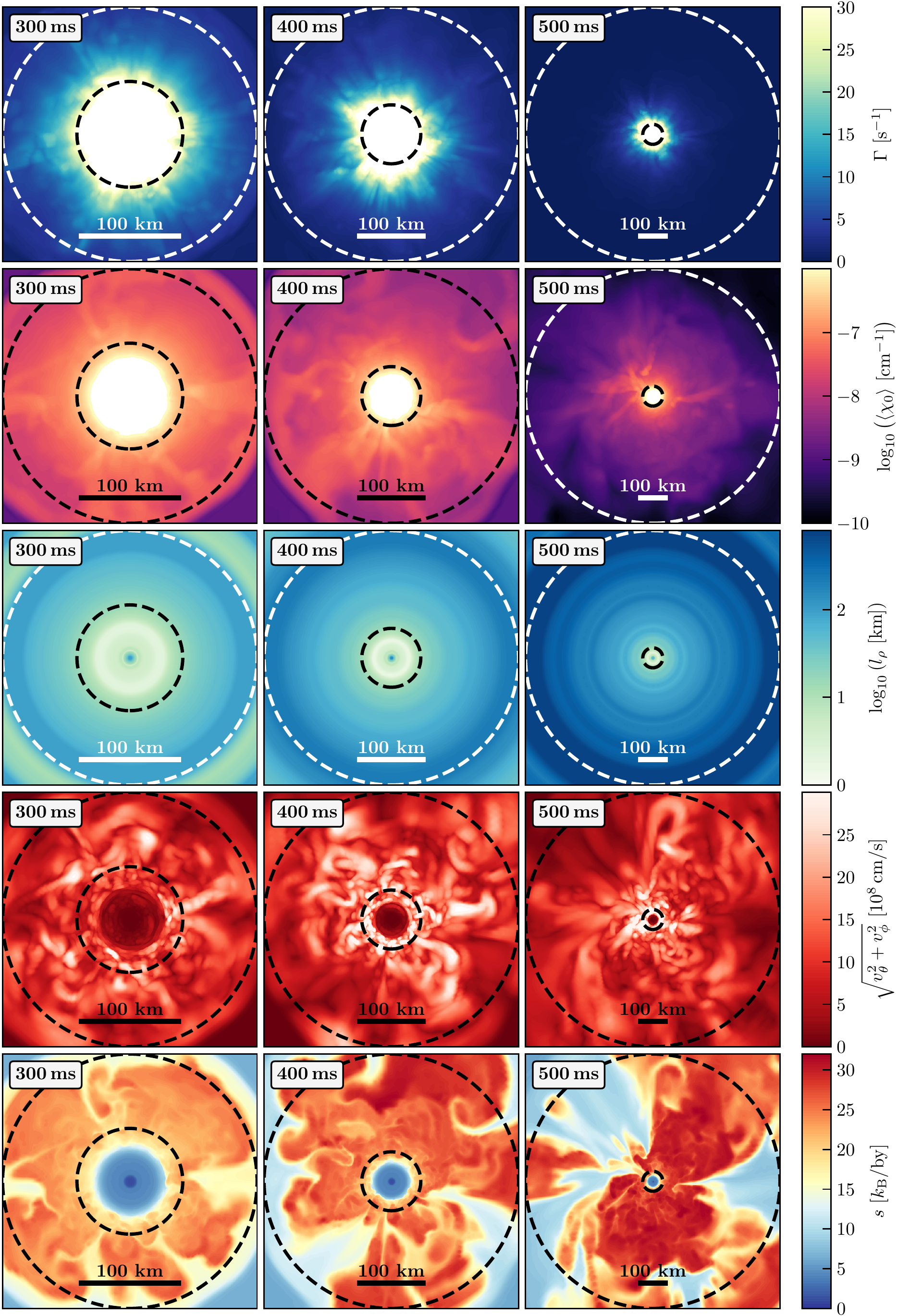}
        \caption{Cross-sectional cuts in the ($x$,$y$)-plane of model s20 with full neutrino transport
and uniform $2^\circ$ angular resolution at 300\,ms, 400\,ms, and 500\,ms after bounce. From \emph{top}
to \emph{bottom}, the damping rate associated with the neutrino drag, $\Gamma$ (see Eq.~(\ref{eq:gamma4})),
the spectrally-averaged neutrino opacity (in the fluid frame) summed over all species, $\langle\chi_0\rangle$,
the characteristic hydrodynamic length scale, $l_{\rho}$ (see Eq.~(\ref{eq:l_rho})), the nonradial velocity,
$(v_\theta^2 + v_\phi^2)^{0.5}$, and the entropy per baryon, $s$, are shown. The inner and outer dashed
circles indicate the angle-averaged gain radius and the minimum shock radius, respectively.}
        \label{fig:drag}
\end{figure*}

For the conditions in the gain layer the production of $\nu_e$ and $\bar\nu_e$
plays a subordinated role compared to their absorption. The first summand in
Eq.~(\ref{eq:g2}) is therefore negligible. From Eqs.~(\ref{eq:mrr}) and
(\ref{eq:mnornor}) it is clear that in the limit of radially streaming neutrinos
at large distances (i.e., $\mu_\mathrm{min}\to 1$) the nonradial components
vanish and therefore, to first order in $v/c$, radiation drag acts only on the
radial velocity component. Nonradial effects depend on Doppler shifts of the
neutrino quantities and have dropped out in Eq.~(\ref{eq:radforce}), because
they are of higher than linear order in $v/c$.  During the accretion phase of
the stalled supernova shock the spectral shape parameters $\alpha$ are typically
between 2 and 3 \citep{Mirizzi2016,Tamborra2012}. The case of $\alpha = 2$
corresponds to a Maxwell-Boltzmann spectrum, $\alpha\approx 2.3$ describes a
Fermi-Dirac spectrum with vanishing degeneracy parameter ($\psi = 0$), and
$\alpha > 2.3$ yields pinched spectra with a faster decline beyond the maximum
than the Fermi-Dirac distribution.

\begin{table}
        \caption{Local evaluation of the neutrino-drag number, Dr, at different radii
$R_\mathrm{ev}$ for model s20.
        }
        \centering
        \newcommand{\ccc}[1]{\multicolumn{1}{c}{#1}}
        \newcommand{\cccr}[1]{\multicolumn{1}{c|}{#1}}
        \begin{tabular}{lrrrr>{\bfseries}rrrr}
                \hline
                \hline
                \ccc{$t_\mathrm{pb}$} & \ccc{$R_\mathrm{ev}$} & \ccc{$l_{\rho}$} & \ccc{$v_9$} &
\ccc{$\Gamma$} & \ccc{Dr} & \ccc{$\kappa_{\nu_e}$} & \ccc{$\kappa_{\bar{\nu}_e}$} & \ccc{$\kappa_{\nu_x}$} \\
                \ccc{$\mathrm{[ms]}$} & \ccc{$\mathrm{[km]}$} & \ccc{$\mathrm{[km]}$} & &
\ccc{$\mathrm{[s^{-1}]}$} & & \multicolumn{3}{c}{$\mathrm{[10^{-9}\,cm^{-1}]}$} \\
                \hline
                200 & 76 & 29 & 1.19 & 16.7 & 24 & 63.2 & 52.1 & 24.7 \\
                & 78 & 31 & 1.19 & 15.7 & 25 & 58.0 & 48.4 & 22.9 \\
                \cline{2-9}
                & $R_0$ & 31 & 1.19 & 16.4 & 24 & 61.4 & 50.9 & 24.1 \\
                & avg & 30 & 1.18 & 16.2 & 24 & 59.9 & 49.8 & 23.6 \\
                \hline
                300 & 64 & 23 & 1.32 & 20.5 & 28 & 40.7 & 35.3 & 14.5 \\
                & 100 & 55 & 1.03 & 8.1 & 23 & 12.7 & 11.5 & 4.9 \\
                & 121 & 94 & 0.94 & 5.4 & 19 & 9.4 & 8.5 & 3.7 \\
                \cline{2-9}
                & $R_0$ & 45 & 1.11 & 9.5 & 26 & 15.0 & 13.6 & 5.8 \\
                & avg & 60 & 1.07 & 9.1 & 20 & 15.1 & 13.7 & 5.8 \\
                \hline
                400 & 56 & 19 & 1.75 & 30.1 & 30 & 46.6 & 40.4 & 16.1 \\
                & 100 & 48 & 1.44 & 8.5 & 35 & 10.1 & 9.2 & 3.9 \\
                & 150 & 101 & 1.06 & 3.4 & 31 & 4.3 & 3.9 & 1.7 \\
                & 188 & 215 & 0.90 & 1.9 & 22 & 2.8 & 2.6 & 1.1 \\
                \cline{2-9}
                & $R_0$ & 69 & 1.28 & 5.3 & 35 & 6.3 & 5.8 & 2.5 \\
                & avg & 113 & 1.16 & 4.8 & 21 & 6.2 & 5.6 & 2.4 \\
                \hline
                500 & 50 & 17 & 2.34 & 34.1 & 41 & 41.0 & 36.1 & 14.1 \\
                & 100 & 52 & 1.55 & 7.6 & 39 & 6.5 & 6.0 & 2.5 \\
                & 150 & 84 & 1.35 & 2.9 & 55 & 2.6 & 2.4 & 1.0 \\
                & 200 & 115 & 1.25 & 1.4 & 79 & 1.3 & 1.2 & 0.5 \\
                & 250 & 185 & 1.15 & 0.8 & 82 & 0.7 & 0.7 & 0.3 \\
                & 300 & 217 & 1.03 & 0.4 & 118 & 0.4 & 0.4 & 0.2 \\
                & 350 & 320 & 0.96 & 0.2 & 123 & 0.3 & 0.3 & 0.1 \\
                & 400 & 420 & 0.89 & 0.1 & 142 & 0.2 & 0.2 & 0.1 \\
                & 450 & 636 & 0.82 & 0.1 & 125 & 0.1 & 0.1 & 0.1 \\
                & 500 & 722 & 0.77 & 0.1 & 165 & 0.1 & 0.1 & 0.1 \\
                \cline{2-9}
                & $R_0$ & 182 & 1.09 & 0.5 & 112 & 0.5 & 0.5 & 0.2 \\
                & avg & 438 & 0.95 & 0.5 & 45 & 0.5 & 0.5 & 0.2 \\
                \hline
        \end{tabular}
        \begin{minipage}{\columnwidth}
        \tablecomments{
$t_\mathrm{pb}$ is the post-bounce time, $R_\mathrm{ev}$ the radius of the evaluation,
$l_{\rho}$ the local characteristic hydrodynamical length scale (Eq.~(\ref{eq:l_rho})),
$v_9 = v/(10^9\,\mathrm{cm\,s^{-1}})$ the typical local turbulent (nonradial) velocity,
$\Gamma$ the local neutrino-damping rate (Eq.~(\ref{eq:gamma4})), Dr the corresponding
local neutrino-drag numbers (Eq.~(\ref{eq:dr1})), and
$\kappa_{\nu_i} \equiv \langle\chi_0\rangle_{\nu_i}$ are the spectrally-averaged local
neutrino opacities in the fluid frame for $\nu_i = \nu_e, \bar{\nu}_e, \nu_x$.
All quantities are averaged over radial shells of 10\,km
thickness around $R_\mathrm{ev}$ and time-averaged over 5\,ms.
Also listed are the values at radius $R_0$, i.e., half-way between the mean gain radius and 
the minimum shock radius as given in Table~\ref{tab:drag1}, and averages (``avg'') over
the entire shell of $[R_\mathrm{gain}+10\,\mathrm{km},R_\mathrm{sh,min}-10\,\mathrm{km}]$.
For the last case the values of $\Gamma$ and $\kappa_{\nu_i}$
are identical with those in Table~\ref{tab:drag1}.}
        \end{minipage}
        \label{tab:drag2}
\end{table}

\newpage
\subsection{Model-based evaluation}
\label{sec:evaluation_s20}

In order to quantify the neutrino-drag effects in numerical simulations, we
analyse our 3D model s20 with full neutrino transport and uniform angular
resolution of $2^\circ$ (see Table~\ref{tab:models}).
Since \textsc{Prometheus-Vertex} contains a Newtonian hydrodynamics solver, we 
consider ---in the spirit of the remark in the footnote in Section~\ref{sec:equations}--- 
only the third term on the rhs of Eq.~(\ref{eq:radforce}).
With their ray-by-ray approach to neutrino transport, the simulations provide results 
compatible with the symmetry assumptions for the neutrino-radiation field considered in 
Section~\ref{sec:evaluation}.

For the quantitative analysis, we apply several simplifications, which ease the 
evaluation and are acceptable in view of other approximations connected to the 
ray-by-ray transport, which evolves
only the radial components of the neutrino fluxes and assumes the neutrino phase-space
distributions to be axially symmetric around the radial direction. First, we 
focus on the radial term of the neutrino drag, which dominates the nonradial ones, 
as can be seen by comparing Eq.~(\ref{eq:prr}) with Eq.~(\ref{eq:pnornor}) and 
Eq.~(\ref{eq:mrr}) with Eq.~(\ref{eq:mnornor}).
Second, we do not transform the radiation quantities, in particular the 
radiation-pressure tensor, $\mathbf{P}(\epsilon)$, into the laboratory frame. Instead, we 
evaluate the drag terms with the pressure tensor computed in the comoving frame 
of the fluid by the transport solver, in line with the available opacities $\chi_0$.
This simplification is justified because on the one hand the typical fluid velocities 
in the postshock layer are at most 5--10\,\% of the speed of light, and on the other 
hand effects of the transformation between lab frame and comoving frame act in opposite 
directions in convective downdrafts and outflows and therefore partly compensate each 
other in the angle-averaged quantities considered here. Third, we use only the nonradial
fluid velocities in the postshock layer as a proxy of the turbulent velocities in order 
to facilitate the direct comparison with the numerical viscosity and Reynolds number
along the lines of their evaluation in Sect.~\ref{sec:turbulence}.

Therefore, using Eq.~(\ref{eq:dragacc}), we compute the damping rate associated 
with the neutrino drag according to
\begin{eqnarray}
\Gamma &=& \frac{|G_\mathrm{drag}|}{\rho\,v} = \nonumber\\
&=& \frac{1}{\rho\,c} \sum_{\nu_i}
\int_0^\infty\mathrm{d}\epsilon\left[\chi_{0,\nu_i} + \epsilon
\frac{\partial\chi_{0,\nu_i}}{\partial\epsilon}\right]P_{\nu_i}(\epsilon)\,,
\label{eq:gamma4}
\end{eqnarray}
with $P_{\nu_i}(\epsilon)$ given by the transport solver as integral of the 
direction-dependent neutrino intensities according to Eq.~(\ref{eq:prad1}).
The neutrino-drag number, Dr, can then be calculated with Eq.~(\ref{eq:dr1}). 
To directly compare Dr with the numerical Reynolds numbers, Re, of our simulation, 
we employ the same definitions for the characteristic turbulent velocity, $v$, and 
for the length scale of the largest turbulent eddies,
$l$, as in Section~\ref{sec:MKJ}, namely
\begin{equation}
v^2 = \frac{1}{4\pi}\int\mathrm{d}\Omega\,(v_{\theta}^2+v_{\phi}^2)\,,
\label{eq:charcteristicvelocity2}
\end{equation}
\begin{equation}
l = R_\mathrm{sh} - R_\mathrm{gain}\,,
\label{eq:largesteddysize3}
\end{equation}
where $R_\mathrm{sh}$ and $R_\mathrm{gain}$ are the angle-averaged shock radius and gain 
radius, respectively. Note that we use a different naming convention here to avoid 
confusion of the length scale, $l$, with the neutrino luminosity, $L$.

In Table~\ref{tab:drag1} we list the neutrino-drag numbers, averaged over the gain
layer of model s20, at different post-bounce times, $t_\mathrm{pb}$, as well as the 
numerical Reynolds numbers, Re, and numerical viscosities, $\nu_\mathrm{N}$, 
obtained with our MKJ method based on the energy dissipation
rate (see Section~\ref{sec:MKJ}). As in Section~\ref{sec:turbulence}, all hydrodynamical 
quantities are measured at radius $R_0$ (Eq.~(\ref{eq:r0})) half-way between the 
average gain radius and the minimum shock radius, $R_\mathrm{sh,min}$, and
smoothed by performing an average over a radial shell of $R_0\pm5\,\mathrm{km}$ and
an additional time average over the interval $t_\mathrm{pb}\pm2.5\,\mathrm{ms}$.
The neutrino-drag numbers, Dr, are computed according to Eq.~(\ref{eq:dr1}) with the
neutrino-damping rate $\Gamma$ from Eq.~(\ref{eq:gamma4}) and with the same values 
of $v$ and $l$ as used in the evaluation of the numerical Reynolds numbers.
$L_{\nu_i}$ are the neutrino luminosities (of species $\nu_i = \nu_e, \bar{\nu}_e, \nu_x$)
in the lab frame and $\kappa_{\nu_i} \equiv \langle\chi_0\rangle_{\nu_i}$ the
spectrally-averaged neutrino opacities in the comoving frame of the stellar fluid.
The neutrino-related quantities, $\Gamma$, $L_{\nu_i}$, and $\kappa_{\nu_i}$, are spatially
averaged over a radial shell extending from $R_\mathrm{gain}+10\,\mathrm{km}$ to
$R_\mathrm{sh,min}-10\,\mathrm{km}$ and time-averaged over 
$t_\mathrm{pb}\pm2.5\,\mathrm{ms}$.

Since the matter density $\rho$, the neutrino opacities, and the neutrino radiation
quantities, all of which enter the computation of the damping rate $\Gamma$ 
(see Eq.~(\ref{eq:gamma4})) and drag number Dr (Eq.~(\ref{eq:dr1})), decline
with increasing radius, we also investigate the radial dependence of the neutrino 
drag in the gain layer. Thus we intend to test ambiguities in the determination of
the neutrino-drag effects in terms of the characteristic parameter Dr.
Figure~\ref{fig:drag} displays some of the relevant quantities
in a cross-sectional plane at different post-bounce times to demonstrate their
spatial variations in connection with the convective vortex flows. In order to 
perform a radius-dependent analysis, we define a characteristic local hydrodynamic
length scale, constrained by the width of the gain layer, as
\begin{equation}
l_{\rho} = \min\left(\frac{\bar{\rho}}{\left|\mathrm{d}\bar{\rho}/\mathrm{d}r\right|}\,,
\,R_\mathrm{sh} - R_\mathrm{gain}\right),
\label{eq:l_rho}
\end{equation}
where $\bar{\rho}$ is the angle-averaged matter density. Moreover, we use local, 
shell-averaged values of the nonradial (turbulent) velocity, $v$, and of the damping rate, 
$\Gamma$ (Eq.~(\ref{eq:gamma4})), 
in Eq.~(\ref{eq:dr1}) for estimating local values of Dr. The corresponding results
are listed in Table~\ref{tab:drag2}, where all local quantities are time-averaged over 5\,ms
and radially averaged over shells of 10\,km thickness around the locations of evaluation, 
$R_\mathrm{ev}$.

The data for the neutrino-drag number Dr in Tables~\ref{tab:drag1} and \ref{tab:drag2} 
are consistent with our order-of-magnitude estimate of Eq.~(\ref{eq:dr2}). Typical
values are several 10 up to more than 100 in the gain layer. Naturally, the drag 
numbers become particularly small when computed with the radial width of the entire gain
layer in Table~\ref{tab:drag1}, but also the local values of Table~\ref{tab:drag2}
are only a factor of $\sim$2 bigger, except at large radii at 500\,ms, where the
neutrino opacities and neutrino pressure $P(r)$, and correspondingly the 
neutrino-damping rate, become small. A comparison of the numerical Reynolds numbers, Re,
and the neutrino-drag numbers, Dr, in Table~\ref{tab:drag1}
confirms that on the scale of large and largest
turbulent eddies in our 2$^\circ$ simulation of model s20,
the neutrino drag clearly dominates the effects of numerical viscosity.
Only on scales smaller than $l \le \sqrt{\nu_\mathrm{N}/\Gamma}$ does damping by
numerical viscosity become compatible with the influence of the neutrino drag on 
fluid motions.
This corresponds to vortex flows of less than 10--20\,km diameter near the gain radius
($R_\mathrm{gain}\sim 50$--70\,km) and of less than 50--60\,km near the shock at a 
radius of about 200\,km.

Neutrino-damping rates of $\sim5$--10\,s$^{-1}$ in the gain layer suggest typical
damping timescales of large-scale fluid motions on the order of 100--200\,ms. Such
timescales may be short enough to slightly delay or hamper the onset of
postshock convection. A quantitative assessment of such effects and their 
potential consequences for the onset of an explosion would require a comparison
with the model-specific growth timescales of convection or SASI in the postshock layer.
However, with large-scale pre-collapse perturbations in the convective silicon- and oxygen-burning
shells of the progenitor \citep[e.g.,][]{Couch2013a,Mueller2015a,Couch2015a,Mueller2016,
Yoshida2019,Yadav2019}, the driving force for the growth of instabilities
\citep[e.g.,][]{Mueller2017,Kazeroni2019} should be sufficiently strong to render
neutrino drag a subdominant effect.

Our conclusions should not be severely affected by the simplifications made 
in our analysis as mentioned at the beginning of this
section. Although considering nonradial velocities might moderately
underestimate the neutrino-drag number (because the radial velocities may be
somewhat higher), a rigorous evaluation of the neutrino drag requires to
sum up all components of the vector-matrix product in the third term on the
rhs of Eq.~(\ref{eq:radforce}) with nonvanishing offdiagonal matrix components. 
This would increase the neutrino-damping rate 
and reduce the drag number, thus countersteering the velocity dependence.
However, in order to account for the effects of the
neutrino drag fully quantitatively, one needs a transport treatment that does not
only include all velocity-dependent terms at least up to first order in $v/c$,
but in addition it also requires a scheme that reaches beyond the ray-by-ray 
approximation and therefore provides reliable values for the offdiagonal 
elements of the neutrino pressure tensor, too. All of these terms must be 
included with proper frame dependences in the momentum equation of the stellar 
plasma as discussed in Section~\ref{sec:equations}.

\bibliographystyle{aasjournal}
\bibliography{paper}

\end{document}